\documentclass[11pt,a4paper]{article}
\pdfoutput=1
\usepackage{amssymb,amsmath,amsfonts, mathtools, mathrsfs}
\usepackage[utf8]{inputenc} 
\usepackage[dvipsnames]{xcolor}

\newif\ifnatbibsort\natbibsorttrue

\relax

\ifnatbibsort\RequirePackage[numbers,sort&compress]{natbib}\else\RequirePackage[numbers,compress]{natbib}\fi
\RequirePackage[colorlinks=true
,urlcolor=blue
,anchorcolor=blue
,citecolor=blue
,filecolor=blue
,linkcolor=blue
,menucolor=blue
,pagecolor=blue
,linktocpage=true
,pdfproducer=medialab
,pdfa=true
]{hyperref}

\usepackage{graphicx}
\usepackage{caption}
\usepackage{tensor}
\usepackage{subfigure}
\usepackage{enumerate}
\usepackage{dsfont}
\usepackage{verbatim}
\usepackage{amsfonts,euscript,amssymb,stmaryrd,braket}
\usepackage{tikz}
\usetikzlibrary{arrows,decorations.markings,patterns}
\usepackage{slashed}

\def\clock{{\count0=\time
		\divide\count0 60
		\ifnum\count0<10 0\fi\the\count0
		\multiply\count0 -60 \advance\count0 \time
		:\ifnum\count0<10 0\fi \the\count0
}}
\newcommand{\timestamp}{{\small\vbox{\hbox{\tt\jobname.tex}
			\hbox{\the\day/\the\month/\the\year, \clock}}}}

\newcommand{\nn}{\nonumber}
\newcommand{\bea}{\begin{eqnarray}}
\newcommand{\eea}{\end{eqnarray}}

\newcommand{\be}{\begin{equation}}
\newcommand{\ee}{\end{equation}}

\makeatletter
\let\old@startsection=\@startsection
\let\oldl@section=\l@section
\renewcommand{\@startsection}[6]{\old@startsection{#1}{#2}{#3}{#4}{#5}{#6\mathversion{bold}}}
\renewcommand{\l@section}[2]{\oldl@section{\mathversion{bold}#1}{#2}}
\makeatother

\numberwithin{equation}{section}

\usepackage{color}

\def \RR {{\mathbb R}}

\def\ri {{\rm i}}
\def\rd {{\rm d}}
\def\e {{\rm e}}

\begin{document}
	\renewcommand{\thefootnote}{\arabic{footnote}}

	\overfullrule=0pt
	\parskip=2pt
	\parindent=12pt
	\headheight=0in \headsep=0in \topmargin=0in \oddsidemargin=0in

	\vspace{ -3cm} \thispagestyle{empty} \vspace{-1cm}
	\begin{flushright} 
		\footnotesize
		\textcolor{red}{\phantom{print-report}}
	\end{flushright}

\begin{center}
	\vspace{.0cm}

	{\Large\bf \mathversion{bold}
	Entanglement Hamiltonian for the massless Dirac field 
	}
	\\
	\vspace{.25cm}
	\noindent
	{\Large\bf \mathversion{bold}
	on a segment with an inhomogeneous background 
	}
	\\

	\vskip  0.8cm
	{
		Erik Tonni$^{\,a}$
		and 
            Stefano Trezzi$^{\,b}$
	}
	\vskip  1.cm
	
	\small
	{\em
		$^{a}\,
            $SISSA and INFN Sezione di Trieste, via Bonomea 265, 34136, Trieste, Italy 
		\vskip .3cm
            $^{b}\,
            $Departament de Física Quàntica i Astrofísica, Institut de Ciències del Cosmos
            \\
        Universitat de Barcelona, Martí i Franquès 1, E-08028 Barcelona, Spain 
	}
	\normalsize

\end{center}

\vspace{0.3cm}

\begin{abstract}

We study the entanglement Hamiltonian of an interval 
for the massless Dirac field  in an inhomogeneous background on a segment
where the same boundary condition at both its endpoints is imposed,
and in its ground state.
We focus on a class of metrics that are Weyl equivalent to the flat metric through a Weyl factor that depends only on the spatial coordinate.
The explicit form of the entanglement Hamiltonian is written as the sum of a local and a bilocal term.
The weight function of the local term allows us to study a contour function for the entanglement entropies.
For the model obtained from the continuum limit of the rainbow chain, the analytic expressions are compared with exact numerical results from the lattice, 
showing an excellent agreement.
\end{abstract}

\newpage
\tableofcontents

\section{Introduction}

The geometric bipartite entanglement between complementary spatial regions can be explored through the corresponding reduced density matrices or, equivalently, the corresponding entanglement Hamiltonians, which provide the modular Hamiltonian of the bipartition
\cite{Haag:1992hx, Casini:2009sr, EislerPeschel:2009review}.
Consider a quantum system in a state characterised by its density matrix $\rho$
and a geometric bipartition of its space identified by a region $A$ and its complement $B$.
By assuming that the Hilbert space $\mathcal{H}$ can be factorised accordingly, namely $\mathcal{H} = \mathcal{H}_A \otimes \mathcal{H}_B$, 
the reduced density matrices of $A$ and $B$ are defined as  
$\rho_A \equiv \textrm{Tr}_{\mathcal{H}_B} \rho$ 
and $\rho_B \equiv \textrm{Tr}_{\mathcal{H}_A} \rho$ respectively,
with the normalisation condition $\textrm{Tr}_{\mathcal{H}_A} \rho_A = \textrm{Tr}_{\mathcal{H}_B} \rho_B =1$. 
Since $\rho_A$ is hermitian and positive semidefinite, 
when the zero eigenvalue is not in its spectrum 
it can be written as $\rho_A \propto \e^{-K_A} $, 
where the hermitian  operator $K_A$ is the entanglement Hamiltonian 
of the region $A$. 
The spectrum of $\rho_A$, which is known as the  entanglement spectrum,
provides important quantities like the R\'enyi entropies $S_A^{(n)} \equiv \tfrac{1}{1-n} \, \log \!\big[ \textrm{Tr}_{\mathcal{H}_A} \rho_A^n \big]$, 
labelled by the integer $n \geqslant 2$ (the R\'enyi index), and the entanglement entropy $S_A \equiv - \textrm{Tr}_{\mathcal{H}_A} (\rho_A \log \rho_A)$. 
The entanglement entropy can be obtained also from the R\'enyi entropies 
through the replica limit, i.e.
the analytic continuation $S_A = \lim_{n \to 1} S_A^{(n)}$;
hence these quantities are often called entanglement entropies. 
When $\rho$ is a pure state, we have that 
$S_A^{(n)} = S_B^{(n)}$ and $S_A = S_B$.
The entanglement Hamiltonian contains more information than the quantities obtained purely from the entanglement spectrum.

In quantum field theories, the analytic expression of the entanglement Hamiltonian in terms of the fundamental fields is known in very few cases. 
In a generic number of spacetime dimensions, these include the seminal result of Bisognano and Wichmann \cite{Bisognano:1975ih,Bisognano:1976za}, which holds for any relativistic quantum field theory in the vacuum when $A$ is the half space $x>0$, and states that $K_A$ is proportional to the generator of the boosts in the $x$-direction. 
Focusing on a conformal field theory (CFT) in the vacuum,
this fundamental result and the proper conformal map provide  $K_A$ 
when $A$ is a spherical region \cite{Hislop:1981uh, Casini:2011kv}.
In a two-dimensional CFT, where the conformal symmetry is infinite dimensional, 
other explicit expressions can be found when $A$ is an interval
\cite{Wong:2013gua, Cardy:2016fqc}.
A universal result for $K_A$ is also known 
when the CFT is defined 
either on the half line with a conformally invariant boundary condition
or on the segment 
with the same conformally invariant boundary condition 
imposed at both boundaries,
hence the model is a boundary conformal field theory (BCFT),
when  $A$ is an interval adjacent to the boundary 
\cite{Cardy:2016fqc, Tonni:2017jom}.
In these cases, the entanglement Hamiltonian $K_A$ is a local operator. 
In particular, $K_A$ is written 
as the integral over $A$ of the energy density multiplied by a weight function that depends on the global geometry of the space,
on the endpoints of $A$
and on the state of the entire system. 
In this BCFT setup, $K_A$ for an interval $A$ adjacent to the boundary 
has been written explicitly also for a specific class 
of spatially inhomogeneous backgrounds where the background metric is Weyl equivalent to the flat metric 
through a Weyl factor that depends only 
on the spatial coordinate and is even about the centre of the segment \cite{Tonni:2017jom}.
When the underlying BCFT model is the free massless Dirac field, whose central charge is $c=1$,
this class of spatially inhomogeneous backgrounds has interesting applications. 
Indeed, it captures the continuum limit 
of the low energy excitations of some free fermionic systems around a varying energy scale.
Some interesting examples of such models are 
the rainbow chain
\cite{Vitagliano:2010db,Ram_rez_2014,Ramirez:2015yfa,Rodr_guez_Laguna_2016, Rodriguez-Laguna:2016roi},
the free Fermi gas in a harmonic trap \cite{Campostrini:2009ema,Campostrini_2010,Campostrini_2010_2,Campostrini:2010pv, Dubail:2016tsc} 
and an XXZ chain with a gradient \cite{Eisler:2017iay}. 
Furthermore, these inhomogeneous backgrounds are also related to optical metrics \cite{Lewenstein:2012,Boada:2010sh,Rodriguez-Laguna:2016kri,Mula:2020udv},
arising in the study of light propagation through media 
with a spatially varying refractive index.
In this manuscript, the numerical checks 
are performed in the rainbow chain,
where a volume law for the entanglement entropy 
is observed in the regime of large inhomogeneity for certain bipartitions.

A qualitatively different class of entanglement Hamiltonians is provided by the cases where $K_A$ 
contains local and bilocal operators.
All the examples where analytic expressions are known 
involve the massless Dirac field, and all of them are inspired by the case where $A$ is the union of an arbitrary finite number of disjoint intervals in the line
and the massless Dirac field is in its ground state \cite{Casini:2009vk}.
In this class, we also find $K_A$ of an interval $A$ 
when the massless Dirac field is at finite temperature and on a circle \cite{Blanco:2019xwi, Fries:2019ozf}.
As for cases where the underlying model is not invariant under translations, 
it is worth mentioning $K_A$ for the massless Dirac field in its ground state when $A$ is either an interval in the half line \cite{Mintchev:2020uom} or the union of two equal intervals 
in the line at the same distance from a pointlike defect \cite{Mintchev:2020jhc}.

\begin{figure}[t!]
\vspace{-.4cm}
\hspace{1.75cm}
\includegraphics[width=.8\textwidth]{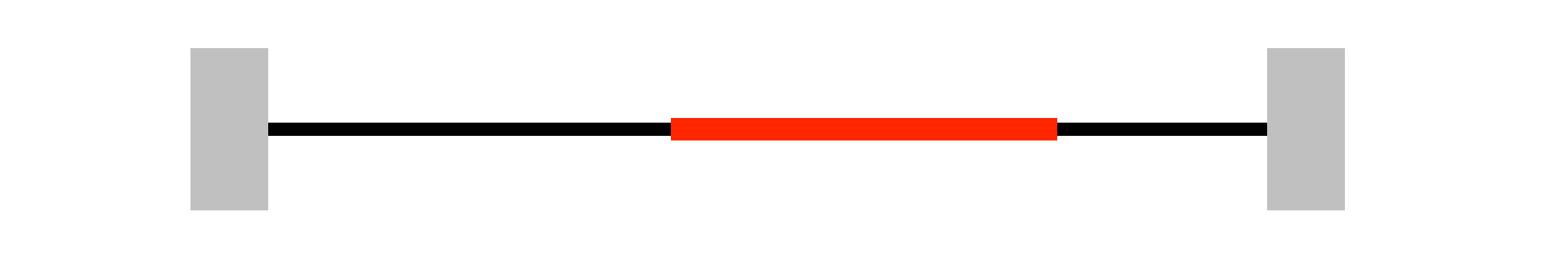}
\vspace{-.3cm}
\caption{Bipartition of the segment given by an interval $A$ (red region) in a generic position. 
The same boundary condition is imposed at both endpoints of the segment. 
}
\label{fig-interval-segment}
\end{figure}

The main result of this manuscript is another 
example where $K_A$ for the massless Dirac field 
is known analytically and it is made 
by a local and a bilocal term. 
We consider the BCFT model defined by the massless Dirac field 
in the segment where the same conformally invariant boundary condition is imposed at both endpoints.
In the case where the whole system is in the ground state, 
we find the entanglement Hamiltonian $K_A$ 
of an interval $A$ in a generic position 
in the segment, 
not adjacent to the boundary 
(see Fig.\,\ref{fig-interval-segment}, 
where $A$ corresponds to the red region),
for the class of spatially inhomogeneous backgrounds 
mentioned above.  

Another quantity that we find worth investigating is the contour function 
for the entanglement entropies \cite{Botero_2004,ChenVidal2014}. 
This scalar function is a spatial density 
for the entanglement entropies
constrained by additional requirements guaranteeing a proper behaviour of this quantity under local unitary transformations, subregion inclusions and possible symmetries \cite{ChenVidal2014}. 
Although the contour function for the entanglement entropies 
does not provide a complete characterisation of the bipartite entanglement like the entanglement Hamiltonian, 
it contains more information than the entanglement spectrum
because its definition involves also the eigenvectors of $K_A$.
In homogeneous CFT, some contour functions for the entanglement entropies have been written through the weight function of the local term in the corresponding entanglement Hamiltonian 
\cite{ChenVidal2014, Coser:2017dtb, Wong:2013gua, Cardy:2016fqc}.
In the inhomogeneous BCFT mentioned above, 
this observation has been applied to 
the bipartition in Fig.\,\ref{fig-interval-segment} 
when $A$ is adjacent to the boundary \cite{Tonni:2017jom}.

In this manuscript, for the ground state of the 
inhomogeneous BCFT given by the massless Dirac field on the segment 
and the bipartition shown in Fig.\,\ref{fig-interval-segment},
we study the contour function for the entanglement entropies provided by 
the weight function of the local term in the entanglement Hamiltonian.
The entanglement entropies of $A$ are computed by integrating this contour function, 
recovering the result of \cite{Rodriguez-Laguna:2016roi} 
obtained through the twist fields method \cite{Calabrese:2004eu}.

A comparison with the exact numerical  results 
obtained in a lattice model is performed for the rainbow chain.
The rainbow chain \cite{Vitagliano:2010db}
is an inhomogeneous free fermionic chain in a finite segment 
whose hopping coefficients 
depend on a single real and positive parameter in such a way that, for certain bipartitions, 
the entanglement entropy of a block obeys 
the volume law in the regime of large inhomogeneity
\cite{Ram_rez_2014,Ramirez:2015yfa,Rodr_guez_Laguna_2016,Rodriguez-Laguna:2016roi}.
The continuum limit of the rainbow chain provides the rainbow model in the continuum. 
The entanglement Hamiltonian in the rainbow chain  
and its continuum limit are investigated by employing the Peschel formula 
\cite{Peschel:2003rdm, EislerPeschel:2009review}
and the continuum limit procedure discussed in 
\cite{Arias:2016nip, Eisler:2019rnr, DiGiulio:2019cxv,Eisler:2022rnp}.
Our analysis is different from the numerical study 
performed in \cite{Tonni:2017jom} 
for a block $A$ adjacent to an endpoint of the segment,
where only the local term occurs.
The entanglement entropies and their contour functions in the rainbow chain for the bipartition in Fig.\,\ref{fig-interval-segment} are also explored.
The analytic predictions in the continuum 
are obtained by specialising the aforementioned BCFT results 
for the entanglement Hamiltonian 
and the contour function for the entanglement entropies 
to the case of the rainbow model.

The outline of the paper is as follows. 
In Sec.\,\ref{sec-EH-homo}, focusing on the homogeneous background,
we construct the entanglement Hamiltonian of the interval $A$ 
and the corresponding contour function for the entanglement entropies. 
In Sec.\,\ref{sec-EH-non-homo}, 
these quantities are investigated in the specific class of spatially inhomogeneous backgrounds of interest.
Within this class, the explicit expressions for the special case of the rainbow model are reported (see Sec.\,\ref{subsec-CFT-rainbow-model}).
The rainbow chain is explored in Sec.\,\ref{sec-EH-rainbow-chain}, 
where a  continuum limit procedure is applied 
to the entanglement Hamiltonian of a block of contiguous sites and to the contour function for the entanglement entropies, finding a remarkable agreement with the corresponding analytic expressions in the rainbow model.
Conclusions are drawn in Sec.\,\ref{sec-conclusions}, 
where possible directions for future studies are also mentioned.

\section{Entanglement Hamiltonian of an interval in a segment}
\label{sec-EH-homo}

In this section we discuss the entanglement Hamiltonian $K_A$ of an interval $A$ for the free massless Dirac field on a segment with the same boundary condition imposed at both its endpoints, when $A$ is not adjacent to the boundary (see Fig.\,\ref{fig-interval-segment}).

The free massless Dirac field $\psi$ on the infinite strip $\mathbb{S} \equiv [-L, L] \times \RR$  of width $2L$,
parameterised by the complex coordinate $z= x + \ri t_{\textrm{\tiny E}}$
(the Euclidean time $t_{\textrm{\tiny E}} \in \RR$ is related to the Lorentzian time $t$ by $t= -\ri t_{\textrm{\tiny E}}$)
with $x\in [-L, L]$, is described by the following 
two-dimensional Euclidean action 
\be
\label{action-dirac-strip-homo}
    S[\psi] 
    \propto
    \int_{\mathbb{S}} \Big(\, 
    \psi^\ast_2 \!
    \stackrel{\leftrightarrow}{\partial_{\bar{z}}} \! \psi_2 
    + 
    \psi^\ast_1 \!
    \stackrel{\leftrightarrow}{\partial_{z}} \! \psi_1
    \,\Big) 
    \, \rd z \, \rd \bar{z}
\ee
where $\psi_1$ and $\psi_2$ denote the left and right chiral components of the Dirac field, 
which become the antiholomorphic and holomorphic components of the field in Euclidean signature, respectively.
We focus on the  case where the same boundary condition is imposed at both boundaries of the strip $\mathbb{S}$.
This model is a BCFT with central charge $c=1$.

Since this model is a BCFT, it is convenient to introduce a conformal map sending the strip $\mathbb{S}$ onto the right half plane $\mathbb{H}_{\textrm{\tiny R}}$,
parameterised by the complex coordinate 
$\hat{z} = \hat{x}+ \ri  \hat{t}_{\textrm{\tiny E}}$, 
with $\hat{x} \geqslant 0$ and $\hat{t}_{\textrm{\tiny E}} \in \RR$.
The boundary conditions  on $\mathbb{S}$ are obtained 
from the one imposed at the boundary of $\mathbb{H}_{\textrm{\tiny R}}$.
The conformal map that will be employed in our analysis also depends on the bipartition of the segment $[-L, L]$, which is determined by an interval in a generic position, and will be discussed below (see (\ref{map-strip-to-rhp})).

Considering the massless Dirac field $\hat{\psi}$ on the half line parameterised by the real coordinate $\hat{x} \geqslant 0$,
the vanishing of the energy flow through the boundary is imposed 
to guarantee the global energy conservation
\cite{Cardy:1984bb, Cardy:1986gw, Cardy:1989ir},
and this condition is satisfied for either 
\begin{equation}
\label{vector-bc-RHP}
    \hat{\lambda}_1\big|_{\hat{x}=0}  
    = \, \e^{\ri \alpha_{\textrm{\tiny v}}}\, 
    \hat{\lambda}_2 \big|_{\hat{x}=0} 
    \;\;\;\qquad\;\;\;
    \alpha_{\textrm{\tiny v}}\in \big[0, 2\pi\big)
\end{equation}
or
\begin{equation}
\label{axial-bc-RHP}
        \hat{\chi}_1\big|_{\hat{x}=0}  
    = \, \e^{-\ri \alpha_{\textrm{\tiny a}}}\, 
    \hat{\chi}_2^\ast \big|_{\hat{x}=0} 
    \;\;\;\qquad\;\;\;
    \alpha_{\textrm{\tiny a}}\in \big[0, 2\pi\big)
\end{equation}
for any value of $\hat{t}_{\textrm{\tiny E}} \in \RR$,
where the components of the Dirac field have been denoted by $\hat{\lambda}_j$ and $\hat{\chi}_j$ to distinguish the two cases. 
The boundary conditions (\ref{vector-bc-RHP}) or (\ref{axial-bc-RHP})
provide two different models 
that are called the vector phase and the axial phase respectively.
Indeed, at quantum level, 
either the charge or the helicity is conserved 
\cite{Liguori:1997vd}. 
In particular, the charge is conserved in the vector phase,
while the helicity is conserved in the axial phase.

In order to shorten the forthcoming expressions, 
following  \cite{Mintchev:2020uom},
we find it convenient to treat  the boundary conditions 
(\ref{vector-bc-RHP}) and (\ref{axial-bc-RHP}) in a unified way. 
This is done by introducing the antiholomorphic field $\hat{\psi}_1$ and the holomorphic field $\hat{\psi}_2$, which correspond to 
$\hat{\lambda}_1$ and $\hat{\lambda}_2$ in the vector phase 
and to $\hat{\chi}^\ast_1$ and $\hat{\chi}_2$ in the axial phase, respectively.
In terms of the fields $\hat{\psi}_1$ and $\hat{\psi}_2$, the boundary conditions (\ref{vector-bc-RHP}) and (\ref{axial-bc-RHP}) read
\begin{equation}
            \hat{\psi}_1\big(\!- \!\ri \hat{t}_{\textrm{\tiny E}}\big)
    = \, 
    \e^{\ri \alpha}\, 
    \hat{\psi}_2^\ast \big( \ri \hat{t}_{\textrm{\tiny E}}\big)
    \;\;\;\qquad\;\;\;
    \alpha \in \big[0, 2\pi\big)
\end{equation}
where $\alpha = \alpha_{\textrm{\tiny v}}$ 
in the vector phase and 
$\alpha = \alpha_{\textrm{\tiny a}}$ 
in the axial phase.

Consider the bipartition of the half line 
determined by the interval $\hat{A}\equiv \big[ \,\hat{a}, \hat{b} \,\big]$,
which is not adjacent to the boundary for $\hat{a}>0$. 
When the massless Dirac field is in its ground state, 
the entanglement Hamiltonian of this bipartition reads \cite{Mintchev:2020uom}
\be
\label{EH-interval-half-line}
\widehat{K}_{\hat{A}} 
\,=\,
2\pi
\int_{\hat{a}}^{\hat{b}} \!
\hat{\beta}_{\textrm{\tiny loc}}(\hat{x} ) \, 
\widehat{\mathcal{E}}(\hat{x} )\, \rd \hat{x} 
+
2\pi 
\int_{\hat{a}}^{\hat{b}} \!
\hat{\beta}_{\textrm{\tiny biloc}}(\hat{x}) \, \widehat{T}^{\,\textrm{\tiny $(\alpha)$}}_{\textrm{\tiny biloc}}(\hat{x}, \hat{x}_{\textrm{c}} ) \, \rd \hat{x}
\ee
where the local operator $\widehat{\mathcal{E}}(\hat{x} )$ 
is the normal ordered version of the energy density on the half line, 
namely 
\be
    \label{T00-lambda-def}
    \widehat{\mathcal{E}}(\hat{x}) 
    \,\equiv\,
    \,\frac{\textrm{i}}{2}
    :\! \!
    \Big[ \Big ((\partial_{\hat{x}} \hat{\psi}^\ast_1)\, \hat{\psi}_1 - 
    \hat{\psi}^\ast_1\, (\partial_{\hat{x}} \hat{\psi}_1) \Big )(\hat{x})
    - \Big((\partial_{\hat{x}} \hat{\psi}^\ast_2)\,  \hat{\psi}_2 - \hat{\psi}^\ast_2\, (\partial_{\hat{x}} \hat{\psi}_2)\Big) (\hat{x})
    \Big]\!\! : 
\ee
and the bilocal operator is defined as follows
\be
    \label{T-bilocal-def}
    \widehat{T}^{\,\textrm{\tiny $(\alpha)$}}_{\textrm{\tiny biloc}}(\hat{x}, \hat{y}) 
    \,\equiv\,
    \frac{\textrm{i}}{2}\;
    \bigg\{ \,
    e^{\textrm{i} \alpha}
    \!:\!\!\Big[\, \hat{\psi}^\ast_1(\hat{y}) \,  \hat{\psi}_2(\hat{x}) + \hat{\psi}^\ast_1(\hat{x}) \,  \hat{\psi}_2(\hat{y}) \Big]\!\!: 
    - \;
    e^{-\textrm{i} \alpha}
    \!:\!\! \Big[\, \hat{\psi}^\ast_2(\hat{y}) \,  \hat{\psi}_1(\hat{x}) + \hat{\psi}^\ast_2(\hat{x}) \,  \hat{\psi}_1(\hat{y}) \Big] \!\!: \!
    \bigg\} \;.
\ee
This bilocal operator is evaluated at $\hat{y} = \hat{x}_{\textrm{c}}$
in (\ref{EH-interval-half-line}),
where $\hat{x}_{\textrm{c}} \in \hat{A}$ is the point conjugate to $\hat{x}$, 
given by 
\be
\label{x-tilde-conj-def}
\hat{x}_{\textrm{c}}(\hat{x}) \equiv \frac{\hat{a} \, \hat{b}}{ \hat{x} }\;.
\ee
In the entanglement Hamiltonian (\ref{EH-interval-half-line}), 
the weight function of the local term is
\be
\label{beta-loc-half-line}
\hat{\beta}_{\textrm{\tiny loc}}(\hat{x}) 
=  
\frac{ \big( \,\hat{b}^2 - \hat{x}^2\, \big) \big( \,\hat{x}^2 - \hat{a}^2\, \big)  }{2 \big(\, \hat{b} - \hat{a}\, \big) \big( \hat{a}\,\hat{b} + \hat{x}^2 \big) }
\ee
while the weight function of the bilocal term reads
\be
\label{beta-biloc-half-line}
\hat{\beta}_{\textrm{\tiny biloc}}(\hat{x}) 
=  
\frac{ \hat{\beta}_{\textrm{\tiny loc}}( \hat{x}_{\textrm{c}} )  }{ \hat{x} + \hat{x}_{\textrm{c}}}
=
\frac{
\hat{a} \, \hat{b} \,\big( \, \hat{b}^2 - \hat{x}^2 \,\big) \big( \, \hat{x}^2 - \hat{a}^2 \,\big) 
}{
2 \big( \, \hat{b} - \hat{a} \,\big) \, \hat{x} \, \big( \, \hat{a}\,\hat{b} + \hat{x}^2 \,\big)^2
}\;.
\ee

In a neighbourhood of the endpoints of the interval,
the expansion of the weight function of the local term in (\ref{beta-loc-half-line})
is $\hat{\beta}_{\textrm{\tiny loc}}(\hat{x}) = \hat{x}-\hat{a} +O\big((\hat{x}-\hat{a})^2\big)$ as $\hat{x}\to \hat{a}$ and
$\hat{\beta}_{\textrm{\tiny loc}}(\hat{x}) = \hat{b}-\hat{x} +O\big((\hat{b}-\hat{x})^2\big)$ as $\hat{x}\to \hat{b}$,
hence $\hat{\beta}_{\textrm{\tiny loc}}(\hat{x})$ satisfies the behaviour expected  from the Bisognano-Wichmann result;
while 
the asymptotic behaviours 
for the weight function of the bilocal term in 
(\ref{beta-biloc-half-line}) 
are given by 
$\hat{\beta}_{\textrm{\tiny biloc}}(\hat{x}) = \tfrac{\hat{b}}{\hat{a}(\hat{a}+\hat{b})} (\hat{x}-\hat{a}) +O\big(( \hat{x}-\hat{a})^2\big)$ as $\hat{x}\to \hat{a}$ and
$\hat{\beta}_{\textrm{\tiny biloc}}(\hat{x}) = 
\tfrac{\hat{a}}{\hat{b}(\hat{a}+\hat{b})} (\hat{b}-\hat{x})  
+O\big((\hat{b}-\hat{x})^2\big)$ as $\hat{x}\to \hat{b}$.

In terms of the complex coordinates $z$ 
and $\hat{z}$ 
parameterising $\mathbb{S}$ and $\mathbb{H}_{\textrm{\tiny R}}$ respectively,
we employ the following holomorphic map 
sending $\mathbb{S} $ onto $\mathbb{H}_{\textrm{\tiny R}}$ 
\be
\label{map-strip-to-rhp}
\hat{w}(z) 
\equiv  
\hat{a}\; \frac{\tan\! \big[ \frac{\pi (z+L)}{4L} \big]}{\tan\! \big[ \frac{\pi (a+L)}{4L} \big]}
\equiv  
\frac{ \hat{a} }{ a_L }\, z_L
\ee
where $\hat{w}(z) \in \mathbb{H}_{\textrm{\tiny R}}$ and the parameter $a$ corresponds to the first endpoint of the interval $A$ in the segment.
This map satisfies $\hat{w}(a) =  \hat{a} $ with $\hat{a} >0$, 
and in the last step it is written through the shorthand notation given by 
\be
\label{L-notation}
z_L \,\equiv\, \tan\! \left[ \frac{\pi (z+L)}{4L} \right]
\;\;\;\;\qquad\;\;\;
a_L \,\equiv\, \tan\! \left[ \frac{\pi (a+L)}{4L} \right]
\;\qquad\;
b_L \,\equiv\, \tan\! \left[ \frac{\pi (b+L)}{4L} \right] \,.
\ee
The conformal map (\ref{map-strip-to-rhp}) sends
the segment having $\textrm{Im}(z) = 0$ in $\mathbb{S} $ onto the half line having $\textrm{Im}( \hat{z} ) = 0$ in $\mathbb{H}_{\textrm{\tiny R}}$;
hence the image of the second endpoint of $A$ is real and reads
\be
\label{tilde-b-bL-def}
\hat{b} \equiv \hat{w}(b) = \frac{ \hat{a} }{ a_L }\, b_L \,.
\ee
The inverse and the first derivative of the map in (\ref{map-strip-to-rhp}) 
are given respectively by
\be
\label{inverse-derivative-w}
z(\hat{w}) = \frac{4L}{\pi} \, \arctan \! \bigg( \frac{ a_L}{ \hat{a} }\, \hat{w}  \bigg) - L
\;\;\;\qquad\;\;\;
\hat{w}'(z) = \frac{ \hat{a} }{ a_L }\; \frac{\pi}{4L\, \big(\! \cos\!\big[ \frac{\pi(z+L)}{4L}\big] \big)^2} \;.
\ee

Other conformal maps relating 
a strip and the half plane have been employed e.g. in 
\cite{Cardy:2016fqc, Rodriguez-Laguna:2016roi, Tonni:2017jom, Estienne:2023ekf}
to explore the bipartite entanglement in critical systems with a boundary.

A vertical line  in $\mathbb{S} $ parameterised by $x_0 + \ri  \, t_{\textrm{\tiny E}} \in \mathbb{S}$ with $ x_0 \in [-L, L]$ and $t_{\textrm{\tiny E}}  \in \RR$,
is mapped by (\ref{map-strip-to-rhp}) 
onto a circular arc whose endpoints are $\hat{z} = \pm \, \ri \,\hat{w}(0)$,
where $\hat{w}(0) = \hat{a}/a_L $ is positive.
The circle supporting this arc is given by $( \hat{x} - C_0 )^2 + \hat{t}^{\, 2}_{\textrm{\tiny E}} = R_0^2$ with $\hat{x} \geqslant 0$ and
\be
C_0 \equiv \frac{ \hat{w}(x_0)^2 - \hat{w}(0)^2   }{2\, \hat{w}(0)}
\;\;\;\qquad\;\;\;
R_0 \equiv \frac{ \hat{w}(x_0)^2 + \hat{w}(0)^2   }{2\, \hat{w}(0)}\;.
\ee
In particular, the central vertical line corresponding to $x_0$ is mapped
onto the half circle in $\mathbb{H}_{\textrm{\tiny R}}$ centred at the origin $\hat{z} = 0$ with radius equal to $\hat{w}(0)$.
As a further consequence, 
the left and right boundaries of $\mathbb{S} $,
corresponding to $x_0 = -L$ and $x_0 = L$ respectively,  
are mapped
onto the vertical segment given by $\hat{z} = \ri \, \hat{t}_{\textrm{\tiny E}} $ with $ - \hat{w}(0) < \hat{t}_{\textrm{\tiny E}}  < \hat{w}(0)$
and onto its complement on the boundary of $ \mathbb{H}_{\textrm{\tiny R}}$,
respectively. 
Another property of the conformal map (\ref{map-strip-to-rhp}) is that the image of 
a horizontal segment  in the strip $\mathbb{S}$ made by the points $x + \ri \, \hat{t}_{\textrm{\tiny E},0} $
is the half circle $\hat{x}^2 + \big( \hat{t}_{\textrm{\tiny E}} - C_1\big)^2 = R_1^2$
centred  on the boundary of $\mathbb{H}_{\textrm{\tiny R}}$, where
\be
C_1 \equiv \frac{ \hat{w}(L + \ri \, \hat{t}_{\textrm{\tiny E},0} )  + \hat{w}(-L + \ri \, \hat{t}_{\textrm{\tiny E},0} )  }{2}
\;\;\;\qquad\;\;\;
R_1 \equiv \frac{ \big| \hat{w}(L + \ri \, \hat{t}_{\textrm{\tiny E},0} )  - \hat{w}(-L + \ri \, \hat{t}_{\textrm{\tiny E},0} ) \big|  }{2}\;.
\ee

The conformal map (\ref{map-strip-to-rhp}) and the boundary condition  
(\ref{vector-bc-RHP}) or (\ref{axial-bc-RHP}) imposed at the boundary of $\mathbb{H}_{\textrm{\tiny R}}$ provide the boundary conditions 
at  the boundaries of $\mathbb{S}$.
In particular, from the second expression in \eqref{inverse-derivative-w}, we find
\be
    \label{w-prime-boundaries-S-1}
    \left.\hat{w}'(z)^{-1/2}\right|_{z = -L + \ri t_{\textrm{\tiny E}}} \!\! = 
    \sqrt{\frac{4  L \, a_L}{\pi \, \hat{a}}} \, \cosh\!\left(
    \frac{\pi \,  t_{\textrm{\tiny E}}}{4 L} \right) 
    \qquad
    \left.\bar{\hat{w}}'(\bar{z})^{-1/2}\right|_{\bar{z} 
    = 
    -L - \ri t_{\textrm{\tiny E}}} \!\! = 
    \sqrt{\frac{4  L \, a_L}{\pi \, \hat{a}}} \, \cosh\!\left(
    \frac{\pi \,  t_{\textrm{\tiny E}}}{4 L} \right)
\ee
and 
\be
    \label{w-prime-boundaries-S-2}
    \left.\hat{w}'(z)^{-1/2}\right|_{z = L + \ri t_{\textrm{\tiny E}}} \!\! = 
    -\,\ri \, \sqrt{\frac{4 L \, a_L}{\pi \, \hat{a}}} \, 
    \sinh\!\left(
    \frac{\pi \, t_{\textrm{\tiny E}}}{4  L} \right) 
    \qquad
    \left.\bar{\hat{w}}'(\bar{z})^{-1/2}\right|_{\bar{z} = L - \ri t_{\textrm{\tiny E}}} \!\! = \ri \, \sqrt{\frac{4 L \, a_L}{\pi \, \hat{a}}} \, 
    \sinh\!\left(
    \frac{\pi \, t_{\textrm{\tiny E}}}{4  L} \right)
\ee
which are real and purely imaginary, respectively. 
Since the conformal weights of the antiholomorphic field $\psi_1$ and of the holomorphic field $\psi_2$ 
are $(0, \frac{1}{2})$ and $(\frac{1}{2}, 0)$ respectively
\cite{Ginsparg:1988ui,DiFrancesco:1997nk},
under the holomorphic map $z \mapsto \hat{w}(z)$
they transform as 
$\hat{\psi}_1(\bar{\hat{w}}) = 
\bar{\hat{w}}'(\bar{z})^{-1/2} \, \psi_1(\bar{z}) $ and $\hat{\psi}_2(\hat{w}) = \hat{w}'(z)^{-1/2} \, \psi_2(z)$.
By combining these transformations with (\ref{w-prime-boundaries-S-1}) and (\ref{w-prime-boundaries-S-2}), 
the boundary conditions \eqref{vector-bc-RHP} 
and \eqref{axial-bc-RHP} are mapped respectively into
\be
    \label{strip-bc-vector}
    \lambda_1(- L - \ri \, t_{\textrm{\tiny E}})  
    = \, \e^{\ri \alpha_{\textrm{\tiny v}}}\, 
    \lambda_2 (- L + \ri \, t_{\textrm{\tiny E}}) 
    \;\;\;\qquad\;\;\;
    \lambda_1 (L - \ri \, t_{\textrm{\tiny E}})  
    = \, - \, \e^{\ri \alpha_{\textrm{\tiny v}}}\, 
    \lambda_2 (L + \ri \, t_{\textrm{\tiny E}})
\ee
in the vector phase and into
\be
    \label{strip-bc-axial}
    \chi_1(- L - \ri \, t_{\textrm{\tiny E}})
    = \, \e^{-\ri \alpha_{\textrm{\tiny a}}}\, 
    \chi^\ast_2 (- L + \ri \, t_{\textrm{\tiny E}})
    \;\;\;\qquad\;\;\;
    \chi_1(L - \ri \, t_{\textrm{\tiny E}})  
    = \e^{-\ri \alpha_{\textrm{\tiny a}}}\, 
    \chi^\ast_2 (L + \ri \, t_{\textrm{\tiny E}})
\ee
in the axial phase, 
for any $t_{\textrm{\tiny E}} \in \RR$.
Although different signs occur in the two r.h.s.'s  
of (\ref{strip-bc-vector}), 
associated with the two boundaries of the strip, 
the two conditions in (\ref{strip-bc-vector})
are obtained from a single boundary condition 
characterising the vector phase 
in $\mathbb{H}_{\textrm{\tiny R}}$, 
and in this sense they correspond to 
the same boundary condition
imposed at both boundaries of $\mathbb{S}$.

The entanglement Hamiltonian $K_{A}$ of the interval  $A \equiv [a,b]$ in the segment $[-L, L]$ for the massless Dirac field with the same boundary condition imposed at both endpoints of the segment,
given by either (\ref{strip-bc-vector}) or (\ref{strip-bc-axial}),
is obtained from the entanglement Hamiltonian of the massless Dirac field on the half line (see (\ref{EH-interval-half-line})) 
and the transformation law of the fields involved in its expression. This gives
\be
    \label{EH-segment-homo}
    K_{A} 
    \,=\,
    2\pi
    \int_{a}^{b} \!
    \beta_{\textrm{\tiny loc}}(x ) \, \mathcal{E}(x )\, \rd x 
    +
    2\pi 
    \int_{a}^{b} \!
    \beta_{\textrm{\tiny biloc}}(x) \, T^{\,\textrm{\tiny $(\alpha)$}}_{\textrm{\tiny biloc}}(x, x_{\textrm{c}} ) \, \rd x
\ee
where the operators $\mathcal{E}(x)$ and $T^{\,\textrm{\tiny $(\alpha)$}}_{\textrm{\tiny biloc}}(x,y)$
are defined respectively
as (\ref{T00-lambda-def}) and (\ref{T-bilocal-def}) with the hatted fields replaced by the corresponding ones on the strip (without the hat, in our notation).

The weight function of the local term in (\ref{EH-segment-homo}) 
is obtained from 
the local term in the modular Hamiltonian (\ref{EH-interval-half-line})
by employing the fact that the conformal dimension of the energy-momentum tensor is equal to $2$. 
Thus, from (\ref{beta-loc-half-line}) and (\ref{map-strip-to-rhp}),
for this weight function we find 
\be
\label{beta-loc-homo}
\beta_{\textrm{\tiny loc}}( x) 
\,\equiv\,
\frac{ \hat{\beta}_{\textrm{\tiny $\mathbb{S}$,loc}}\big( \hat{w}(x) \big) }{ \hat{w}'(x)  }
\,=\,
\frac{4L}{\pi} \bigg[  \cos\! \bigg( \frac{\pi (x+L)}{4L} \bigg) \bigg]^2
\frac{\big( b_L^2 - x_L^2\big) \big( x_L^2 - a_L^2 \big)}{ 2\big(b_L - a_L\big) \big( a_L \,b_L + x_L^2\big)}
\ee
where $ \hat{\beta}_{\textrm{\tiny $\mathbb{S}$,loc}} (\hat{x})$ 
is defined as $ \hat{\beta}_{\textrm{\tiny loc}} (\hat{x})$ 
in (\ref{beta-loc-half-line})
with $\hat{b}$ and $\hat{a}$ 
replaced by $\hat{w}(b)$ and $\hat{w}(a)$ respectively.
In fact, $\hat{\beta}_{\textrm{\tiny $\mathbb{S}$,loc}} (\hat{x}) = \hat{\beta}_{\textrm{\tiny loc}} (\hat{x})$
because $\hat{w}(a)=\hat{a}$ and $\hat{w}(b)=\hat{b}$,
from (\ref{map-strip-to-rhp}) and (\ref{tilde-b-bL-def}) respectively. 
In the last expression of (\ref{beta-loc-homo}),  
the notation introduced in (\ref{L-notation}) has been employed.
The weight function (\ref{beta-loc-homo}) can be written more explicitly as follows
\be
\label{beta-loc-homo-v1}
\beta_{\textrm{\tiny loc}}( x) 
\,=\,
\frac{2L}{\pi}\; 
\frac{ 
\big[ \sin\!\big( \tfrac{\pi b}{2L}\big) - \sin\!\big( \tfrac{\pi x}{2L}\big)  \big] \,  \big[  \sin\!\big( \tfrac{\pi x}{2L}\big)  -  \sin\!\big( \tfrac{\pi a}{2L}\big) \big]
}{
\sin\!\big( \tfrac{\pi (b-a)}{2L}\big)  + \big[  \cos\!\big( \tfrac{\pi b}{2L}\big) - \cos\!\big( \tfrac{\pi a}{2L}\big)  \big] \, \sin\!\big( \tfrac{\pi x}{2L}\big) 
} \;.
\ee
We remark that $\beta_{\textrm{\tiny loc}}(a) = \beta_{\textrm{\tiny loc}}(b) = 0$.
The asymptotic behaviour of (\ref{beta-loc-homo})-(\ref{beta-loc-homo-v1}) 
close to the endpoints of $A$ is the one expected from the 
Bisognano-Wichmann result, namely
\be
\begin{array}{ll}
\beta_{\textrm{\tiny loc}}( x) = (x-a) + O\big((x-a)^2\big) 
\hspace{1.5cm}
& x \to a
\\
\rule{0pt}{.6cm}
\beta_{\textrm{\tiny loc}}( x) = -\, (x-b) + O\big((x-b)^2\big) 
\hspace{1.5cm}
& x \to b\,.
\end{array}
\ee
We find it worth anticipating here that in Sec.\,\ref{sec-EH-rainbow-chain} 
the weight function in (\ref{beta-loc-homo})-(\ref{beta-loc-homo-v1}) 
is obtained from exact numerical results for  a homogeneous free fermionic chain 
in the segment
through a continuum limit procedure 
(see the solid blue curves in Fig.\,\ref{fig-beta-loc-A} 
and in the top panels of Fig.\,\ref{fig-beta-AB}).

It is instructive to consider some limiting regimes of 
the weight function in (\ref{beta-loc-homo})-(\ref{beta-loc-homo-v1}).

The limiting case where the interval is adjacent to the boundary
is recovered by taking e.g. $b \to L$ in (\ref{beta-loc-homo})-(\ref{beta-loc-homo-v1}), finding the result of \cite{Tonni:2017jom}
\be
\lim_{b \,\to\, L} \beta_{\textrm{\tiny loc}}( x) = 
\frac{2L}{\pi} \; 
\frac{\sin\!\big[\pi x/(2L)\big] - \sin\!\big[\pi a/(2L)\big] }{ \cos\!\big[\pi a/(2L)\big] } \;.
\ee

The well known parabolic profile occurring in the entanglement Hamiltonian 
of the interval in the infinite line \cite{Hislop:1981uh, Casini:2011kv} is obtained in the limit $L \to +\infty$.
Indeed, for (\ref{beta-loc-homo})-(\ref{beta-loc-homo-v1}) in this limiting regime
we find  $\beta_{\textrm{\tiny loc}}( x) = \frac{(b-x)(x-a)}{b-a} + O(1/L^2)$.

The case of the interval in the half line studied in \cite{Mintchev:2020uom} can be recovered by first
setting  $a = \hat{a} - L$, $b = \hat{b} - L$ and $x = \hat{x} - L$ 
in (\ref{beta-loc-homo})-(\ref{beta-loc-homo-v1}) 
and then taking $L \to +\infty$.
This gives 
$\beta_{\textrm{\tiny loc}}( x)   =  \hat{\beta}_{\textrm{\tiny loc}}(\hat{x}) +O(1/L^2)$ as $L \to +\infty$,
i.e. the $O(1)$ term is (\ref{beta-loc-half-line}), as expected.

In order to write the weight function of the bilocal term in (\ref{EH-segment-homo}),
we have to introduce the point $x_{\textrm{c}}(x) \in A$ conjugate to $x\in A$,
which can be obtained from (\ref{x-tilde-conj-def}) through the inverse function $z(\hat{w})$ in (\ref{inverse-derivative-w})
as follows
\begin{equation}
\label{x-conj}
x_{\textrm{c}}(x) = z(\hat{x}_{\textrm{c}} )
=
\frac{4L}{\pi} \, \arctan \! \big(  a_L  \,\hat{b} /  \hat{w}(x)    \big) - L
=
\frac{4L}{\pi} \, \arctan \! \big(  a_L  \, b_L / x_L   \big) - L
\end{equation}
where it has been used that 
$\hat{x} = \hat{w}(x) =  \hat{a} \, x_L/ a_L $,
from (\ref{map-strip-to-rhp}),
and (\ref{tilde-b-bL-def}).
By employing the notation defined in (\ref{L-notation}),
the last expression in (\ref{x-conj}) can be written as
\be
\label{x-conj-homo-L-notation}
x_{\textrm{c},L}(x)  = \frac{  a_L  \, b_L}{ x_L } \;.
\ee
From (\ref{x-conj}), notice  that 
$x_{\textrm{c}}(a) = b$, $x_{\textrm{c}}(b) = a$, and $x_{\textrm{c}}(0) =  \frac{4 L}{\pi}  \arctan(a_L \, b_L)-L$.

There exists a special point $x_{\textrm{sc}} \in A$ which is self-conjugate, 
meaning that it coincides with its conjugate point.
It  is determined by the condition $x_{\textrm{c}}(x_{\textrm{sc}} )  = x_{\textrm{sc}} $,
which  can be written by exploiting (\ref{x-conj-homo-L-notation}), 
finding 
\be
\label{x-sc-homo}
x_{\textrm{sc},L}^2 = a_L  \, b_L 
\ee
where $x_{\textrm{sc},L} \equiv \tan[\pi(x_{\textrm{sc}} +L)/(4L)]$,
according to (\ref{L-notation}). 
From (\ref{x-sc-homo}), one obtains an explicit expression for the point $x_{\textrm{sc}}$, 
namely $x_{\textrm{sc}} = \tfrac{4 L}{\pi}  \arctan\!\big(\sqrt{a_L \, b_L} \,\big) - L$.

In the symmetric case, 
the centre of $A$ coincides with the centre of the segment 
(i.e. $a=-b$ with $0< b <L$)
and, from (\ref{x-conj}) and (\ref{x-sc-homo}), 
we find that 
$x_{\textrm{c}}(x) = -\,x $ and $x_{\textrm{sc}} = 0 $ respectively. 

The expression (\ref{x-conj})
in the limiting case of the interval in the half line can be explored as done above for $\beta_{\textrm{\tiny loc}}(x)$, 
namely by setting $a = \hat{a} - L$, $b = \hat{b} - L$, $x = \hat{x} - L$ and $x_{\textrm{c}} = \hat{x}_{\textrm{c}} - L$ in (\ref{x-conj}) first,
and then sending $L \to +\infty$.
This gives $\hat{x}_{\textrm{c}}  \to \hat{a} \, \hat{b} / \hat{x}$ in this limit, as expected from the results reported in \cite{Mintchev:2020uom}.

Finally, in the special case where $A$ is adjacent to the boundary, 
namely when either  $b \to L$ or $a \to -L$, 
for (\ref{x-conj}) we find that either 
$x_{\textrm{c}}(x)  \to L$ or $x_{\textrm{c}}(x)  \to -L$
respectively, for all $x\in A$.

The weight function of the bilocal term in (\ref{EH-segment-homo})
is obtained from the weight function (\ref{beta-biloc-half-line}),
by employing the conformal map (\ref{map-strip-to-rhp}) and the fact that the chiral fermion field has conformal dimension equal to $1/2$,
and reads
\be
\label{beta-biloc-homo}
\beta_{\textrm{\tiny biloc}}( x) 
\,=\,
\sqrt{ \frac{\hat{w}'(x)}{\hat{w}'(x_{\textrm{c}} )} }\;
 \hat{\beta}_{\textrm{\tiny $\mathbb{S}$,biloc}}\big(\hat{w}(x) \big) 
\,=\,
\frac{ \cos\!\big[ \frac{\pi( x_{\textrm{c}} +L)}{4L}\big]  }{ \cos\!\big[ \frac{\pi(x +L)}{4L}\big] } 
\;\frac{
a_L  \, b_L \,\big( b_L^2 - x_L^2 \big) \big( x_L^2 - a_L^2 \big) 
}{
2 \big( b_L - a_L \big) \, x_L \, \big( a_L\,b_L + x_L^2 \big)^2
}
\ee
where $\hat{w}'(x)$ is given in (\ref{inverse-derivative-w}),
$x_{\textrm{c}}$ in (\ref{x-conj})
and $ \hat{\beta}_{\textrm{\tiny $\mathbb{S}$,biloc}} (\hat{x})$ is defined as $ \hat{\beta}_{\textrm{\tiny biloc}} (\hat{x})$ in (\ref{beta-biloc-half-line})
with $\hat{b}$ and $\hat{a}$ replaced by $\hat{w}(b)$ and $\hat{w}(a)$ respectively.
Similarly to (\ref{beta-loc-homo}), 
since the map (\ref{map-strip-to-rhp})
satisfies $\hat{w}(a)=\hat{a}$ and $\hat{w}(b)=\hat{b}$,
we have that $ \hat{\beta}_{\textrm{\tiny $\mathbb{S}$,biloc}} (\hat{x})= \hat{\beta}_{\textrm{\tiny biloc}} (\hat{x})$. 
In (\ref{beta-biloc-homo}) it has been also used that $\frac{\pi( x +L)}{4L} \in (0, \tfrac{\pi}{2})$ for $x\in (-L,L)$.
Notice that, in (\ref{beta-biloc-homo}),
$\hat{w}'(x)^{1/2}$ occurs because also the transformation of the volume element of the integral must be taken into account to get the bilocal term in (\ref{EH-segment-homo}).
From (\ref{x-conj}) and (\ref{L-notation}), we observe that 
\be
\label{J-biloc-cos}
\cos\!\left[ \frac{\pi( x_{\textrm{c}} +L)}{4L}\right] 
=
\frac{x_L}{ \sqrt{a^2_L\, b^2_L + x_L^2 } }
\ee
which leads us to write (\ref{beta-biloc-homo}) 
in terms of the notation introduced in (\ref{L-notation})
as follows
\be
\label{beta-biloc-homo-v2}
\beta_{\textrm{\tiny biloc}}( x) 
\,=\,
\frac{ a_L  \, b_L \,\big( b_L^2 - x_L^2 \big) \big( x_L^2 - a_L^2 \big)  
}{
2 \big( b_L - a_L \big) \, \cos\!\big[ \frac{\pi(x +L)}{4L}\big]  \, \sqrt{ a^2_L\, b^2_L + x_L^2 } \;\big( a_L\,b_L + x_L^2 \big)^{2}}\;.
\ee
A more explicit form for the weight function given in (\ref{beta-biloc-homo}) or  (\ref{beta-biloc-homo-v2}) reads
\bea
\label{beta-biloc-homo-v1}
\beta_{\textrm{\tiny biloc}}( x) 
&=&
\frac{ 1}{ \sqrt{ \tan^2\!\big[ \frac{\pi(a +L)}{4L}\big]  \tan^2\!\big[ \frac{\pi(b +L)}{4L}\big] + \tan^2\!\big[ \frac{\pi(x +L)}{4L}\big]   } } 
\;
\frac{\tan\!\big[ \frac{\pi(a +L)}{4L}\big]  \tan\!\big[ \frac{\pi(b +L)}{4L}\big]
}{
2\Big(  \! \tan\!\big[ \frac{\pi(b +L)}{4L}\big] - \tan\!\big[ \frac{\pi(a +L)}{4L}\big] \Big) 
}
\nn
\\
\rule{0pt}{.9cm}
& &
\times\,
\frac{
\big( \tan^2\!\big[ \frac{\pi(b +L)}{4L}\big] - \tan^2\!\big[ \frac{\pi(x +L)}{4L}\big]  \big) 
\big( \tan^2\!\big[ \frac{\pi(x +L)}{4L}\big] - \tan^2\!\big[ \frac{\pi(a +L)}{4L}\big]  \big) 
}{
\cos\!\big[ \frac{\pi(x +L)}{4L}\big] \,
\Big(  \! \tan\!\big[ \frac{\pi(a +L)}{4L}\big]  \tan\!\big[ \frac{\pi(b +L)}{4L}\big] + \tan^2\!\big[ \frac{\pi(x +L)}{4L}\big] \Big)^2
}\;.
\eea

The weight functions 
(\ref{beta-biloc-homo-v2}) and (\ref{beta-loc-homo}) 
are related as follows
\begin{equation}
\label{beta-loc-biloc-relation-homo}
\beta_{\textrm{\tiny biloc}}(x) 
\,=\,
\frac{\pi}{4L}
\;
\frac{
a_L  \, b_L \, \big(1+ x_L^2\big)^{3/2}
}{ 
\sqrt{a^2_L\,b^2_L + x_L^2} \; 
\big(a_L\,b_L + x_L^2\big)
}
\;\beta_{\textrm{\tiny loc}}(x) \,.
\end{equation}
We find it worth writing an equivalent form for this relation.
By using the first equality both in (\ref{beta-biloc-homo})
and in (\ref{beta-biloc-half-line}), 
we find that 
\begin{equation}
\label{beta-biloc-strip-to-loc}
    \beta_{\textrm{\tiny biloc}}(x) 
\,=\,
\sqrt{ \frac{\hat{w}'(x)}{\hat{w}'(x_{\textrm{c}} )} }\;
 \hat{\beta}_{\textrm{\tiny biloc}}\big(\hat{w}(x) \big) 
 \,=\,
 \sqrt{ \frac{\hat{w}'(x)}{\hat{w}'(x_{\textrm{c}} )} }\;
 \frac{\hat{\beta}_{\textrm{\tiny loc}}\big( \hat{a} \hat{b}/ \hat{w}(x) \big)}{
 \hat{w}(x) + \hat{a} \hat{b}/ \hat{w}(x)}
\end{equation}
where we can use that, 
from (\ref{map-strip-to-rhp}), (\ref{tilde-b-bL-def}) 
and (\ref{x-conj-homo-L-notation}), 
the following relation holds
\be
\label{id-w-xc}
\hat{w} \big(x_{\textrm{c}}(x)\big)
 \,=\, 
 \frac{\hat{a} \, b_L}{x_L}
 \,=\,
  \frac{\hat{b} \, a_L}{x_L}
   \,=\,
   \frac{\hat{a}\, \hat{b}}{\hat{w}(x)} \;.
\ee
From (\ref{beta-loc-homo}) we get 
$\hat{\beta}_{\textrm{\tiny loc}}
\big( \hat{w} \big(x_{\textrm{c}}\big) \big) 
= \hat{w}' (x_{\textrm{c}})\,
\beta_{\textrm{\tiny loc}} ( x_{\textrm{c}} ) $ and,
combining this relation with 
(\ref{beta-biloc-strip-to-loc}) and (\ref{id-w-xc}),
we find
\begin{equation}
\label{beta-biloc-strip-to-loc-step2}
    \beta_{\textrm{\tiny biloc}}(x) 
    \,=\, 
    \sqrt{\hat{w}'(x) \, \hat{w}'(x_{\textrm{c}})} \;
    \frac{\beta_{\textrm{\tiny loc}}(x_{\textrm{c}})}{
    \hat{w}(x) + \hat{w}(x_{\textrm{c}})}
    \,=\,
    \frac{\sqrt{\hat{w}'(x) \, \hat{w}'(x_{\textrm{c}})}}{\hat{a}/a_L} \;
    \frac{\beta_{\textrm{\tiny loc}}(x_{\textrm{c}})}{
    x_L + x_{\textrm{c},L} }\;.
\end{equation}
In the last expression of (\ref{beta-biloc-strip-to-loc-step2}), by employing (\ref{L-notation}), 
(\ref{inverse-derivative-w}) and (\ref{J-biloc-cos}), 
we observe that 
\begin{equation}
\label{beta-biloc-strip-to-loc-step3}
    \frac{\sqrt{\hat{w}'(x) \, \hat{w}'(x_{\textrm{c}})}}{\hat{a}/a_L} 
    \,=\,
    \frac{\pi}{4L}
    \;
    \frac{\sqrt{a^2_L\,b^2_L + x_L^2}}{
    \sin\!\big[ \frac{\pi(x +L)}{4L}\big]}\;.
\end{equation}
Finally, from (\ref{beta-biloc-strip-to-loc-step2}) and (\ref{beta-biloc-strip-to-loc-step3}),
we conclude that (\ref{beta-loc-biloc-relation-homo}) can be written as follows
\begin{equation}
\label{beta-biloc-strip-to-loc-step4}
    \beta_{\textrm{\tiny biloc}}(x) 
    \,=\,
    \frac{\pi}{4L}
    \;
    \frac{\sqrt{a^2_L\,b^2_L + x_L^2}}{\sin\!\big[ \frac{\pi(x +L)}{4L}\big]}
    \;\frac{\beta_{\textrm{\tiny loc}}(x_{\textrm{c}})}{x_L + x_{\textrm{c},L}}
\end{equation}
whose structure is similar 
to the relation between the weight functions 
found in \cite{Mintchev:2020uom}.

At the endpoints of $A$, the weight function
in (\ref{beta-biloc-homo}), (\ref{beta-biloc-homo-v2}) or (\ref{beta-biloc-homo-v1}) vanishes in a linear way; indeed, 
we have that $\beta_{\textrm{\tiny biloc}}(a) = \beta_{\textrm{\tiny biloc}}(b) = 0$
and
\begin{eqnarray}
    \label{BW-behaviour-homo-a}
\beta_{\textrm{\tiny biloc}}(x) 
&=&
\frac{\pi}{4 L} \, \frac{\cos(\frac{\pi b}{2 L})}{\cos(\frac{\pi a}{2 L}) \, \cos[\frac{\pi (b+a)}{4 L}]} 
\, (x-a) + O\big((x-a)^2\big) 
\hspace{1.5cm}
x \to a
\\
\label{BW-behaviour-homo-b}
\rule{0pt}{.8cm}
\beta_{\textrm{\tiny biloc}}(x) 
&=&
\frac{\pi}{4 L} \, \frac{\cos(\frac{\pi a}{2 L})}{\cos(\frac{\pi b}{2 L}) \, \cos[\frac{\pi (b+a)}{4 L}]} \, (b-x) + O\big((x-b)^2\big) 
\hspace{1.5cm}
x \to b \,.
\end{eqnarray}

Let us discuss the weight function $\beta_{\textrm{\tiny biloc}}(x)$ 
in (\ref{beta-biloc-homo}), (\ref{beta-biloc-homo-v2}) or (\ref{beta-biloc-homo-v1}) 
in the same limiting regimes considered above for 
$\beta_{\textrm{\tiny loc}}(x)$.

When $A$ is adjacent to the boundary \cite{Tonni:2017jom}, the entanglement Hamiltonian $K_{A}$ becomes local. 
Indeed, considering $\beta_{\textrm{\tiny biloc}}(x)$ 
for any given $x$ that does not coincide with 
the endpoints of $A$
and taking its limit for either $b \to L$ or $a \to -L$,
we find either $O(b-L)$ or $O(a+L)$ respectively.

Another limiting case where $K_{A}$ becomes local
corresponds to the interval $A$ in the infinite line \cite{Hislop:1981uh, Casini:2011kv};
indeed, $\beta_{\textrm{\tiny biloc}}(x) = \tfrac{\pi (b-x)(x-a)}{4L\,(b-a)} + O(1/L^2)$ as $L \to +\infty$.

Instead, a limiting regime where  $\beta_{\textrm{\tiny biloc}}(x)$ 
remains non trivial corresponds to the case of the interval 
in the half line explored in \cite{Mintchev:2020uom},
which can be recovered by applying the limiting procedure 
discussed above for $\beta_{\textrm{\tiny loc}}(x)$.
Thus, setting  $a = \hat{a} - L$, $b = \hat{b} - L$ 
and $x = \hat{x} - L$ 
in $\beta_{\textrm{\tiny biloc}}(x)$  first 
and then taking $L \to +\infty$ in the resulting expression,
we find $\beta_{\textrm{\tiny biloc}}( x)   =  \hat{\beta}_{\textrm{\tiny biloc}}(\hat{x}) +O(1/L^2)$
 as $L \to +\infty$,
which is the weight function (\ref{beta-biloc-half-line}) of the bilocal term 
in the entanglement Hamiltonian of an interval $\hat{A}\equiv \big[ \,\hat{a}, \hat{b} \,\big]$ in the half line, as expected. 

As also done for the weight function of the local term, 
we anticipate that in Sec.\,\ref{sec-EH-rainbow-chain}
the weight function of the bilocal term
in (\ref{beta-biloc-homo}), (\ref{beta-biloc-homo-v2}) or (\ref{beta-biloc-homo-v1}) 
agrees with the exact  numerical results obtained 
for the homogeneous free fermionic chain on the segment 
(see the solid blue curves in Fig.\,\ref{fig-beta-biloc-A} 
and in the bottom panels of Fig.\,\ref{fig-beta-AB}).

The weight function of the local term in 
(\ref{beta-loc-homo})-(\ref{beta-loc-homo-v1}) 
allows us to evaluate the entanglement entropies 
for the setup that we are investigating. 
Indeed, following \cite{ChenVidal2014, Coser:2017dtb, Cardy:2016fqc}, 
let us introduce the contour function 
for the entanglement entropies as
\be
\label{contour-function-cft-homo}
C^{(n)}_A(x) 
\equiv
\frac{c}{12} \left(1 + \frac{1}{n} \right) \frac{1}{\beta_{\textrm{\tiny loc}}( x) } 
\;\;\;\;\qquad\;\;\;
x\in A
\ee
where the integer $n \geqslant 2$ is the R\'enyi index while the case $n=1$ corresponds to the entanglement entropy,
$c=1$ for the massless Dirac field,
and $\beta_{\textrm{\tiny loc}}( x)$ is  the weight function of the local term given in (\ref{beta-loc-homo})-(\ref{beta-loc-homo-v1}).
A candidate for the entanglement entropies of the interval $A$ in the segment 
is obtained by integrating the contour function (\ref{contour-function-cft-homo}) over the interval $A_\epsilon \subsetneq A$,
where $A_\epsilon \equiv [\,a + \epsilon \,, b -\epsilon\,]$,
with $\epsilon \simeq 0^+$ defined as the UV cutoff \cite{Ohmori:2014eia, Cardy:2016fqc}.
This integral can be performed by employing the definition of $\beta_{\textrm{\tiny loc}}( x) $ in (\ref{beta-loc-homo}), 
finding 
\bea
\label{integ-contour}
\mathcal{I}(r,s) 
&\equiv &
\int_{ r }^{s }  
\frac{ \rd x }{\beta_{\textrm{\tiny loc}}( x) }
\, =
\int_{ \hat{w}(r) }^{ \hat{w}(s) }  
\frac{ \rd \hat{w} }{ \hat{\beta}_{\textrm{\tiny $\mathbb{S}$,loc}}(\hat{w}) }  
\, =
\bigg( 
\log \! \bigg[ 
\frac{\big(\hat{w}(y) -\hat{w}(a)\big)\,\big(\hat{w}(y) + \hat{w}(b)\big)  }{  \big(\hat{w}(b)- \hat{w}(y) \big)\,\big(\hat{w}(y) + \hat{w}(a)\big) } 
\bigg]\,
\bigg)\bigg|^{y=s}_{y=r}
\nn
\\
\rule{0pt}{.8cm}
&=&
 \, \log \! \bigg[ 
\frac{ \big( b_L - r_L \big) \, \big( b_L + s_L \big) \; \big(  r_L + a_L \big) \, \big( s_L - a_L \big)  }{ \big( b_L + r_L \big) \, \big( b_L - s_L \big) \; \big(  r_L  - a_L\big) \, \big( s_L + a_L \big)  }
\bigg]\,
\eea
in terms of (\ref{map-strip-to-rhp}) and (\ref{L-notation}), where $a< r< s< b$.
The expression for the R\'enyi entropies is obtained by first specialising (\ref{integ-contour}) to the endpoints of $A_\epsilon$ 
and then expanding as $\epsilon \to 0^+$.
By employing \eqref{map-strip-to-rhp}, this gives
\bea
\label{integ-contour-homo-eps}
\mathcal{I}(a+ \epsilon \, , b - \epsilon) 
&=&
2\,
\log \! \Bigg[ \, \frac{2}{\epsilon}\,\sqrt{\frac{\hat{w}(a) \, \hat{w}(b) }{ \hat{w}'(a) \, \hat{w}'(b) }} \; \frac{ \hat{w}(b)-\hat{w}(a) }{ \hat{w}(a)+\hat{w}(b) } \, \Bigg]
+O(\epsilon)
\\
\rule{0pt}{.9cm}
& = & 
2 \,
\log \! \Bigg( \, \frac{8L}{\pi\, \epsilon} \, \sqrt{a_L \, b_L}  \; 
\cos\! \left[ \frac{\pi (a+L)}{4L} \right] \, \cos\! \left[ \frac{\pi (b+L)}{4L} \right] \;
\frac{b_L - a_L}{a_L + b_L}
\,  \Bigg)
+O(\epsilon)\,.
\nn
\eea
From (\ref{contour-function-cft-homo}) and (\ref{integ-contour-homo-eps}), for the R\'enyi entropies of an interval $A$ in the segment we find 
\be
\label{renyi-entropies-homo}
S_A^{(n)} \,=\,
 \frac{n+1}{6\, n} \,
\log \! \Bigg( \, \frac{4L}{\pi\, \epsilon} \; 
\sqrt{ \cos\! \left( \frac{\pi a}{2L} \right) \, \cos\! \left( \frac{\pi b}{2L} \right) }\;
\frac{ \sin\!\big[ \frac{\pi( b-a)}{4L}\big]  }{ \cos\!\big[ \frac{\pi(a+b)}{4L}\big] } 
\,  \Bigg)
+O(1)
\ee
which is the result obtained in \cite{Rodriguez-Laguna:2016roi}
by employing that $S_A^{(n)}$ can be expressed as 
the two-point function of branch-point twist fields \cite{Calabrese:2004eu}
located at the endpoints of the interval $A$.
We remark that the procedure employed above to obtain the entanglement entropies (\ref{renyi-entropies-homo}) 
is based on the contour function (\ref{contour-function-cft-homo}) and therefore it is 
different from the twist fields method.

It is instructive to consider (\ref{renyi-entropies-homo})
in some relevant limiting regimes. 

In the limit $L \to +\infty$, the argument of the logarithm in (\ref{renyi-entropies-homo}) simplifies to $\tfrac{b-a}{\epsilon}$
and  the R\'enyi entropies of an interval in the infinite line are recovered, as expected. 

Another regime that is worth considering for the massless Dirac field 
corresponds to the interval in the half line, whose R\'enyi entropies have been discussed in \cite{Mintchev:2020uom}.
This limit can be explored by setting  $a = \hat{a} - L$, $b = \hat{b} - L$ and $x = \hat{x} - L$ in (\ref{renyi-entropies-homo})  first 
and then taking $L \to +\infty$, as done above for $\beta_{\textrm{\tiny loc}}( x) $ and $\beta_{\textrm{\tiny biloc}}( x)$.
Following these steps, 
for the argument of the logarithm in (\ref{renyi-entropies-homo}) 
one obtains $2\,\sqrt{\hat{a} \,\hat{b}} \, ( \hat{b} - \hat{a} ) / [( \hat{a} + \hat{b}) \epsilon]  + O(1/L^2)$;
hence, the result obtained in  \cite{Mintchev:2020uom} is recovered. 

The limit of (\ref{renyi-entropies-homo}) as the interval $A$ becomes adjacent 
to the boundary is more subtle because 
the regularisation procedure discussed in \cite{Ohmori:2014eia, Cardy:2016fqc} 
requires that $b+\epsilon < L$ and $a-\epsilon > -L$.
However, by setting  e.g. $b = L - \epsilon$ in (\ref{renyi-entropies-homo})
and expanding as $\epsilon \to 0^+$, 
the leading term of the argument of the logarithm is
$\sqrt{ \tfrac{8 L}{\pi \epsilon} \cos\! \big( \tfrac{\pi a}{2 L}\big) \big[ 1+ O(\epsilon)\big]}$; hence, the result obtained in \cite{Tonni:2017jom} for the interval adjacent to the boundary is recovered 
(up to a factor of $2$ in the definition of the UV cutoff).

We also find it insightful to introduce the following contour function for the entanglement entropies over the entire spatial segment
\be
\label{contour-function-cft-homo-AB}
C^{(n)}_{A,B}(x) 
\equiv
\frac{c}{12} \left(1 + \frac{1}{n} \right) \frac{1}{ \big| \beta_{\textrm{\tiny loc}}( x) \big|} 
\;\;\;\;\qquad\;\;\;
x\in A \cup B
\ee
which becomes (\ref{contour-function-cft-homo}) when $x\in A$.
A candidate for the entanglement entropies of $B$ is obtained by integrating (\ref{contour-function-cft-homo-AB}) in $B_\epsilon \subsetneq B$,
where $B_\epsilon \equiv [\,-L\,, a - \epsilon\,] \cup [\, b +\epsilon\,, L\,]$.
As a consistency check of (\ref{contour-function-cft-homo-AB}), by employing (\ref{integ-contour})
we find that $S_A^{(n)} = S_B^{(n)}$ at leading order when $\epsilon \to 0^+$
(in this computation it has been used that 
$\mathcal{I}(-L+\epsilon_0, L-\epsilon_0) \to 0$ as $\epsilon_0 \to 0^+$),
as expected from the fact that the entire system is in a pure state.
Notice that, while the identity $S_A^{(n)} = S_B^{(n)}$ is straightforward when the twist fields method is employed, it provides a non trivial consistency check when the entanglement entropies are obtained from their contour function.

\section{Entanglement Hamiltonian in inhomogeneous backgrounds}
\label{sec-EH-non-homo}

In this section we extend the analysis of Sec.\,\ref{sec-EH-homo} 
about the entanglement Hamiltonian 
and the contour function of the entanglement entropies
to the specific class of spatially inhomogeneous backgrounds
defined by (\ref{metric}).
The expressions for these quantities for a generic background 
belonging to this class 
are given in Sec.\,\ref{subsec-generic-background}, 
while in Sec.\,\ref{subsec-CFT-rainbow-model} 
they are specialised  to the rainbow model, 
providing the predictions for the continuum limit 
of the numerical analysis in the rainbow chain, 
discussed in Sec.\,\ref{sec-EH-rainbow-chain}.

\subsection{Generic background}
\label{subsec-generic-background}

We consider the class of two-dimensional manifolds 
given by the strip $\mathbb{S}$ introduced in Sec.\,\ref{sec-EH-homo} 
equipped with a metric of the following form
\be
\label{metric}
    ds^2 = \e^{2\sigma(x)} \, \rd t_{\textrm{\tiny E}}^2 + \rd x^2 = \e^{2\sigma(x)} \, (\rd t_{\textrm{\tiny E}}^2 + \rd \tilde{x}^2) = \e^{2\sigma(x)} \, \rd \zeta \, \rd \bar{\zeta}\,
\ee
where we restrict to the cases having $\sigma(-x)=\sigma(x)$ and 
$\tilde{x}(x)$ is defined as follows 
\be
\label{weyl-factor}
    \tilde{x}(x) \equiv \int_0^x \! \e^{-\sigma(y)} \, \rd y \;\;\;\;\qquad\;\;\;\; \tilde{x}'(x) = \e^{-\sigma(x)} \;\;\;\;\qquad\;\;\;\; \zeta \equiv \tilde{x} + \ri \, t_{\textrm{\tiny E}} \;.
\ee
Notice that $\tilde{x}(x)$ is an odd function. 
The inverse function of $\tilde{x}(x)$ 
will be denoted by $\tilde{f}$ in the following, 
namely $x =\tilde{f}(\tilde{x})$.
The Ricci scalar of the metric in (\ref{metric}) is 
$\mathcal{R} = - 2 \, \big[\sigma'(x)^2 + \sigma''(x) \big]$.
From the second equality in (\ref{metric}), notice that the Riemannian manifold that we are considering is conformally equivalent to the flat strip $\widetilde{\mathbb{S}}$, described by the coordinates $(\tilde{x} , t_{\textrm{\tiny E}})$ in Euclidean signature and  whose width is given by $2\tilde{L}$,
with $\tilde{L} \equiv \tilde{x}(L)$.

The inhomogeneous backgrounds characterised by (\ref{metric}) occur in the continuum limit of one-dimensional free fermionic systems 
with a Fermi velocity that depends on the position, i.e. 
$v_{\textrm{\tiny F}}(x) = J(x) = \e^{\sigma(x)}$
(see also the Hamiltonian (\ref{inhomo-chain-ham})).
This includes the rainbow chain 
\cite{Vitagliano:2010db,Ram_rez_2014,Ramirez:2015yfa,Rodr_guez_Laguna_2016,Rodriguez-Laguna:2016roi} 
(see also \cite{SamosSaenzdeBuruaga:2019vkx,Ramirez:2020ppf,Mula:2020udv,Mula:2022lsj,Santalla:2022ygq,Byles:2023hjx,Szabo:2024rca,Bonsignori:2024gky} and the discussions in Sec.\,\ref{subsec-CFT-rainbow-model} and Sec.\,\ref{sec-EH-rainbow-chain}),
the Fermi gas trapped in a harmonic potential 
\cite{Jaksch:1998zz,Campostrini:2009ema,Campostrini_2010_3,Campostrini_2010_2,Campostrini:2010pv,Vicari_2012,Eisler_2013,Marino_2014,Calabrese_2015},
the gradient chain
\cite{Smith_1971,Saitoh_1973,Eisler_2009,Bonsignori:2024gky} 
and other interesting models discussed e.g. in 
\cite{Allegra:2015hxc, Dubail:2016tsc}.
Furthermore, the Lorentzian metrics obtained by setting 
$t = -\textrm{i} t_{\textrm{\tiny E}}$ 
in (\ref{metric}) 
are also called optical metrics \cite{Lewenstein:2012,Boada:2010sh,Rodriguez-Laguna:2016kri,Mula:2020udv}
and they occur in the study of the light propagation 
in a medium with an inhomogeneous index of refraction.

We consider a free massless Dirac field on this curved background in Euclidean signature,
whose action reads
\be
    \label{action-dirac-strip-sigma}
    S[\psi] 
    \propto 
    \int_{\mathbb{S}} \e^\sigma \Big(\, 
    \psi^\ast_2 \!
    \stackrel{\leftrightarrow}{\partial_{\bar{\zeta}}} \! \psi_2 
    + 
    \psi^\ast_1 \!
    \stackrel{\leftrightarrow}{\partial_{\zeta}} \! \psi_1 
    \,\Big) \,
    \rd \zeta \, \rd \bar{\zeta} \;.
\ee
The boundary conditions for the components of the Dirac field are obtained from the ones described in Sec.\,\ref{sec-EH-homo} in the flat strip (see (\ref{strip-bc-vector}) and (\ref{strip-bc-axial}))
through the map $\tilde{x}(x)$ in (\ref{weyl-factor})
and the Weyl transformation in (\ref{metric}).
Here the flat strip to consider is $\widetilde{\mathbb{S}}$, 
whose boundaries are parameterised by $ \mp \,\tilde{L} + \ri \, t_{\textrm{\tiny E}}$
with $ t_{\textrm{\tiny E}} \in \mathbb{R}$.
By performing the change of variable given by $\tilde{x}(x)$ in the boundary conditions (\ref{strip-bc-vector}) and (\ref{strip-bc-axial}) specified for $\widetilde{\mathbb{S}}$,
where $\tilde{x}'(x) $ in (\ref{weyl-factor}) is real and positive, 
and considering the factors provided by the Weyl transformation, one finds that the boundary conditions supporting (\ref{action-dirac-strip-sigma}) are  given by (\ref{strip-bc-vector}) and (\ref{strip-bc-axial}) also in the inhomogeneous case that we are considering.

The entanglement Hamiltonian $K^{\textrm{\tiny $(\sigma)$}}_{A} $ of an interval $A$ in the segment at $t_{\textrm{\tiny E}} = 0$ embedded in the strip $\mathbb{S}$ can be found from the entanglement Hamiltonian $K_{A} $ in the flat strip $\widetilde{\mathbb{S}}$, given in (\ref{EH-segment-homo}), by adapting to this case 
the procedure that provides $K_{A} $ from (\ref{EH-interval-half-line}), discussed in Sec.\,\ref{sec-EH-homo}.
The result is 
\be
    \label{EH-segment-in-homo}
    K^{\textrm{\tiny $(\sigma)$}}_{A} 
    \,=\,
    2\pi
    \int_{a}^{b} \!
    \beta^{\textrm{\tiny $(\sigma)$}}_{\textrm{\tiny loc}}(x) \, \mathcal{E}(x)\, \rd x
    +
    2\pi 
    \int_{a}^{b} \!
    \beta^{\textrm{\tiny $(\sigma)$}}_{\textrm{\tiny biloc}}(x) \, T^{\,\textrm{\tiny $(\alpha)$}}_{\textrm{\tiny biloc}}
    (x, x_{\textrm{c},\sigma} ) 
    \, \rd x
\ee
where the local operator $\mathcal{E}(x)$ and 
the bilocal operator $T^{\,\textrm{\tiny $(\alpha)$}}_{\textrm{\tiny biloc}}(x, y ) $ 
have the same form as the corresponding ones occurring in (\ref{EH-segment-homo}), with the fields in $\widetilde{\mathbb{S}}$ replaced by the corresponding fields in the curved manifold $\mathbb{S}$.

The weight function $\beta^{\textrm{\tiny $(\sigma)$}}_{\textrm{\tiny loc}}(x) $ in the local term of (\ref{EH-segment-in-homo})
can be found 
by adapting to this case the calculation performed above to obtain (\ref{beta-loc-homo}),
discussed in Sec.\,\ref{sec-EH-homo}.
In particular, combining (\ref{beta-loc-homo}), (\ref{weyl-factor}) and the fact that 
the energy-momentum tensor has conformal dimension equal to $2$,
for the weight function $\beta^{\textrm{\tiny $(\sigma)$}}_{\textrm{\tiny loc}}(x) $ we find 
\be
\label{beta-loc-non-homo}
\beta^{\textrm{\tiny $(\sigma)$}}_{\textrm{\tiny loc}}  (x) 
\,\equiv\,
\frac{ \tilde{\beta}_{\textrm{\tiny loc}}\big(\tilde{x}(x) \big) }{ \tilde{x}'(x)  }
\,=\,
\e^{\sigma(x)}\,
\frac{4 \tilde{L}}{\pi} 
\bigg[  \cos\! \bigg( \frac{\pi ( \tilde{x} + \tilde{L})}{4 \tilde{L}} \bigg) \bigg]^2
\frac{\big( \, \tilde{b}_L^2 - \tilde{x}_L^2\big) \big( \tilde{x}_L^2 - \tilde{a}_L^2 \big)}{ 2\big( \, \tilde{b}_L - \tilde{a}_L\big) \big( \tilde{a}_L \,\tilde{b}_L + \tilde{x}_L^2\big)}
\ee
where $\tilde{\beta}_{\textrm{\tiny loc}}(y)$ is defined as (\ref{beta-loc-homo}) 
with $a$, $b$ and $L$ replaced by $\tilde{a} \equiv \tilde{x}(a)$, $\tilde{b}\equiv \tilde{x}(b)$ and $\tilde{L}\equiv \tilde{x}(L)$ respectively through (\ref{weyl-factor}),
and we have introduced
\be
\label{x-tilde-notation-L}
\tilde{x}_L 
\equiv  
\tan\! \bigg[ 
\frac{\pi \big( \tilde{x}(x)+ \tilde{L} \big)}{4 \tilde{L}} 
\bigg]
\;\;\;\qquad\;\;\;
\tilde{a}_L \equiv \tilde{x}_L \big|_{x=a}
\qquad
\tilde{b}_L \equiv \tilde{x}_L \big|_{x=b}
\ee
in terms of the notation defined in  
(\ref{L-notation}).
By employing (\ref{beta-loc-homo-v1}), (\ref{weyl-factor}) and (\ref{x-tilde-notation-L}), 
the weight function (\ref{beta-loc-non-homo}) can also be written 
in the following form
\be
\label{beta-loc-non-homo-v1}
\beta^{\textrm{\tiny $(\sigma)$}}_{\textrm{\tiny loc}}  (x) 
\,=\,
\e^{\sigma(x)}\;
\frac{2\tilde{L}}{\pi}\; 
\frac{ 
\big[ \sin\!\big( \pi \tilde{b}/(2\tilde{L})\big) - \sin\!\big( \pi \tilde{x}/(2\tilde{L})\big)  \big] \,  \big[  \sin\!\big( \pi \tilde{x}/(2\tilde{L})\big)  -  \sin\!\big( \pi \tilde{a}/(2\tilde{L})\big)  \big]
}{
\sin\!\big( \pi (\tilde{b} - \tilde{a})/(2\tilde{L})\big)  + \big[  \cos\!\big( \pi \tilde{b}/(2\tilde{L})\big) - \cos\!\big( \pi \tilde{a}/(2\tilde{L})\big)  \big] \, \sin\!\big( \pi \tilde{x}/(2\tilde{L})\big) 
} \;.
\ee

The weight function $\beta^{\textrm{\tiny $(\sigma)$}}_{\textrm{\tiny biloc}}(x) $ in the bilocal term of (\ref{EH-segment-in-homo})
is obtained by adapting the calculation that gives 
(\ref{beta-biloc-homo}). 
In order to write the result,
let us first employ
(\ref{x-conj}) and (\ref{x-tilde-notation-L})
to define
\be
    \label{x-conj-tilde-def}
    \tilde{x}_{\textrm{c}}(x)
    \equiv
     \frac{4 \tilde{L}}{\pi} \, \arctan \! \bigg( \frac{ \tilde{a}_L  \, \tilde{b}_L}{ \tilde{x}_L }   \bigg)
     - \tilde{L}
\ee
that provides the point $x_{\textrm{c}, \sigma}$ conjugate  to $x$
as follows 
\be
    \label{x-conj-sigma-def}
    x_{\textrm{c}, \sigma}(x) 
    =
    \tilde{f} \big(\tilde{x}_{\textrm{c}}(x) \big)
\ee
through the inverse function of $\tilde{x}(x)$,
introduced in the text below (\ref{weyl-factor}).
Notice that $x_{\textrm{c}, \sigma}(x) \in A$ when $x \in A$.
Indeed, first one observes that
$x_{\textrm{c}, \sigma}(a) = b$, $x_{\textrm{c}, \sigma}(b) = a$
and $\tilde{a}_L < \tilde{x}_L < \tilde{b}_L$ for $a < x < b$.
This implies that the argument of the arctan in (\ref{x-conj-tilde-def}) belongs to $\big(\tilde{a}_L\,, \tilde{b}_L \big)$.
Then, since both arctan and $\tilde{f}$ are monotonic functions, 
we have that  $x_{\textrm{c}, \sigma}(x) \in A$ when $x \in A$.
The self-conjugate point at $x= x_{\textrm{sc}, \sigma}$ 
is defined by the condition 
$x_{\textrm{c}, \sigma}(x_{\textrm{sc}, \sigma}) = x_{\textrm{sc}, \sigma}$,
which is typically a transcendental equation, 
depending on $\sigma(x)$.

An interesting feature of (\ref{x-conj-sigma-def}) concerns 
the special case where $A$ is in the centre  of the segment, 
i.e. $a = - b$ with $0< b < L$.
In this case, since $\sigma(-x) = \sigma(x)$, 
we have that $\tilde{a} = -\tilde{b}$
and $\tilde{a}_L = 1 \, / \, \tilde{b}_L$.
Then, by using also the identity
$\arctan[\cot(y + \pi / 4)] = \pi/4 - y$ for $y \in (-\pi/4, \pi/4)$, for (\ref{x-conj-tilde-def}) 
we find that $\tilde{x}_{\textrm{c}}(x) = - \tilde{x}$,
which can be combined with the fact that $\tilde{f}$ is an odd function, 
finding that (\ref{x-conj-sigma-def}) for $a = - b$
simplifies to 
$x_{\textrm{c}, \sigma}(x) \big|_{a \,= \,- b} = \,- \,x$.
Hence, in this symmetric configuration the self-conjugate point is the central point of the segment, namely
$x_{\textrm{sc}, \sigma} \big|_{a \,= \,- b} = 0$.

The weight function $\beta^{\textrm{\tiny $(\sigma)$}}_{\textrm{\tiny biloc}}(x) $
can be written in terms of 
(\ref{weyl-factor}) and (\ref{x-conj-sigma-def})
as follows  
\be
    \label{beta-biloc-non-homo-gen}
    \beta_{\textrm{\tiny biloc}}^{\textrm{\tiny $(\sigma)$}}(x) 
    = 
    \sqrt{ \frac{\tilde{x}'(x)}{ 
    \tilde{x}'\big( x_{\textrm{c}, \sigma}(x) \big)} }\;
    \tilde{\beta}_{\textrm{\tiny biloc}}\big(\tilde{x}(x) \big) 
\ee
being $\tilde{\beta}_{\textrm{\tiny biloc}}(x)$ defined as (\ref{beta-biloc-homo})  
(see also the equivalent expressions  in (\ref{beta-biloc-homo-v2}) and (\ref{beta-biloc-homo-v1}))
with $a$, $b$ and $L$ replaced by $\tilde{a}$, $\tilde{b}$ and $\tilde{L}$ respectively.
More explicit expressions for the weight function 
(\ref{beta-biloc-non-homo-gen}) are obtained from either (\ref{beta-biloc-homo}) 
or (\ref{beta-biloc-homo-v2}), 
finding respectively
\bea
\label{beta-biloc-non-homo-V1}
  \rule{0pt}{.7cm}
\beta_{\textrm{\tiny biloc}}^{\textrm{\tiny $(\sigma)$}}(x) 
&=&
\e^{[ \sigma(x_{\textrm{c} , \sigma} ) -\sigma(x) ] /2}\;
\frac{ \cos\!\Big[ \frac{\pi( \tilde{x}_{\textrm{c}} + \tilde{L})}{4\tilde{L}}\Big]  }{ \cos\!\Big[ \frac{\pi(\tilde{x} + \tilde{L})}{4 \tilde{L}}\Big] } 
\;\frac{
\tilde{a}_L  \, \tilde{b}_L \,\big( \tilde{b}_L^2 - \tilde{x}_L^2 \big) \big( \tilde{x}_L^2 - \tilde{a}_L^2 \big) 
}{
2 \big( \tilde{b}_L - \tilde{a}_L \big) \, \tilde{x}_L \, \big( \tilde{a}_L\,\tilde{b}_L + \tilde{x}_L^2 \big)^2
}
\\
\label{beta-biloc-non-homo-V2}
\rule{0pt}{.9cm}
&=&
\frac{ \e^{[ \sigma(x_{\textrm{c}, \sigma} ) -\sigma(x) ] /2}\;
\tilde{a}_L  \, \tilde{b}_L \,\big( \tilde{b}_L^2 - \tilde{x}_L^2 \big) \big( \tilde{x}_L^2 - \tilde{a}_L^2 \big) 
}{ 
2 \big( \tilde{b}_L - \tilde{a}_L \big) \, \cos\!\Big[ \frac{\pi(\tilde{x} +\tilde{L})}{4\tilde{L}}\Big] \, 
\sqrt{\tilde{a}^2_L\,\tilde{b}^2_L + \tilde{x}_L^2} \; 
\big( \tilde{a}_L\,\tilde{b}_L + \tilde{x}_L^2 \big)^{2} }
\eea
in terms of (\ref{x-tilde-notation-L}), (\ref{x-conj-tilde-def}) and (\ref{x-conj-sigma-def}).

The  weight function 
$\beta_{\textrm{\tiny biloc}}^{\textrm{\tiny $(\sigma)$}}(x)$ in 
(\ref{beta-biloc-non-homo-V2})
is related to $\beta_{\textrm{\tiny loc}}^{\textrm{\tiny $(\sigma)$}}(x)$ in (\ref{beta-loc-non-homo}) 
as follows
\begin{equation}
\label{beta-loc-biloc-relation-v3}
    \beta_{\textrm{\tiny biloc}}^{\textrm{\tiny $(\sigma)$}}(x) 
\,=\,
 \e^{[ \sigma(x_{\textrm{c}, \sigma} ) -3\sigma(x) ] /2}\;
 \frac{\pi}{4\tilde{L}}
\;
\frac{
\tilde{a}_L  \, \tilde{b}_L \, \big(1+\tilde{x}_L^2\big)^{3/2}
}{ 
\sqrt{\tilde{a}^2_L\,\tilde{b}^2_L + \tilde{x}_L^2} \; 
 \big( \tilde{a}_L\,\tilde{b}_L + \tilde{x}_L^2\big)
 }
 \;\beta_{\textrm{\tiny loc}}^{\textrm{\tiny $(\sigma)$}}(x)
\end{equation}
which becomes (\ref{beta-loc-biloc-relation-homo}) when $\sigma(x)$ vanishes identically, as expected. 
Combining (\ref{beta-biloc-non-homo-gen}) and (\ref{beta-biloc-strip-to-loc-step4}), 
the relation 
(\ref{beta-loc-biloc-relation-v3}) 
can be written also as
\begin{eqnarray}
\label{beta-loc-biloc-relation-v4}
\beta_{\textrm{\tiny biloc}}^{\textrm{\tiny $(\sigma)$}}(x) 
&=&
\e^{[ \sigma(x_{\textrm{c} , \sigma} ) -\sigma(x) ] /2}\;
\frac{\pi}{4 \tilde{L}}
    \;
    \frac{\sqrt{\tilde{a}^2_L\,\tilde{b}^2_L + \tilde{x}_L^2}}{\sin\!\Big[ \frac{\pi(\tilde{x} + \tilde{L})}{4\tilde{L}}\Big]}
    \;\frac{\tilde{\beta}_{\textrm{\tiny loc}}(\tilde{x}_{\textrm{c}})}{\tilde{x}_L + \tilde{x}_{\textrm{c},L}}
    \nonumber
    \\
    \rule{0pt}{1.cm}
&=&
\e^{-[ \sigma(x_{\textrm{c} , \sigma} ) + \sigma(x) ] /2}\;
\frac{\pi}{4 \tilde{L}}\;
    \frac{\sqrt{\tilde{a}^2_L\,\tilde{b}^2_L + \tilde{x}_L^2}}{\sin\!\Big[ \frac{\pi(\tilde{x} + \tilde{L})}{4\tilde{L}}\Big]}
    \;\frac{
    \beta^{\textrm{\tiny $(\sigma)$}}_{\textrm{\tiny loc}}  (x_{\textrm{c}, \sigma}) 
    }{\tilde{x}_L + \tilde{x}_{\textrm{c},L}}
\end{eqnarray}
where the last expression has been obtained by using that,
from the first equality in (\ref{beta-loc-non-homo}) 
evaluated in $x_{\textrm{c}, \sigma}$, we have
\begin{equation}
\tilde{\beta}_{\textrm{\tiny loc}}\big(\tilde{x}(x_{\textrm{c}, \sigma}) \big)
\,=\,
\e^{-\sigma(x_{\textrm{c}, \sigma}) } \,
\beta^{\textrm{\tiny $(\sigma)$}}_{\textrm{\tiny loc}}  (x_{\textrm{c}, \sigma}) 
\end{equation}
whose l.h.s. is $\tilde{\beta}_{\textrm{\tiny loc}}(\tilde{x}_{\textrm{c}})$ 
because $\tilde{x}(x_{\textrm{c}, \sigma}) = \tilde{x}_{\textrm{c}}$,
from (\ref{x-conj-sigma-def}).

By adapting the analysis of \cite{Bonsignori:2024gky} to our setup, 
it could be worth introducing 
\begin{equation}
\label{gamma-weights-def-sigma}
\gamma_{\textrm{\tiny loc}}^{\textrm{\tiny $(\sigma)$}}(x) 
\,\equiv\,
\e^{-\sigma(x)} \,
\beta_{\textrm{\tiny loc}}^{\textrm{\tiny $(\sigma)$}}(x) 
    \;\;\;\;\;\qquad\;\;\;\;\;
\gamma_{\textrm{\tiny biloc}}^{\textrm{\tiny $(\sigma)$}}(x) 
\,\equiv\,
\e^{-[\sigma(x_{\textrm{c}, \sigma}) -\sigma(x)]/2} \,
\beta_{\textrm{\tiny biloc}}^{\textrm{\tiny $(\sigma)$}}(x)\,. 
\end{equation}
For instance, in the case of the rainbow model, whose weight functions are discussed in Sec.\,\ref{subsec-CFT-rainbow-model}, 
the auxiliary weight functions (\ref{gamma-weights-def-sigma}) do not display a singular behaviour, 
unlike the corresponding weight functions.

The derivation of the entanglement entropies $S_A^{(n)}$ 
from their contour function that we discussed in Sec.\,\ref{sec-EH-homo} 
for the homogeneous case can be straightforwardly adapted to the inhomogeneous background that we are investigating.
Following \cite{Tonni:2017jom}, 
from the weight function 
$\beta^{\textrm{\tiny $(\sigma)$}}_{\textrm{\tiny loc}}( x)$ reported in (\ref{beta-loc-non-homo}) or in (\ref{beta-loc-non-homo-v1}), we introduce the contour function for the entanglement entropies given by 
\be
\label{contour-non-homo-A}
C^{(n)}_{A; \textrm{\tiny $(\sigma)$}}(x) 
\equiv
\frac{c}{12} \left(1 + \frac{1}{n} \right) \frac{1}{\beta^{\textrm{\tiny $(\sigma)$}}_{\textrm{\tiny loc}}( x) } 
\;\;\;\;\qquad\;\;\;\;
x \in A
\ee
where $c=1$ for the massless Dirac field.
By employing the expression of $\beta^{\textrm{\tiny $(\sigma)$}}_{\textrm{\tiny loc}}( x)$ in (\ref{beta-loc-non-homo}) and the integral in (\ref{integ-contour}), 
it is straightforward to observe that the integral occurring in the determination of the entanglement entropies from the contour function (\ref{contour-non-homo-A}) reads
\be
\label{integ-contour-non-homo}
\mathcal{I}_{\textrm{\tiny $(\sigma)$}}(r,s) 
\,\equiv 
\int_{ r }^{ s }  
\! \frac{ \rd x }{\beta^{\textrm{\tiny $(\sigma)$}}_{\textrm{\tiny loc}}( x) } 
\, =
\int_{ \tilde{x} (r) }^{ \tilde{x} (s) }  
\!\frac{ \rd y }{ \tilde{\beta}_{\textrm{\tiny loc}}( y ) }
\,=\,
 \, \log \! \Bigg[ 
\frac{ \big( \, \tilde{b}_L - \tilde{r}_L \big) \, \big( \, \tilde{b}_L + \tilde{s}_L \big) \; \big(  \tilde{r}_L + \tilde{a}_L \big) \, \big( \tilde{s}_L - \tilde{a}_L \big)  
}{ 
\big( \, \tilde{b}_L + \tilde{r}_L \big) \, \big( \, \tilde{b}_L - \tilde{s}_L \big) \; \big(  \tilde{r}_L  - \tilde{a}_L\big) \, \big( \tilde{s}_L + \tilde{a}_L \big)  }
\Bigg]
\ee
where $a < r < s < b$ and 
we have defined $\tilde{r}_L \equiv \tilde{x}_L |_{x=r}$ and $\tilde{s}_L \equiv \tilde{x}_L |_{x=s}$,
according to the notation introduced in (\ref{x-tilde-notation-L}).
The entanglement entropies $S_A^{(n)}$ are obtained from  (\ref{integ-contour-non-homo}), 
by setting $r =a+\epsilon$ and $s =b-\epsilon$ first and then
expanding as $\epsilon \to 0^+$, 
as done in Sec.\,\ref{sec-EH-homo} 
to obtain (\ref{renyi-entropies-homo}).
At leading order, 
since $\tilde{r}_L = \tilde{a}_L + \tilde{x}_L'(a) \, \epsilon + O( \epsilon^2)$ 
and $\tilde{s}_L = \tilde{b}_L - \tilde{x}_L'(b) \, \epsilon + O( \epsilon^2)$,
we find
\bea
\label{integ-contour-non-homo-eps}
\mathcal{I}_{\textrm{\tiny $(\sigma)$}}(a+ \epsilon \, , b - \epsilon) 
&=&
2\,
\log \! \Bigg[ \, \frac{2}{\epsilon}\,\sqrt{\frac{  \tilde{a}_L \, \tilde{b}_L }{  \tilde{x}_L'(a) \, \tilde{x}_L'(b) }} \; \, \frac{ \tilde{b}_L  - \tilde{a}_L }{ \tilde{a}_L + \tilde{b}_L  } \, \Bigg]
+O(\epsilon)
\\
\rule{0pt}{.9cm}
& & \hspace{-1.cm} 
=\, 2 \,
\log \! \Bigg( \, \frac{8 \tilde{L}}{\pi\, \epsilon} \, \sqrt{ \frac{ \tilde{a}_L \, \tilde{b}_L }{ \tilde{x}'(a) \, \tilde{x}'(b) } }  \; 
\cos\! \left[ \frac{\pi \big( \tilde{a}+\tilde{L} \big)}{4\tilde{L}} \right] \, \cos\! \left[ \frac{\pi \big( \,\tilde{b}+\tilde{L} \big)}{4\tilde{L}} \right] \;
\frac{ \tilde{b}_L - \tilde{a}_L}{ \tilde{a}_L + \tilde{b}_L}
\,  \Bigg)
+O(\epsilon)
\nn
\eea
where in the last step we also exploited that, from (\ref{x-tilde-notation-L}), we have 
\be
\label{x-tilde-L-prime}
\tilde{x}'_L (x)  = \frac{\pi \; \tilde{x}'(x) }{4\tilde{L}\, \Big(\! \cos\!\Big[ \frac{\pi( \tilde{x}(x)+\tilde{L})}{4 \tilde{L}}\Big] \Big)^2} \;.
\ee
The entanglement entropies of the interval $A$ in the segment for the massless Dirac field 
in the inhomogeneous background characterised by (\ref{metric}) and in its ground state
are obtained from (\ref{contour-non-homo-A}) 
by using (\ref{integ-contour-non-homo-eps}), 
and read
\be
\label{renyi-entropies-non-homo}
S_A^{(n)} \,=\,
 \frac{n+1}{6\, n} \,
\log \! \left(  \frac{4 \tilde{L} \; \e^{[ \sigma(a) + \sigma(b) ]/2} }{\pi\, \epsilon } \; 
\sqrt{ \cos\! \bigg( \frac{\pi \, \tilde{a} }{2\tilde{L}} \bigg) \, \cos\! \bigg( \frac{\pi \,\tilde{b} }{2\tilde{L}} \bigg) }\;\,
\frac{ \sin\!\left[ \frac{\pi(  \tilde{b}- \tilde{a})}{4 \tilde{L} }\right]  }{ \cos\!\left[ \frac{\pi (\tilde{a}+\tilde{b} )}{4\tilde{L}}\right] } 
\,  \right)
+O(1) \,.
\ee
Since this expression coincides with the result obtained in 
\cite{Rodriguez-Laguna:2016roi} through the twist fields method,
this computation provides an important consistency check for 
(\ref{contour-non-homo-A}).

In the case of the homogeneous segment, 
the contour function for the entanglement entropies  on the entire segment has been introduced in (\ref{contour-function-cft-homo-AB}), showing that it is consistent with the identity $S_A^{(n)} = S_B^{(n)}$ at leading order as $\epsilon \to 0^+$,
as expected from the purity of the ground state. 
Similarly, in the inhomogeneous background 
characterised by (\ref{metric}),  
the contour function for the entanglement entropies on the entire segment can be defined through 
the weight function $\beta^{\textrm{\tiny $(\sigma)$}}_{\textrm{\tiny loc}}( x)$ (see (\ref{beta-loc-non-homo}) or (\ref{beta-loc-non-homo-v1})) as 
\be
\label{contour-non-homo-sigma-AB}
C^{(n)}_{A, B; \textrm{\tiny $(\sigma)$}}(x) 
\equiv
\frac{c}{12} \left(1 + \frac{1}{n} \right) \frac{1}{ \big| \beta^{\textrm{\tiny $(\sigma)$}}_{\textrm{\tiny loc}}( x) \big|} 
\;\;\;\;\qquad\;\;\;
x\in A \cup B
\ee
that becomes (\ref{contour-non-homo-A}) for $x\in A$.
The expression (\ref{contour-non-homo-sigma-AB})
provides a contour function of the entanglement entropies for $x \in B$;
hence, $S_B^{(n)}$ can be found by integrating 
over $B_\epsilon \equiv [\,-L\,, a - \epsilon\,] \cup [\, b +\epsilon\,, L\,]$.
As a consistency check, 
by using the property $\sigma(-x) = \sigma(x)$ 
and the fact that $\mathcal{I}(-\tilde{L}+\epsilon_0, \tilde{L}-\epsilon_0) \to 0$  as $\epsilon_0 \to 0^+$,
we verified that $S_A^{(n)} = S_B^{(n)}$ 
at leading order as $\epsilon \to 0^+$.

In the remaining part of this subsection, we explore some limiting regimes for the entanglement Hamiltonian, the contour functions and the entanglement entropies,
by assuming that  $\sigma(x)$ is independent of the parameters $a$, $b$ and $L$.

First, we verify that the main results reported in Sec.\,\ref{sec-EH-homo} are recovered 
when $\sigma(x)$ vanishes identically, as expected. 
From (\ref{weyl-factor}), one observes  
that $\tilde{x}(x)$ becomes simply $x$, hence any tilded quantity becomes the corresponding untilded one.
For the weight function of the local term, 
this implies that (\ref{beta-loc-non-homo})
and (\ref{beta-loc-non-homo-v1})
become (\ref{beta-loc-homo}) and (\ref{beta-loc-homo-v1})
respectively.
This tells us that, 
as for the contour function of the entanglement entropies, 
we have that (\ref{contour-non-homo-A}) 
and (\ref{contour-non-homo-sigma-AB})
give (\ref{contour-function-cft-homo}) 
and (\ref{contour-function-cft-homo-AB})  respectively. 
Considering the weight function of the bilocal term, 
in this limiting regime of homogeneous background,
the expressions (\ref{beta-biloc-non-homo-V1}) and (\ref{beta-biloc-non-homo-V2}) 
simplify to (\ref{beta-biloc-homo}) and (\ref{beta-biloc-homo-v2}) respectively.
Finally, 
as for the entanglement entropies, it is straightforward to find that 
(\ref{renyi-entropies-non-homo})  becomes (\ref{renyi-entropies-homo}).

It is interesting to explore the behaviour of the weight functions 
$\beta_{\textrm{\tiny loc}}^{\textrm{\tiny $(\sigma)$}}(x)$ and $\beta_{\textrm{\tiny biloc}}^{\textrm{\tiny $(\sigma)$}}(x)$ 
close to the endpoints of $A$.
First, by using (\ref{x-tilde-L-prime}) and (\ref{weyl-factor}),
we observe that at the endpoints of $A$
the following identities hold 
\be
\label{identity-beta-loc-BW}
\e^{\sigma(a)}\,
\frac{4 \tilde{L}}{\pi} 
\bigg[  \cos\! \bigg( \frac{\pi ( \tilde{a} + \tilde{L})}{4 \tilde{L}} \bigg) \bigg]^2
\tilde{x}'_L(a) 
\,=\;
\e^{\sigma(b)}\,
\frac{4 \tilde{L}}{\pi} 
\bigg[  \cos\! \bigg( \frac{\pi ( \tilde{b} + \tilde{L})}{4 \tilde{L}} \bigg) \bigg]^2
\tilde{x}'_L(b) 
\,=\, 1 \,.
\ee
Then, as for the expansion of (\ref{beta-loc-non-homo})
as $x \to a$ and $x \to b$, 
by employing (\ref{identity-beta-loc-BW}) we find that
\begin{eqnarray}
    \label{beta-loc-non-homo-ent-pts-a}
    \beta^{\textrm{\tiny $(\sigma)$}}_{\textrm{\tiny loc}}(x) &=& 
    (x-a) + O\big((x-a)^2\big) 
    \hspace{1.5cm}
    x \to a
    \\
    \label{beta-loc-non-homo-ent-pts-b}
    \rule{0pt}{0.5cm}
    \beta^{\textrm{\tiny $(\sigma)$}}_{\textrm{\tiny loc}}(x) &=& 
    (b-x) + O\big((x-b)^2\big) 
    \hspace{1.58cm}
    x \to b
\end{eqnarray}
which is the behaviour expected from the Bisognano-Wichmann theorem, for any $\sigma(x)$ in the class that we are considering. 
As for $\beta^{\textrm{\tiny $(\sigma)$}}_{\textrm{\tiny biloc}}(x)$, 
by combining the relation (\ref{beta-loc-biloc-relation-v3}) with the expansions (\ref{beta-loc-non-homo-ent-pts-a})-(\ref{beta-loc-non-homo-ent-pts-b}), 
we find respectively 
\begin{eqnarray}
    \label{beta-biloc-non-homo-ent-pts-a}
    \beta^{\textrm{\tiny $(\sigma)$}}_{\textrm{\tiny biloc}}(x) 
    &=& 
    \e^{[\sigma(b) -3\sigma(a)]/2}\;
\frac{\pi}{4\tilde{L}}
\;
\frac{
\tilde{b}_L\, \big(1+\tilde{a}_L^2\big)^{3/2}
}{
\tilde{a}_L  \big( \tilde{a}_L + \tilde{b}_L \big)\,
\big(1+ \tilde{b}_L^2\big)^{1/2}
}\;
    (x-a) + O\big((x-a)^2\big) 
    \hspace{.8cm}
    x \to a
    \phantom{xxxxxx}
    \\
    \label{beta-biloc-non-homo-ent-pts-b}
    \rule{0pt}{0.9cm}
    \beta^{\textrm{\tiny $(\sigma)$}}_{\textrm{\tiny biloc}}(x) &=& 
    \e^{[\sigma(a ) -3\sigma(b)]/2}\;
\frac{\pi}{4\tilde{L}}
\;
\frac{
\tilde{a}_L\, \big(1+\tilde{b}_L^2\big)^{3/2}
}{
\tilde{b}_L  \big( \tilde{a}_L + \tilde{b}_L \big)\,
\big(1+ \tilde{a}_L^2\big)^{1/2}
}\;
    (b-x) + O\big((x-b)^2\big) 
    \hspace{.87cm}
    x \to b
\end{eqnarray}
where we have also used that 
$x_{\textrm{c}, \sigma}(a) = b$ and $x_{\textrm{c}, \sigma}(b) = a$
(see the text below (\ref{x-conj-sigma-def})).

We find it worth concluding this discussion by considering two limiting regimes where the entanglement Hamiltonian becomes a local operator. 
They correspond to $A$ adjacent to the boundary and to $L \to +\infty$ with the additional condition that $\tilde{L}_\infty$ diverges in this limit.

As for the former limiting case, 
considering e.g. the limit $b \to L$ in the weight function of the local term 
(see (\ref{beta-loc-non-homo}) and (\ref{beta-loc-non-homo-v1}))
and of the bilocal term 
(see (\ref{beta-biloc-non-homo-V1}) and (\ref{beta-biloc-non-homo-V2})),
we find respectively 
\be
    \label{beloc_inhom_btoL}
    \beta_{\textrm{\tiny loc}}^{\textrm{\tiny $(\sigma)$}}(x) 
    \,\to \, 
    \e^{\sigma(x)} \, \frac{2 \tilde{L}}{\pi} \; 
    \frac{\sin\!\big[\pi \tilde{x}/(2 \tilde{L})\big] - \sin\!\big[\pi \tilde{a}/(2 \tilde{L})\big] }{ \cos\!\big[\pi \tilde{a}/(2 \tilde{L})\big] }
    \;\;\;\qquad\;\;\;
    \beta_{\textrm{\tiny biloc}}^{\textrm{\tiny $(\sigma)$}}(x) \,\to \, 0
\ee
where the second result has been obtained by employing the fact that 
$x_{\textrm{c}, \sigma}(x) \to L$ as $b \to L$ in (\ref{x-conj-sigma-def}).
As for the entanglement entropies in this limit, 
the same considerations made in Sec.\,\ref{sec-EH-homo} 
for the homogeneous case 
(see the text below (\ref{renyi-entropies-homo}))
can be adapted to the expression (\ref{renyi-entropies-non-homo}), 
finding the result of \cite{Tonni:2017jom}
(up to a factor of $2$ in the definition of the UV cutoff), namely
\be
    \label{EEs_inhom_btoL}
    S_A^{(n)} \to \, 
    \frac{n + 1}{12 \, n} \, \log\!
    \bigg[\, \e^{\sigma(a)}\; \frac{8 \, \tilde{L}}{\pi \, \epsilon} 
    \,\cos\!\bigg(\frac{\pi \, \tilde{a}}{2 \, \tilde{L}} \bigg) \bigg]
    + O(1) \,.
\ee

In the limit $L \to +\infty$, we have to distinguish 
between the two cases of divergent and finite $\tilde{L}_\infty$,
where $\tilde{L}_\infty$ denotes the limiting value of $\tilde{L}$ as $L \to +\infty$.
When $\tilde{L}_\infty$ diverges (like e.g. in the rainbow model 
discussed in Sec.\,\ref{subsec-CFT-rainbow-model}),
by adapting to (\ref{beta-loc-non-homo})-(\ref{beta-loc-non-homo-v1})
and (\ref{beta-biloc-non-homo-V1}) and (\ref{beta-biloc-non-homo-V2})
the computation performed in Sec.\,\ref{sec-EH-homo} 
for the homogeneous background, we obtain respectively 
\be
    \beta_{\textrm{\tiny loc}}^{\textrm{\tiny $(\sigma)$}}(x) 
    \,\to\,
    \e^{\sigma(x)} \, \frac{(\tilde{b} - \tilde{x}) \, (\tilde{x} - \tilde{a})}{\tilde{b} - \tilde{a}}
    \;\;\;\qquad\;\;\;
    \beta_{\textrm{\tiny biloc}}^{\textrm{\tiny $(\sigma)$}}(x) \,\to \, 0
\ee
where the second result is obtained by using that 
$x_{\textrm{c}, \sigma}(x) \to \tilde{f}(\tilde{a} + \tilde{b} - \tilde{x})$
in (\ref{x-conj-sigma-def}).
The entanglement entropies (\ref{renyi-entropies-non-homo}) 
in this limiting regime become
\be
    \label{EEs_inhom_Ltoinfty}
    S_A^{(n)} \to \, 
    \frac{n+1}{6 \, n} \, 
    \log\!\bigg(\e^{[\sigma(a) + \sigma(b)]/2} \; 
    \frac{\tilde{b} - \tilde{a}}{\epsilon} \,\bigg)
    + O(1) \,.
\ee
Instead, when $\tilde{L}_\infty$ is finite, 
the entanglement Hamiltonian remains non-local. 
Indeed, the limit of the weight functions 
in (\ref{beta-loc-non-homo})-(\ref{beta-loc-non-homo-v1}) and (\ref{beta-biloc-non-homo-V1})-(\ref{beta-biloc-non-homo-V2}), 
of the contour function in (\ref{contour-non-homo-A}) and of the entanglement entropies in (\ref{renyi-entropies-non-homo}) 
can be obtained simply by replacing 
$\tilde{L}$ with $\tilde{L}_\infty$ in the corresponding expressions.

\subsection{Rainbow model}
\label{subsec-CFT-rainbow-model}

\begin{figure}[t!]
\vspace{-.4cm}
\hspace{-.85cm}
\includegraphics[width=1.1\textwidth]{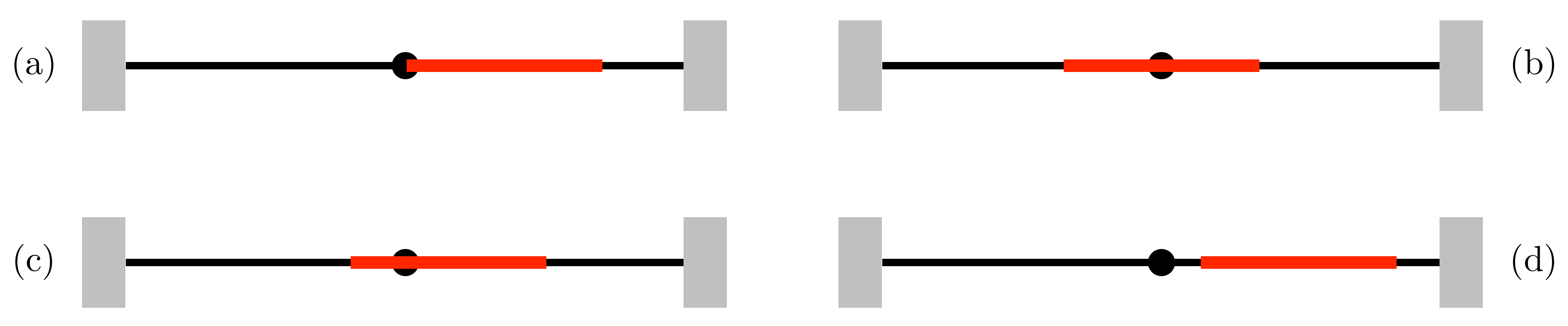}
\vspace{-.4cm}
\caption{Bipartitions of the segment through an interval $A$ (solid red line) not adjacent to the boundaries,
when the  middle point of the segment (black dot) plays a special role, like e.g. in the rainbow model. 
}
\label{fig-configurations}
\end{figure}

In the following, the general results presented 
in Sec.\,\ref{subsec-generic-background} are applied to the rainbow model \cite{Vitagliano:2010db,Ram_rez_2014,Ramirez:2015yfa},
which is defined by the metric (\ref{metric}) with 
\be
\label{sigma-rainbow}
\sigma(x) 
= - \,h |x|
= - \,\lambda \,|x/L|
\;\;\;\;\qquad\;\;\;\;
\lambda \equiv h L
\ee
and by the action (\ref{action-dirac-strip-sigma}) 
in the vector phase (\ref{strip-bc-vector}) 
with $\alpha_{\textrm{\tiny v}} = 3 \pi / 2$, 
as discussed in Sec.\,\ref{sec-EH-rainbow-chain}.
The Ricci scalar of the metric 
given by (\ref{metric}) and (\ref{sigma-rainbow})
is $\mathcal{R}^{\textrm{\tiny (R)}} = 4 h \,\delta(x) - 2 h^2$; 
hence, the curvature is constant and negative everywhere except for the point at $x=0$,
where a curvature singularity occurs,
which corresponds to the black dot in Fig.\,\ref{fig-configurations}. 
Various interesting features of the rainbow model have been explored e.g. in 
\cite{
Rodriguez-Laguna:2016roi,Tonni:2017jom,
SamosSaenzdeBuruaga:2019vkx,Ramirez:2020ppf,Mula:2020udv,Mula:2022lsj,Santalla:2022ygq,Byles:2023hjx,Szabo:2024rca,Bonsignori:2024gky}.

Specialising (\ref{weyl-factor}) to (\ref{sigma-rainbow}), for the rainbow model we find that
\be
\label{x-tilde-rainbow}
\tilde{x}(x) 
= 
\int_0^x \! \e^{h |y|} \, \rd y
\, = \, 
\textrm{sign}(x) \, \frac{\e^{h |x|} - 1}{ h } 
\, = \, 
L\; \textrm{sign}(x/L) \, \frac{\e^{\lambda |x/L|} - 1}{ \lambda } 
\, = \, 
L\, R_\lambda(x/L)
\ee
where, in order to highlight that  
$\tilde{x}(x) /L$ 
is a function of $x/L$ parameterised by $\lambda $,
in the last step we have introduced
\be
\label{r-function-def}
R_\lambda(y) \equiv \, \textrm{sign}(y)  \, \frac{\e^{\lambda |y|} - 1}{ \lambda }
\;\;\;\;\;\qquad \;\;\;\;
-\! 1 \leqslant y \leqslant 1 \;.
\ee
Hence, from the text below (\ref{beta-loc-non-homo}), we have that $\tilde{L} = L\, R_\lambda(1)$ for this model.
Since $\textrm{sign}(\tilde{x}) = \textrm{sign}(x)$,
the inverse of (\ref{x-tilde-rainbow}) reads 
\be
\label{f-tilde-rainbow-def}
\tilde{f}(\tilde{x}) 
\equiv 
\frac{\textrm{sign}(\tilde{x}) }{ h} \, \log\! \big( 1+ h |\tilde{x}|\big)
\,=\,
L\; \frac{\textrm{sign}(\tilde{x}/L) }{ \lambda } \, \log\! \big( 1+ \lambda\, |\tilde{x}/L|\big)
\,=\,
L \, \mathcal{R}_\lambda(\tilde{x}/L)
\ee
where $\mathcal{R}_\lambda$ denotes the inverse function of (\ref{r-function-def}).

From (\ref{sigma-rainbow}) and (\ref{f-tilde-rainbow-def}),
the Weyl factor in the intermediate expression of 
(\ref{metric}) specialised to the rainbow model becomes
\begin{equation}
\label{weyl-factor-AdS}
    \e^{-2 h |x|}  
    \,=\,
    \e^{-2 \lambda \,| \mathcal{R}_\lambda (\tilde{x}/L)|}
    \,=\,
    \frac{1}{\big( 1+ h |\tilde{x}| \big)^2} \;.
\end{equation}
The metric of Euclidean $\textrm{AdS}_2$ 
in Poincar\'e coordinates is $ds^2 = 
(R_\textrm{AdS}/y)^2  \,
\big( \rd t_{\textrm{\tiny E}}^2 + \rd y^2\big)$ with $y>0$. 
Comparing this metric with (\ref{metric})
having the Weyl factor (\ref{weyl-factor-AdS}),
it has been observed in \cite{Rodriguez-Laguna:2016roi} that, 
by introducing the manifold made by two copies 
$\textrm{AdS}_2^{+}$ and $\textrm{AdS}_2^{-}$ of Euclidean $\textrm{AdS}_2$ corresponding to $y>0$ and $y<0$ respectively, the Euclidean manifold supporting 
the rainbow model 
can be seen as $\mathcal{P}_{+} \cup \mathcal{P}_{-}$,
being $\mathcal{P}_{\pm} \subsetneq \textrm{AdS}_2^{\pm}$ 
the two strips defined by 
first setting $R_\textrm{AdS} = 1/h$
and then considering  $y \in \big( 1/h \,, \e^{hL}/h\big]$ 
and $y \in \big[ \! -\e^{hL}/h \,, -1/h\big)$ 
respectively.

By employing (\ref{x-tilde-rainbow}), the expressions (\ref{x-tilde-notation-L}) specialised to the rainbow model become
\be
\label{x-tilde-rainbow-L}
\tilde{x}_L 
=\,
\tan\! \bigg[ \frac{\pi}{4} \bigg( 1+ \frac{R_\lambda(x/L) }{R_\lambda(1) } \bigg) \bigg]
\;\;\;\;\;\qquad\;\;\;\;\;
\tilde{a}_L \equiv \tilde{x}_L \big|_{x=a}
\qquad
\tilde{b}_L \equiv \tilde{x}_L \big|_{x=b}
\ee
which depend on the parameter $\lambda$ through (\ref{r-function-def}).

Consider the rainbow model in the ground state and the bipartition of the segment 
given by the interval $A$. 
Various features of the bipartite entanglement depend on the position of $A$ with respect to the singularity at $x=0$ (corresponding to the black dot in 
Fig.\,\ref{fig-configurations}),
as discussed below.
Since $\sigma(x)$ is an even function,
we can consider $b>0$ and $|a| \leqslant b$
without loss of generality. 
Thus, only four possibilities occur,
and  are shown in Fig.\,\ref{fig-configurations}:
(a) $a=0$ and $b>0$;
(b) $a=-b$ with $b>0$
(i.e. $x=0$ is the middle point of $A$);
(c) $a < 0 < b$ with $|a| \leqslant b$;
(d) $0<a<b$ (i.e. $0 \notin A$).
Notice that, 
while for the entanglement Hamiltonian of $A$  
in the configurations (a), (c) and (d)
the limit where $A$ is adjacent to the boundary 
can be considered, recovering the corresponding result in \cite{Tonni:2017jom}, 
for the configuration (b) this limit is trivial because $B$ vanishes.

As for the weight function of the local term, 
by using (\ref{sigma-rainbow}) and (\ref{x-tilde-rainbow}), we find that (\ref{beta-loc-non-homo}) and (\ref{beta-loc-non-homo-v1}) 
specialised to the rainbow model give respectively 
\be
\label{beta-loc-rainbow-cft}
\beta^{\textrm{\tiny (R)}}_{\textrm{\tiny loc}}  (x) 
\,=\,
\frac{2 L \,R_\lambda(1) }{\pi\, \e^{\lambda |x/L| }} 
\bigg[  \cos\! \bigg( 
\frac{\pi}{4} \bigg( 1+ \frac{R_\lambda(x/L) }{R_\lambda(1) } \bigg)
\! \bigg) \bigg]^2
\frac{\big( \tilde{b}_L^2 - \tilde{x}_L^2\big) \big( \tilde{x}_L^2 - \tilde{a}_L^2 \big)}{ \big( \tilde{b}_L - \tilde{a}_L\big) \big( \tilde{a}_L \,\tilde{b}_L + \tilde{x}_L^2\big)}
\ee
and 
\be
\label{beta-loc-rainbow-cft-V1}
 \beta^{\textrm{\tiny (R)}}_{\textrm{\tiny loc}}  (x) 
\,=\,
\frac{2 L \,R_\lambda(1) }{\pi\, \e^{\lambda |x/L| } } \;
\frac{ 
\big[ \sin\!\big( \pi \tilde{b}/(2\tilde{L})\big) - \sin\!\big( \pi \tilde{x}/(2\tilde{L})\big)  \big] \,  \big[  \sin\!\big( \pi \tilde{x}/(2\tilde{L})\big)  -  \sin\!\big( \pi \tilde{a}/(2\tilde{L})\big)  \big]
}{
\sin\!\big( \pi (\tilde{b} - \tilde{a})/(2\tilde{L})\big)  + \big[  \cos\!\big( \pi \tilde{b}/(2\tilde{L})\big) - \cos\!\big( \pi \tilde{a}/(2\tilde{L})\big)  \big] \, \sin\!\big( \pi \tilde{x}/(2\tilde{L})\big) 
}
\ee
where $\tilde{x}/\tilde{L} = R_\lambda(x/L)/ R_\lambda(1) $.
Thus,  the ratio $ \beta^{\textrm{\tiny (R)}}_{\textrm{\tiny loc}}  (x) / L$
is a  function of $x/L$ parameterised by $a/L$, $b/L$ and $\lambda$,
which provides the solid curves in Fig.\,\ref{fig-beta-loc-A} and in the top panels of Fig.\,\ref{fig-beta-AB}.

The weight function for the bilocal term can be written by first introducing 
the position $x_{\textrm{c}, \textrm{\tiny R} }$ of the point conjugate to a generic point  $x$.
From (\ref{x-tilde-rainbow}) and (\ref{f-tilde-rainbow-def}), 
we find that (\ref{x-conj-tilde-def}) and (\ref{x-conj-sigma-def}) specialised to the rainbow model give  respectively
\be
\label{x-conj-tilde-rainbow}
\frac{ \tilde{x}_{\textrm{c}}(x) }{L}
\equiv
R_\lambda(1)
\bigg[\,
\frac{4}{\pi} \, \arctan \! \bigg( \frac{ \tilde{a}_L  \, \tilde{b}_L}{ \tilde{x}_L }   \bigg) - 1\,
\bigg]
\;\;\;\qquad\;\;\;
x_{\textrm{c}, \textrm{\tiny R} }(x) = L \, \mathcal{R}_\lambda\big(\tilde{x}_{\textrm{c}}(x) /L\big)
\ee
hence also $x_{\textrm{c},  \textrm{\tiny R} }(x) / L$ is a function of $x/L$ parameterised by $a/L$, $b/L$ and $\lambda$.
The conjugate point $x_{\textrm{c}, \textrm{\tiny R}}(x)$ 
in (\ref{x-conj-tilde-rainbow}) provides the black solid curves in Fig.\,\ref{fig-matrixplot-case-a}, Fig.\,\ref{fig-matrixplot-case-c}, and Fig.\,\ref{fig-matrixplot-AB-case-cc}.

By specialising (\ref{beta-biloc-non-homo-V2}) to the rainbow model and employing (\ref{x-conj-tilde-rainbow}), we find that the weight function of the bilocal term in the rainbow model is 
\be
\label{beta-biloc-rainbow}
\beta_{\textrm{\tiny biloc}}^{\textrm{\tiny (R)}}(x) 
\,=\,
\frac{ \e^{- \lambda ( |x_{\textrm{c},  \textrm{\tiny R} } /L | - |x/L | ) / 2}\;
\tilde{a}_L  \, \tilde{b}_L \,\big( \tilde{b}_L^2 - \tilde{x}_L^2 \big) \big( \tilde{x}_L^2 - \tilde{a}_L^2 \big) 
}{ 
2 \big( \tilde{b}_L - \tilde{a}_L \big) \, \cos\!\Big[ \frac{\pi}{4} \Big( 1+ \tfrac{R_\lambda(x/L) }{R_\lambda(1) } \Big) \Big] \, \sqrt{\tilde{a}^2_L\,\tilde{b}^2_L + \tilde{x}_L^2} \; \big( \tilde{a}_L\,\tilde{b}_L + \tilde{x}_L^2 \big)^{2} }
\ee
whose r.h.s. is a function of $x/L$ parameterised by $a/L$, $b/L$ and $\lambda$.
The analytic expression (\ref{beta-biloc-rainbow}) provides the solid curves in all the panels of Fig.\,\ref{fig-beta-biloc-A} and in the bottom panels of Fig.\,\ref{fig-beta-AB}.

When the singular point $x=0$ belongs to $A$
(i.e. $ a< 0 < b $), 
it is worth considering also its conjugate point $x_{\textrm{c}, \textrm{\tiny R}}(0) $.
From (\ref{x-conj-tilde-rainbow}), by using that $\tilde{x}_L |_{x=0} =1 $, we find
\be
\label{x-conj-0-rainbow}
\frac{ x_{\textrm{c},  \textrm{\tiny R} }(0) }{L}
\,=\,
\mathcal{R}_\lambda
\bigg(
\frac{4 \, R_\lambda(1)}{\pi} \, \arctan \! \big( \tilde{a}_L  \, \tilde{b}_L \big)  - R_\lambda(1)
\bigg)
\ee
which gives the vertical dash-dotted coloured lines in the bottom left panel of Fig.\,\ref{fig-beta-biloc-A}.

As for the auxiliary functions introduced in (\ref{gamma-weights-def-sigma}) for a generic $\sigma(x)$, in the special case of the rainbow model, from (\ref{sigma-rainbow}), they become respectively 
\be
    \label{gamma-weights-rainbow}
    \gamma_{\textrm{\tiny loc}}^{\textrm{\tiny (R)}}(x) \,=\,
    \e^{h |x|} \,
    \beta_{\textrm{\tiny loc}}^{\textrm{\tiny (R)}}(x) \;\;\;\;\;\qquad\;\;\;\;\; \gamma_{\textrm{\tiny biloc}}^{\textrm{\tiny (R)}}(x) \,=\,
    \e^{h (|x_{\textrm{c}, \textrm{\tiny R}}| - |x|)/2} \,
    \beta_{\textrm{\tiny biloc}}^{\textrm{\tiny (R)}}(x) \,.
\ee
These auxiliary weight functions do not contain absolute values and are smooth over the entire segment, 
unlike $\beta_{\textrm{\tiny loc}}^{\textrm{\tiny (R)}}(x)$ and $\beta_{\textrm{\tiny biloc}}^{\textrm{\tiny (R)}}(x)$, 
which display isolated singular points.
In the special case where $a=-b$ with $b>0$ (see the top right panel of Fig.\,\ref{fig-configurations}), 
since $x_{\textrm{c}, \textrm{\tiny R}}(x) = -x$ we have that $\beta^{\textrm{\tiny (R)}}_{\textrm{\tiny biloc}}(x) = \gamma_{\textrm{\tiny biloc}}^{\textrm{\tiny (R)}}(x)$ 
(see the solid curves in the top right panel of Fig.\,\ref{fig-beta-biloc-A}).

The contour function for the entanglement  entropies in the rainbow model can be easily written by specialising (\ref{contour-non-homo-A}) and (\ref{contour-non-homo-sigma-AB}) to the weight function in 
(\ref{beta-loc-rainbow-cft})-(\ref{beta-loc-rainbow-cft-V1}).
We find 
\be
\label{contour-non-homo-raimbow-A}
C^{(n)}_{A; \textrm{\tiny (R)}}(x) 
\equiv
\frac{1}{12} \left(1 + \frac{1}{n} \right) \frac{1}{\beta^{\textrm{\tiny (R)}}_{\textrm{\tiny loc}}( x) } 
\;\;\;\;\qquad\;\;\;\;
x \in A
\ee
in the interval $A$, while over the entire segment it reads
\be
\label{contour-non-homo-rainbow-AB}
C^{(n)}_{A, B; \textrm{\tiny (R)}}(x) 
\equiv
\frac{c}{12} \left(1 + \frac{1}{n} \right) \frac{1}{ \big| \beta^{\textrm{\tiny (R)}}_{\textrm{\tiny loc}}( x) \big|} 
\;\;\;\;\qquad\;\;\;
x\in A \cup B \,.
\ee
The contour function in (\ref{contour-non-homo-raimbow-A}) provides the solid curves in Fig.\,\ref{fig-contour-A}, and the contour function in (\ref{contour-non-homo-rainbow-AB}) gives the solid curves in Fig.\,\ref{fig-contour-AB} and Fig.\,\ref{fig-contour-n3-AB}.
From (\ref{beta-loc-rainbow-cft})-(\ref{beta-loc-rainbow-cft-V1}), 
it is straightforward to observe that  $L \, C^{(n)}_{A; \textrm{\tiny (R)}}(x)$ 
and $L \, C^{(n)}_{A, B; \textrm{\tiny (R)}}(x)$ 
are functions of the dimensionless ratio $x/L$, parameterised by the dimensionless parameters 
$a/L$, $b/L$ and $\lambda = h L$.

Finally, as for the entanglement entropies of the interval $A$ in the rainbow model, 
by specialising (\ref{renyi-entropies-non-homo})  to this model through 
(\ref{sigma-rainbow}) and (\ref{x-tilde-rainbow}), 
we find 
\be
\label{renyi-entropies-rainbow-chain}
S_A^{(n)} \,=\,
 \frac{n+1}{6\, n} \,
\log \! \left(  \frac{4 \tilde{L} \; \e^{-h ( |a| + |b| )/2} }{\pi\, \epsilon } \; 
\sqrt{ \cos\! \bigg( \frac{\pi \, \tilde{a} }{2\tilde{L}} \bigg) \, \cos\! \bigg( \frac{\pi \,\tilde{b} }{2\tilde{L}} \bigg) }\;
\frac{ \sin\!\left[ \frac{\pi(  \tilde{b}- \tilde{a})}{4 \tilde{L} }\right]  }{ \cos\!\left[ \frac{\pi (\tilde{a}+\tilde{b} )}{4\tilde{L}}\right] } 
\,  \right)
+
O(1)
\ee
which provides the solid curves in Fig.\,\ref{fig-entropy-A}, once the $O(1)$ term has been specified, as discussed in Sec.\,\ref{sec-EH-rainbow-chain} (see (\ref{entropies-constant-term-rainbow})).
The argument of the logarithm in (\ref{renyi-entropies-rainbow-chain}) depends on the dimensionless quantities $\epsilon/L$, $a/L$, $b/L$ and $\lambda = h L$.

In the remaining part of this subsection, we describe 
some interesting limiting regimes of the analytic expressions 
for the rainbow model reported above. 
The limits where 
$h \to 0$ (i.e. the limit of homogeneous background),
where $A$ becomes adjacent to the boundary, 
and where $A$ is the interval in the line
have been discussed in the final part of 
Sec.\,\ref{subsec-generic-background} for a generic background, 
and it is straightforward to specialise 
the expressions derived there to the rainbow model, 
by using (\ref{sigma-rainbow}).
In the following, we focus on the limit of 
large $h$, i.e. large $\lambda$ when $L$ is kept fixed,
which is the most interesting limiting regime for the rainbow model.

In this limit of large inhomogeneity, 
let us first observe that (\ref{x-tilde-rainbow-L}) 
becomes 
\be
\label{r-small-function-def}
\tilde{x}_L = 1+ \frac{\pi}{2}\, r_\lambda(x/L) + \dots
\;\;\;\;\;\qquad \;\;\;\;
r_\lambda(y) \equiv \, \textrm{sign}(y)  \, \e^{-\lambda (1-|y|)} 
\;\; \qquad \;\;
-\! 1 \leqslant y \leqslant 1
\ee
where $r_\lambda(x/L)  \ll 1$,
and the dots denote subleading terms.
This observation allows us to write 
the weight function of the local term in (\ref{beta-loc-rainbow-cft}) 
at large $\lambda$ as 
\be
\label{beta-loc-rainbow-cft-large-lambda}
\beta^{\textrm{\tiny (R)}}_{\textrm{\tiny loc}}  (x) 
=
\frac{\e^{\lambda(1-|x/L|)}  }{h} 
\;\mathcal{Q}_{A,\lambda}(x/L)
+
\dots
\ee
where we have introduced the following ratio
\begin{equation}
\label{mathcal-Q-def}
    \mathcal{Q}_{A,\lambda}(y)
    \equiv
    \frac{ \big[ r_\lambda(b/L) - r_\lambda(y)   \big] \, \big[ r_\lambda(y) - r_\lambda(a/L)   \big]  }{ r_\lambda(b/L) - r_\lambda(a/L)   } \;.
\end{equation}

It is very instructive to specify this result to the types of bipartitions displayed in Fig.\,\ref{fig-configurations}.
We recall that these configurations, where $|a| \leqslant b$,
correspond to all the possible choices 
because $\sigma(x)$ is an even function. 
In particular, in our analysis we will consider bipartitions of type (a) and (b) for $x \in A$ (see Fig.\,\ref{fig-beta-loc-A}), and of type (c) for $x \in A \cup B$ (see Fig.\,\ref{fig-beta-loc-A} and Fig.\,\ref{fig-beta-AB}). The behaviour of $\beta^{\textrm{\tiny (R)}}_{\textrm{\tiny loc}}  (x)$ for a configuration of type (d) is qualitatively the same as the one for type (a).
When the singularity is contained in $A$ (i.e. $a < 0 <b$)
it is convenient to introduce the interval
$A_0 \subsetneq A$ given by $ A_0 \equiv (- |a|, |a|)$ when  $|a| \leqslant b$,
which is symmetric with respect to the origin.
Let us denote by $A_\ast$ the complement of $A_0$ in $A$ that does not contain the entangling points, by $B_\ast$ its mirror image 
with respect to the origin, namely
$B_\ast \equiv \big\{\! -x \;\big|\; x \in A_\ast \big\}$,
and by $B_0$ the complement of $B_\ast$ in $B$ that does not contain the entangling points.
For a bipartition of type (a) and (d) in Fig.\,\ref{fig-configurations} we have $A_0 = \emptyset$, 
for type (b) we have $A_\ast = \emptyset$, 
whereas for type (c) we have both $A_0 \neq \emptyset$ and $A_\ast \neq \emptyset$. 
In particular, for type (c) these domains are given by  
$A_0 = (-|a|,|a|)$, $A_\ast = (|a|, b)$, $B_\ast = (-b,-|a|)$ 
and $B_0 = [-L,-b) \cup (b,L]$.
The relevance of $A_\ast$ and $B_\ast$ is due to the fact that 
the limit of large $h$ while $L$ is kept fixed 
in (\ref{beta-loc-rainbow-cft-large-lambda}) gives
\be
\label{beta-loc-rainbow-cft-large-lambda-case1}
\big| \beta^{\textrm{\tiny (R)}}_{\textrm{\tiny loc}}  (x) \big|
=
\frac{1 }{h}  + \dots
\;\;\;\;\;\qquad\;\;\;\;
x \in A_\ast \cup B_\ast
\ee
meaning that $\beta^{\textrm{\tiny (R)}}_{\textrm{\tiny loc}}  (x)$ displays a plateau in these regions. 
Since for a bipartition of type (b) we have that $A_\ast = \emptyset$, such a plateau does not occur in this case.
In the ground state of the rainbow chain in the regime of large inhomogeneity, $A_\ast$ and $B_\ast$ play an important role, 
as discussed in Sec.\,\ref{sec-EH-rainbow-chain} 
(see Fig.\,\ref{fig-rainbows-types}).

Instead, an exponential behaviour 
in $x$ occurs in this limit when $x \notin A_\ast \cup B_\ast$ and is not close to the endpoints of $A$.
For instance, for a bipartition of type (b),
we have $r_\lambda(a/L) = - \, r_\lambda(b/L) $ and $r_\lambda(x/L) < r_\lambda(b/L)$ for any $x\in A$; 
hence, from (\ref{beta-loc-rainbow-cft-large-lambda}) 
one obtains
\be
\label{beta-loc-rainbow-cft-large-lambda-case2}
\beta^{\textrm{\tiny (R)}}_{\textrm{\tiny loc}}  (x) 
=
\frac{\e^{h(b-|x|)} }{2\,h}  + \dots
\;\;\;\;\qquad\;\;\;\;
a = - b
\;\;\qquad\;\;
x\in A_0 \,.
\ee
As for a bipartition of type (c), 
from (\ref{beta-loc-rainbow-cft-large-lambda}) we get 
\be
\label{beta-loc-rainbow-cft-large-lambda-case3}
\beta^{\textrm{\tiny (R)}}_{\textrm{\tiny loc}}  (x) 
=
\frac{\e^{h(|a|-|x|)} }{h}  + \dots
\;\;\;\;\qquad\;\;\;\;
a< 0< b \;\qquad\; |a| < b
\;\;\qquad\;\;
x\in A_0
\ee
and 
\be
\label{beta-loc-rainbow-cft-large-lambda-case3-bis}
\big|
\beta^{\textrm{\tiny (R)}}_{\textrm{\tiny loc}}  (x) 
\big|
=
\frac{\e^{h(|x|-b)} }{h}  + \dots
\;\;\;\;\qquad\;\;\;\;
a< 0< b \;\qquad\; |a| < b
\;\;\qquad\;\;
x\in B_0 \,.
\ee
We highlight the exponential behaviour of
$\beta^{\textrm{\tiny (R)}}_{\textrm{\tiny loc}}(x) $ 
in (\ref{beta-loc-rainbow-cft-large-lambda-case2}), (\ref{beta-loc-rainbow-cft-large-lambda-case3}) and (\ref{beta-loc-rainbow-cft-large-lambda-case3-bis})
as $h \to +\infty$.
Instead, in the cases given by the bipartitions of type (a) and (d) in Fig.\,\ref{fig-configurations}, 
where $A_0 = \emptyset$,
such an exponential behaviour of $\beta^{\textrm{\tiny (R)}}_{\textrm{\tiny loc}}  (x)$ does not occur in $A$.

As for the conjugate point, 
assuming $|a|  \neq |b| $ and 
taking $h \to +\infty$ in the first expression in 
(\ref{x-conj-tilde-rainbow}),
we find
\begin{equation}
\label{tilde-xc-rainbow-large-h}
    \tilde{x}_{\textrm{c}}(x) 
    =
    \frac{1}{h} 
    \Big[\,
    \textrm{sign}(a)\, \e^{h|a|}
    + \textrm{sign}(b)\, \e^{h|b|}
    - \textrm{sign}(x)\, \e^{h|x|}
    \,\Big] + \dots
\end{equation}
where (\ref{r-small-function-def}) has been employed,
and the dots denote subleading terms. 
By using (\ref{tilde-xc-rainbow-large-h}) in 
the second expression in (\ref{x-conj-tilde-rainbow}), 
the leading term of the conjugate point in this limit is
\begin{equation}
\label{xc-rainbow-large-h}
    x_{\textrm{c}, \textrm{\tiny R} }(x) =
    \textrm{sign}\big(\tilde{x}_{\textrm{c}}(x)\big)\;
    \textrm{max}\big\{ |a| \,,  |b| \,, |x| \big\} + \dots
\end{equation}
which is independent of $x$ when $x\in A$ 
and whose sign can be read from (\ref{tilde-xc-rainbow-large-h}). 
For the symmetric configuration (i.e. $a = -b$ with $b>0$), 
we have that $x_{\textrm{c}, \textrm{\tiny R}}(x) = -x $, 
as shown in the text below (\ref{x-conj-sigma-def}) 
for a generic inhomogeneous background belonging to the class (\ref{metric}) that we are considering.

In the large inhomogeneity regime, 
by using (\ref{r-small-function-def}), for the weight function of the bilocal term in (\ref{beta-biloc-rainbow}) we find
that the leading term can be written in terms of (\ref{mathcal-Q-def}) as 
\be
\label{beta-biloc-rainbow-cft-large-lambda}
\beta^{\textrm{\tiny (R)}}_{\textrm{\tiny biloc}}  (x) 
\,=\,
\frac{ \pi }{ 4 } \;
\e^{- \lambda ( |x_{\textrm{c},  \textrm{\tiny R} } /L | - |x/L | ) / 2}  
\;\mathcal{Q}_{A,\lambda}(x/L)
+
\dots \;.
\ee

As done above for $\beta^{\textrm{\tiny (R)}}_{\textrm{\tiny loc}}  (x)$, 
in the following we specify this result to the types of bipartitions shown in Fig.\,\ref{fig-configurations}. 
In particular, we will again consider bipartitions of type (a) and (b) for $x \in A$ (see Fig.\,\ref{fig-beta-biloc-A}), and of type (c) for $x \in A \cup B$ (see Fig.\,\ref{fig-beta-biloc-A} and Fig.\,\ref{fig-beta-AB}). The behaviour of $\beta^{\textrm{\tiny (R)}}_{\textrm{\tiny biloc}}  (x)$ for a configuration of type (d) is qualitatively the same as the one for type (a).
In a bipartition of type (b), we find that (\ref{beta-biloc-rainbow-cft-large-lambda})
simplifies to 
 \begin{equation}
 \label{beta-biloc-rb-large-h}
\beta^{\textrm{\tiny (R)}}_{\textrm{\tiny biloc}}  (x) 
\,=\, \frac{\pi}{8} \,
\e^{- h (L - b)} + \dots
\;\;\;\;\qquad\;\;\;\;
a = - b
\;\;\qquad\;\;
x\in A_0
\end{equation}
which is independent of $x$, like (\ref{beta-loc-rainbow-cft-large-lambda-case1}). 
Remarkably, in the regime of large inhomogeneity, 
a plateau occurs in $\beta^{\textrm{\tiny (R)}}_{\textrm{\tiny biloc}}(x)$
for the configuration where 
$\beta^{\textrm{\tiny (R)}}_{\textrm{\tiny loc}}(x)$ 
does not display any plateau in $A$.
However, the height of this plateau is exponentially 
suppressed as $h$ grows.
In all the other cases, $\beta^{\textrm{\tiny (R)}}_{\textrm{\tiny biloc}}  (x)$ has an exponential behaviour in $x$ in this regime. For instance, considering the bipartitions of type (a), (c) and (d), 
where $|a| < b$ with $b>0$, we find that 
(\ref{beta-biloc-rainbow-cft-large-lambda}) simplifies to
\be
\label{beta-biloc-rainbow-cft-large-lambda-Fig2}
\big| \beta^{\textrm{\tiny (R)}}_{\textrm{\tiny biloc}}(x) \big| 
\,=\,
\frac{ \pi }{ 4 } \;\e^{- h \,\alpha(x)}  
+
\dots \;\;\;\;\qquad\;\;\;\;
|a| < b \;\qquad\; b > 0
\;\;\qquad\;\;
\ee
where 
$ \alpha(x)= 
\big( L - |x|\big) + \tfrac{1}{2} \big( b - |x| \big)$ 
for the bipartitions of type (a) and (d) in $x \in A_\ast$ and of type (c) in $x \in A_\ast \cup B_\ast$,
while for a bipartition of type (c) in $x\in A_0$ and in $x \in B_0$ we have that 
$\alpha(x)= \big( L - |a|\big) + \tfrac{1}{2} \big( b - |x| \big)$ 
and $\alpha(x)= 
\big( L - |x|\big) - \big( |x| - b \big)$ respectively.
Since $\alpha(x)>0$ for any $x\in A$, 
we conclude that $\beta^{\textrm{\tiny (R)}}_{\textrm{\tiny biloc}}(x)$ 
goes to zero as $h \to +\infty$ for $x\in A$.
In a bipartition of type (b), where $A_\ast = \emptyset$, such an exponential behaviour in $x \in A$ 
does not occur for $\beta^{\textrm{\tiny (R)}}_{\textrm{\tiny biloc}}(x)$, but it still vanishes as $h \to +\infty$ 
because the height of the plateau (\ref{beta-biloc-rb-large-h}) is exponentially suppressed in $h$.

As for the auxiliary weight functions (\ref{gamma-weights-rainbow}), from (\ref{beta-loc-rainbow-cft-large-lambda}) and (\ref{beta-biloc-rainbow-cft-large-lambda}) we see that their expansions in the large $h$ regime read
\begin{equation}
\gamma^{\textrm{\tiny (R)}}_{\textrm{\tiny loc}}  (x) 
=
\frac{\e^{\lambda}  }{h} 
\,\mathcal{Q}_{A,\lambda}(x/L)
+
\dots
\;\;\;\;\qquad\;\;\;
    \gamma^{\textrm{\tiny (R)}}_{\textrm{\tiny biloc}}(x) 
\,=\,
\frac{ \pi }{ 4 } \,\mathcal{Q}_{A,\lambda}(x/L)
+
\dots
\end{equation}
in terms of the ratio (\ref{mathcal-Q-def}),
which therefore determines the behaviour in $x$ of both 
auxiliary weight functions in this regime.

As for the entanglement entropies for the rainbow model 
in (\ref{renyi-entropies-rainbow-chain}),
their limits 
for $A$ adjacent to the right boundary 
and $A$ in the line
are obtained by specialising  to (\ref{sigma-rainbow}) 
the expressions reported in (\ref{EEs_inhom_btoL}) and (\ref{EEs_inhom_Ltoinfty}) 
for a generic $\sigma(x)$, respectively.
Some of them have been largely discussed in the literature on the rainbow chain \cite{Vitagliano:2010db,Ram_rez_2014,Ramirez:2015yfa,Rodriguez-Laguna:2016roi}.

In the following, we consider the limit of 
(\ref{renyi-entropies-rainbow-chain}) as $h \to +\infty$.
%
When $|a| \neq |b|$, we find 
\be
\label{entropies-rainbow-large-h}
    S_A^{(n)} 
    =
    \frac{(n+1)\,h}{12 \, n} \;
    \textrm{max} \big\{ \,|b| - |a|\, , |a| - |b| \,\big\}
    -
    \frac{n+1}{6 \, n} \; \log (h\, \epsilon)
    +\dots
\ee
where the dots denote subleading terms.
Thus, for the bipartitions of type (a) and (d) 
in Fig.\,\ref{fig-configurations}
the entanglement entropies $S_A^{(n)}$ display a scaling behaviour given by the length of $A$, namely a volume law.
The Gibbs entropy $S_{\textrm{\tiny th}}$ of a
system of finite length $\ell$ coming from the Stefan-Boltzmann law for a two-dimensional CFT with central charge $c$ is 
$S_{\textrm{\tiny th}} = \pi\, c\, \ell/(3\beta)$
\cite{Cardy:2010fa, Bloete:1986qm, Affleck:1986bv}.
Comparing this result specialised to $c=1$ and $\ell = b-a$ 
with $S_A$ in the rainbow model for the bipartitions of type (a) and (d)
obtained from (\ref{entropies-rainbow-large-h}),
one finds the following effective (large) temperature \cite{Ramirez:2015yfa}
\be
    \label{rainbow-temperature}
    T_{\textrm{\tiny R}} = \frac{h}{2 \pi} \;.
\ee
As for the bipartition of type (c), the scaling of $S_A^{(n)}$ in (\ref{entropies-rainbow-large-h}) is given by the length of $A_\ast$, i.e. by $b-|a|< b-a$;
hence, (\ref{rainbow-temperature}) is obtained by comparing $S_A$ with 
the Gibbs entropy of a system of effective length $\ell = b-|a|$.
Instead, in the symmetric configuration where $|b| = |a|$
(i.e. the bipartition of type (b) in Fig.\,\ref{fig-configurations}),
while the result (\ref{entropies-rainbow-large-h}) still holds
(the first term in the r.h.s. does not occur 
and only the term proportional to $\log (h\, \epsilon)$ remains),
$S_A$ does not scale as a length anymore, 
and therefore it cannot be interpreted as a thermal entropy.

We find it worth remarking that,
since the entanglement entropies $S_A^{(n)}$ 
can be found from the contour function (\ref{contour-non-homo-raimbow-A}),
these observations can be made also by employing 
the asymptotics given by 
(\ref{beta-loc-rainbow-cft-large-lambda-case1}), 
(\ref{beta-loc-rainbow-cft-large-lambda-case2}) 
and (\ref{beta-loc-rainbow-cft-large-lambda-case3}).

The analogy between the mixed state $\rho_A$ obtained from the ground state of the rainbow model in the large $h$ regime and the thermal state at effective temperature $T_{\textrm{\tiny R}}$ given by (\ref{rainbow-temperature})
performed at the level of the density matrices allows us to refine the previous discussion.
Indeed, by employing the above considerations about the behaviour of the weight functions $\beta^{\textrm{\tiny (R)}}_{\textrm{\tiny loc}}(x)$ 
and $\beta^{\textrm{\tiny (R)}}_{\textrm{\tiny biloc}}(x)$ as $h \to +\infty$
(see (\ref{beta-loc-rainbow-cft-large-lambda})-(\ref{beta-loc-rainbow-cft-large-lambda-case3}) and (\ref{beta-biloc-rainbow-cft-large-lambda})-(\ref{beta-biloc-rainbow-cft-large-lambda-Fig2}) respectively),
for the entanglement Hamiltonian $K_A$ 
(see (\ref{EH-segment-in-homo}) specialised to the rainbow model)
in this regime  we find 
$K_A = \tfrac{1}{T_{\textrm{\tiny R}}} \int_A \mathcal{E}(x) \, \rd x +\dots $
for the bipartitions of type (a) and (d) in Fig.\,\ref{fig-configurations},
in terms of the energy density $\mathcal{E}(x)$ 
and
the large effective temperature (\ref{rainbow-temperature}),
where the dots denote subleading terms. 
Hence, in these cases the reduced density matrix becomes the density matrix of a thermal state. 
Instead, for the bipartitions of type (c), 
the expansion of $K_A$ does not give 
$\tfrac{1}{T_{\textrm{\tiny R}}} \int_{A_\ast} \mathcal{E}(x) \, \rd x +\dots $,
because of the exponential behaviour of $\beta^{\textrm{\tiny (R)}}_{\textrm{\tiny loc}}(x)$ in $x \in A_0$ given by (\ref{beta-loc-rainbow-cft-large-lambda-case3}). 
Also for a bipartition of type (b) we find that $\rho_A$ is not approximated by a thermal density matrix 
in the large inhomogeneity regime, since in this case $\beta^{\textrm{\tiny (R)}}_{\textrm{\tiny loc}}(x)$ displays 
the exponential behaviour (\ref{beta-loc-rainbow-cft-large-lambda-case2}) 
over the entire interval $A$ (except close to the entangling points, where a linear behaviour occurs). 
In all bipartitions of Fig.\,\ref{fig-configurations}, the weight function 
$\beta^{\textrm{\tiny (R)}}_{\textrm{\tiny biloc}}(x)$ is exponentially suppressed in $h$ at large inhomogeneity, and therefore
its contribution is negligible with respect to $\beta^{\textrm{\tiny (R)}}_{\textrm{\tiny loc}}(x)$. 
Thus, the thermal behaviour 
at the effective temperature (\ref{rainbow-temperature})
for $\rho_A$ in the regime of large inhomogeneity 
is determined by the local term of $K_A$,
and it is observed only when $A_0 = \emptyset$,
namely when $S_A$ satisfies a proper volume law, 
which scales with the length of $A$.

\section{Entanglement Hamiltonian of a block  in the rainbow chain}
\label{sec-EH-rainbow-chain}

In this section we explore the quantities discussed in the previous section for the rainbow chain,
providing numerical checks of the analytic expressions reported in Sec.\,\ref{subsec-CFT-rainbow-model}.

Consider the class of inhomogeneous free fermionic chains characterised by the following Hamiltonian
(see e.g. \cite{Ramirez:2015yfa,Rodriguez-Laguna:2016roi,Mula:2020udv,Vitagliano:2010db,Ram_rez_2014})
\be
\label{inhomo-chain-ham}
H = \,- \! \sum_{m \,\in \, \mathbb{M} } \! J_m 
\big( c_m^\dagger \,c_{m+1} + \textrm{h.c.} \big)
\ee
where $c_m^\dagger$ and $c_m$ are the usual creation and annihilation operators of a fermionic particle at the $m$-th site,
$\mathbb{M}$ contains the discrete values labelling the sites of the chain and $J_m > 0$ 
are the hopping parameters characterising the inhomogeneity of the free fermionic chain. 
The models described by (\ref{inhomo-chain-ham}) in the continuum limit 
are related to the ones considered 
in Sec.\,\ref{subsec-generic-background}; 
indeed, in the continuum limit,  
the hopping parameters $J_m$ provide 
the function $J(x) =\e^{\sigma(x)}$,
where $\sigma(x)$ is the function occurring in the metric (\ref{metric}).

The rainbow chain \cite{Vitagliano:2010db,Ram_rez_2014,Ramirez:2015yfa,Rodr_guez_Laguna_2016,Rodriguez-Laguna:2016roi}
belongs to the class defined by (\ref{inhomo-chain-ham}); indeed, its Hamiltonian is
\be
\label{ham-rainbow-chain}
H \,=\, 
 - \frac{J_0}{2} \, c_{\textrm{\tiny $\tfrac{1}{2}$} }^\dagger  c_{\textrm{\tiny $-\tfrac{1}{2}$} }
 - \frac{J_0}{2} \sum_{m\,=\,\tfrac{1}{2}}^{L-\tfrac{3}{2}}
 \e^{-hm}
 \Big[\,
 c_m^\dagger \, c_{m+1} +  c_{-m}^\dagger \, c_{-(m+1)} 
 \,\Big]
 + \textrm{h.c.} 
\ee
where $J_0 > 0$ provides the energy scale,
$h \geqslant 0$ is the parameter characterising the inhomogeneity of the hopping amplitudes
and $2L$ is the  total number of sites.
The boundary conditions are given by the following requirements \cite{Ramirez:2015yfa}
\begin{equation}
        \label{BC_RM_lattice}
    c_{\pm (L + 1/2)} = 0 \,.
\end{equation}

\begin{figure}[t!]
\vspace{-.4cm}
\hspace{-1.4cm}
\includegraphics[width=1.17\textwidth]{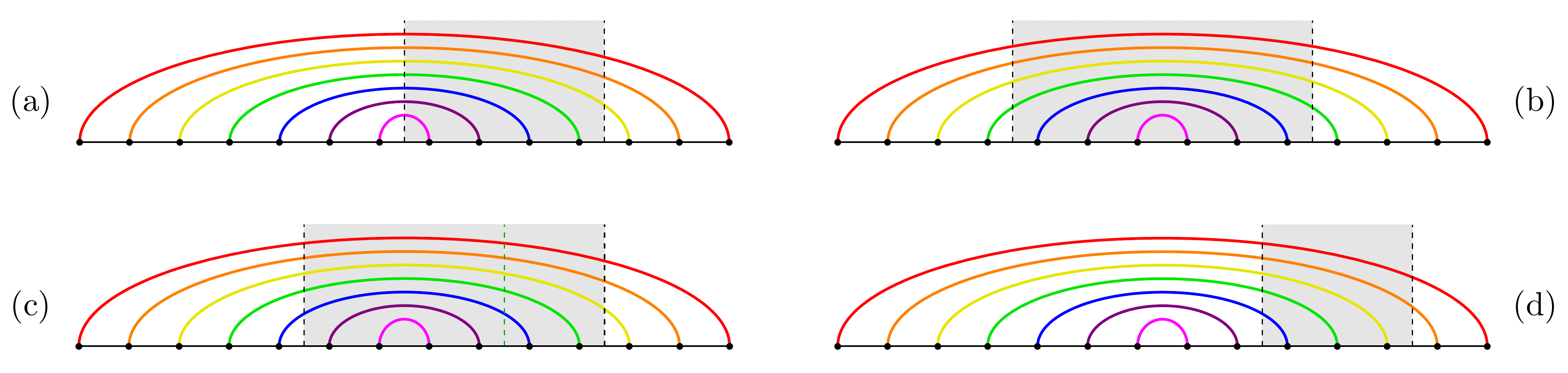}
\vspace{-.4cm}
\caption{Rainbow state for the bipartitions in Fig.\,\ref{fig-configurations}.
The grey strip corresponds to  $A$,
and the vertical dashed green line identifies 
$A_\ast \subsetneq A$ for the configuration of type (c).
}
\label{fig-rainbows-types}
\end{figure}

A Jordan-Wigner transformation brings (\ref{ham-rainbow-chain}) into the Hamiltonian of an inhomogeneous spin-$\tfrac{1}{2}$ XX chain.
The case $h=0$ corresponds to the homogeneous free fermionic chain 
in a segment made by $2L$ sites. 
In the regime of strong inhomogeneity, 
the ground state of (\ref{ham-rainbow-chain}) becomes a valence bond state (the rainbow state) that can be explored 
by applying the strong disorder renormalization group algorithm of Dasgupta and Ma \cite{Dasgupta_1980}
(see also \cite{Ram_rez_2014, Vitagliano:2010db, Fisher_1994}).
In the XX version of the Hamiltonian, 
this valence bond state is constructed 
from singlets between the sites labelled  by $m$ and $-m$ carrying an energy proportional to $J_0 \,\e^{-2mh}$,
with $m = 1/2, \dots, L-1/2$ (see Fig.\,\ref{fig-rainbows-types}, where each coloured curve corresponds to a singlet).
In the segment supporting the rainbow chain 
(\ref{ham-rainbow-chain}),
we consider the bipartition given by a block $A$ 
in generic position, made by  $L_A$ contiguous sites, 
and its complement $B$, containing $L_B = 2L - L_A$ sites.
Notice that $B$ is made by contiguous sites
only when $A$ is adjacent to one of the boundaries of the rainbow chain.
In the strong inhomogeneity regime, 
the entanglement entropy $S_A$ of a certain bipartition is obtained simply by counting the number of bonds having one endpoint in $A$ and one in $B$.
Hence, when $A$ is adjacent to the boundary and it contains at most $L$ sites, the entanglement entropy satisfies a volume law. 
This important feature is also observed for some configurations 
where the block $A$ is not adjacent to the boundary, 
as discussed below in more detail
(in particular, see the bipartitions of type (a) and (d) 
in Fig.\,\ref{fig-configurations} and Fig.\,\ref{fig-rainbows-types}).
We remark that, while the entanglement entropy 
of a gapped lattice model in one spatial dimension 
characterised by local interactions and in its ground state always 
satisfies an area law \cite{Hastings:2007iok},
this behaviour is quite unusual in lattice models at 
the critical point. 
In the rainbow chain, which is a critical lattice model, the volume law is induced 
by the particular form of the hopping parameters.

The rainbow state for the bipartitions of the rainbow chain shown in Fig.\,\ref{fig-configurations}
is displayed in Fig.\,\ref{fig-rainbows-types},
where the different colours  of the singlets denote the different energies 
needed to break them, and 
the grey rectangle delimited by the vertical dashed black lines 
corresponds to the block $A$. 
The linear scaling of $S_A$ in the regime of strong inhomogeneity 
is determined by the singlets having one endpoint in $A$ 
and the other one in $B$.
Hence, in the configurations of type (a) and (d) the entanglement entropy satisfies a genuine volume law. 
In the configuration of type (c), where the endpoints $a$ and $b$ of $A$ 
are such that $a < 0 < b$ and $|a| < b$, 
it is worth identifying the block $A_\ast \subsetneq A$ 
(see the text above (\ref{beta-loc-rainbow-cft-large-lambda-case1}) and the vertical dashed green line in the bottom left panel of Fig.\,\ref{fig-rainbows-types});
indeed, $S_A$ scales like the length $L_{A_\ast}$ of $A_\ast$,
which is $L_{A_\ast} < L_A$ and therefore we do not have a proper volume law scaling in this case.
Also in the configuration of type (b)
a volume law scaling does not occur
because in this case all the singlets of the rainbow state have both their supporting sites either in $A$ or in $B$;
hence, the entanglement entropy  of this bipartition 
vanishes in the regime of large inhomogeneity \cite{Vitagliano:2010db}.

The weak inhomogeneity regime in the rainbow chain leads to study its continuum limit, which provides the rainbow model \cite{Ramirez:2015yfa} 
(see Sec.\,\ref{subsec-CFT-rainbow-model}). 
This is done by first introducing the lattice spacing $\mathsf{a}$ 
and then taking the limit $\mathsf{a} \to 0$, $h \to 0$ and $L \to +\infty$, while $h/\mathsf{a}$ and $L\,\mathsf{a}$ are kept fixed; hence, $\lambda = h L$ is independent of the lattice spacing.
The position in the segment is labelled by $x = \mathsf{a} \, m$.
In this limiting procedure, it is convenient to rename 
$h/\mathsf{a}$ and $L\,\mathsf{a}$ as $h$ and $L$ respectively. 
Thus, in the continuum limit $h$ and $L$ have the dimensions of  inverse length and length respectively. 
The two chiral fields $\psi_1(x) $ and $\psi_2(x) $
in the continuum limit are introduced as the slow varying modes of $c_m$ expanded around $\pm k_{\textrm{\tiny F}} $ at half filling $k_{\textrm{\tiny F}} = \pi / (2 \mathsf{a})$, namely
\begin{equation}
    \label{osc_spinor}
    \frac{c_m}{\sqrt{\mathsf{a}}} 
    \simeq
    \e^{\ri \, k_{\textrm{\tiny F}} \, x} \, 
    \psi_1(x) 
    + \e^{-\ri \, k_{\textrm{\tiny F}} \, x} \, 
    \psi_2(x) \,.
\end{equation}
Combining this expansion with the boundary conditions (\ref{BC_RM_lattice}), 
in the continuum limit one finds the following boundary conditions
at the endpoints of the segment \cite{Ramirez:2015yfa} 
\be
\label{bdy-condition-rainbow-chain-psi}
    \psi_2(\pm L) 
    =
    \mp \,\ri \, 
    \psi_1(\pm L) 
\ee
which couple the fields with different chirality at the boundaries. Comparing (\ref{bdy-condition-rainbow-chain-psi}) 
with (\ref{strip-bc-vector}) and (\ref{strip-bc-axial}) 
at $t_{\textrm{\tiny E}} =0$, 
we observe that the boundary conditions of the rainbow model correspond to the case $\alpha_{\textrm{\tiny v}} = 3 \pi / 2$ of the vector phase.

Since the fermionic model (\ref{ham-rainbow-chain}) is quadratic and its ground state is Gaussian, 
the crucial quantity to consider in order to explore the bipartite entanglement 
is the correlation matrix $C$ in the ground state, whose generic element is 
$C_{m,n}  \equiv \langle c_m^\dagger c_n \rangle$.
This correlation matrix can be found by first writing (\ref{ham-rainbow-chain}) as $ H = \sum_{m,n} H_{m,n} \,c_m^\dagger \, c_n $ with $H_{m,n} = J_m \, \delta_{n, m +\textrm{sign}(m)} + J_n \, \delta_{m, n +\textrm{sign}(n)}$, $J_m = - J_0 \, \e^{-h \, |m|}/2$ and $m,n \in \big\{ -\! \big(L-\tfrac{1}{2}\big) , \dots , L - \tfrac{1}{2} \big\}$. Here, $\delta_{r,s}$ is the Kronecker delta, and $H_{m,n} = -J_0/2$ for either $(m,n)=\big(\tfrac{1}{2} , - \tfrac{1}{2}\big)$ or $(m,n)=(-\tfrac{1}{2} , \tfrac{1}{2})$.
In our numerical analysis
(where the map 
$m \mapsto  m + \big(L-\tfrac{1}{2}\big) + 1$ 
has been used in order to work with positive integers for the labels of the sites),
we set $J_0 = 1$.
The generic element of the ground state correlation matrix 
is obtained through the numerical diagonalisation of the matrix $H_{m,n}$
as $C_{m,n} = \sum_{q \in Q_0} v_q(m)\, v_q(n)$,
where $v_q(m)$  is the real eigenvector associated with  
the eigenvalue $\mathcal{E}_q$ of $H_{m,n}$
and $Q_0 \equiv \big\{ q \; \big|\; \mathcal{E}_q <0 \big\}$
(see e.g. \cite{Mula:2022lsj}).

The entanglement entropy $S_A$ 
and the R\'enyi entropies $S_A^{(n)}$ 
in the free fermionic chain (\ref{ham-rainbow-chain}) are obtained through the standard procedure
\cite{Peschel:1999xeo,Chung_2000,Chung:2001zz,Peschel:2003rdm,Peschel:2004qkm,EislerPeschel:2009review,Cheong_2003}. 
In particular, one first constructs
the reduced correlation matrix $C_A$ by restricting the ground state correlation matrix $C$ to the subsystem $A$, 
namely $(C_A)_{i,j} \equiv C_{i,j} $ for $i,j \in A$,
and then computes the eigenvalues $\zeta_k$ of $C_A$ numerically. 
This spectrum provides $S_A$ and $S_A^{(n)}$ as  
\be
\label{EE-lattice}
S_A = \sum_{k=1}^{L_A} f_1(\zeta_k) 
\;\;\;\;\qquad\;\;\;\;
S_A^{(n)} =   \sum_{k=1}^{L_A} f_n(\zeta_k) 
\ee
where 
\be
\label{f1-fn-def}
f_1(x) \equiv - \,x \log(x) - (1-x) \log(1-x)
\;\;\;\;\qquad\;\;\;\;
f_n(x) \equiv \frac{1}{1-n} \, \log\!\big[ x^n +(1-x)^n\big]
\ee
that satisfy $f_n(x)  \to f_1(x) $ as $n \to 1$;
hence, $S_A^{(n)} \to  S_A$ as $n \to 1$.
The complementary region $B$ is made by two disjoint blocks of contiguous sites when $A$ is not adjacent to the boundary, 
and its entanglement entropies are obtained in the same way from the reduced correlation matrix $C_B$, whose generic element is $(C_B)_{i,j} \equiv C_{i,j} $ for $i,j \in B$.

A crucial role in our work is played by 
a specific construction of the contour functions 
for the entanglement entropy and the R\'enyi entropies
that has been first proposed for 
the homogeneous free fermionic chain in \cite{ChenVidal2014}.
These contour functions provide a more refined description 
of the bipartite entanglement with respect to 
the corresponding entanglement entropies.
Indeed, 
also the eigenvectors $\phi_k(i)$ of the reduced correlation matrix $C_A$ 
are involved in their construction 
and not only the eigenvalues $\zeta_k$.
Since the eigenvectors $\phi_k(i) $ of $C_A$
are normalised by the condition $\sum_{i \in A}\big| \phi_k(i) \big|^2 = 1$ for every $1 \leqslant k \leqslant L_A$, it is natural to introduce 
\be
\label{mode-participation-function-def}
p_k(i) \equiv \big| \phi_k(i) \big|^2
\;\;\;\;\;\;\qquad\;\;\;\;\;
i \in A
\;\;\;\qquad\;\;\;
1 \leqslant k \leqslant L_A
\ee
which are often called mode participation functions 
\cite{Botero_2004}
and 
allow us to write  the entanglement entropies in (\ref{EE-lattice})
respectively as 
\be
S_A = \sum_{i \, \in \, A} \mathcal{C}_A(i)
\;\;\;\;\qquad\;\;\;\;
S_A^{(n)} = \sum_{i \, \in \, A} \mathcal{C}_A^{(n)}(i)
\ee
where $\mathcal{C}_A(i) \geqslant 0$ and $\mathcal{C}_A^{(n)}(i) \geqslant 0$ 
are the contour functions for the entanglement entropy and the R\'enyi entropies respectively.
These contour functions  can be written through (\ref{f1-fn-def}) and the mode participation functions in (\ref{mode-participation-function-def}) as follows
\be
\label{contour-functions-lattice}
\mathcal{C}_A(i) \equiv  \sum_{k=1}^{L_A}  f_1(\zeta_k) \, p_k(i) 
\;\;\;\qquad\;\;\;
 \mathcal{C}_A^{(n)}(i) \equiv  \sum_{k=1}^{L_A} f_n(\zeta_k) \, p_k(i) 
 \;\;\;\;\;\;\qquad\;\;\;\;\;
i \in A \;.
\ee
They provide information about the spatial structure of the bipartite entanglement between $A$ and its complement.
It is worth remarking that the construction of the contour functions for the entanglement entropies 
is highly non-unique. 
Despite the fact that some properties have been proposed \cite{ChenVidal2014},
a complete list of properties characterising the contour functions in a unique way is still not available.
Moreover, infinitely many contour functions can be written 
among the ones associated with mode participation functions $p_k(i)$
satisfying the condition $\sum_ {i \in A} p_k(i) = 1$ with $p_k(i) \geqslant 0$. 
However, it is natural to expect that the most reasonable ones 
are constructed from the eigenvalues and the eigenvectors 
of the reduced correlation matrix.
The definitions of the mode participation function and of the contour functions given in (\ref{mode-participation-function-def})-(\ref{contour-functions-lattice}) for the block $A$
can be adapted also  to the subsystem $B$.
One of the main results of this manuscript consists in providing numerical evidence that the continuum limit of the contour functions (\ref{contour-functions-lattice}) 
with $p_k(i)$ given by (\ref{mode-participation-function-def}) 
is  related to the functions defined in (\ref{contour-non-homo-raimbow-A}) 
and  (\ref{contour-non-homo-rainbow-AB}) through 
the weight function of the local term of the entanglement Hamiltonian 
(see Fig.\,\ref{fig-contour-A}, Fig.\,\ref{fig-contour-AB} and Fig.\,\ref{fig-contour-n3-AB}).
Some contour functions have been constructed also in harmonic lattices \cite{Botero_2004, Fr_rot_2015, Coser:2017dtb}.

The most complete description of the subsystem $A$ is given by its reduced density matrix $\rho_A$ or, equivalently, by the corresponding entanglement Hamiltonian $\mathcal{K}_A$,
provided that $\rho_A$ has non vanishing eigenvalues.
For the rainbow chain (\ref{ham-rainbow-chain}) in its ground state, 
the entanglement Hamiltonian takes the following quadratic form
\be
\label{EH-A-lattice}
\mathcal{K}_A = \! 
\sum_{i,j \in A} \! K_{i,j} \, c_i^\dagger\, c_j
\ee
where the $L_A \times L_A$ matrix $K$ is obtained from the reduced correlation matrix $C_A$ through Peschel's formula 
\cite{Peschel:2003rdm}
\be
\label{peschel-formula}
K = \log\! \big(C_A^{-1} - \mathbf{1} \big)
\ee
which tells us that $K$ and $C_A$ 
have the same real eigenvectors $\phi_k(i)$.
Hence, the matrix $K$ 
can be written through its spectral representation as follows
\be
\label{K-matrix-spectral-dec}
K_{i,j} = \sum_{k =1}^{L_A} \varepsilon_k  \,\phi_k(i) \, \phi_k(j)
\ee
where the single-particle entanglement energies $\varepsilon_k$ read
\be
\label{ss-ent-energies}
\varepsilon_k  \,\equiv \, \log\!\big( 1/\zeta_k -1 \big)
\;\;\;\qquad\;\;\;
1 \leqslant k \leqslant L_A
\ee
in terms of the eigenvalues $\zeta_k$ of $C_A$, 
which provide  the entanglement entropies through (\ref{EE-lattice}). 

We remark that, while the entanglement entropies (\ref{EE-lattice})
are fully determined by the eigenvalues $\zeta_k$ of $C_A$, 
both in the contour functions (\ref{contour-functions-lattice}) 
and in the entanglement Hamiltonian matrix 
(\ref{peschel-formula})-(\ref{K-matrix-spectral-dec})
also the eigenvectors $\phi_k(i) $ of $C_A$ occur.

\begin{figure}[t!]
\vspace{-1.cm}
\hspace{-1.8cm}
\includegraphics[width=1.22\textwidth]{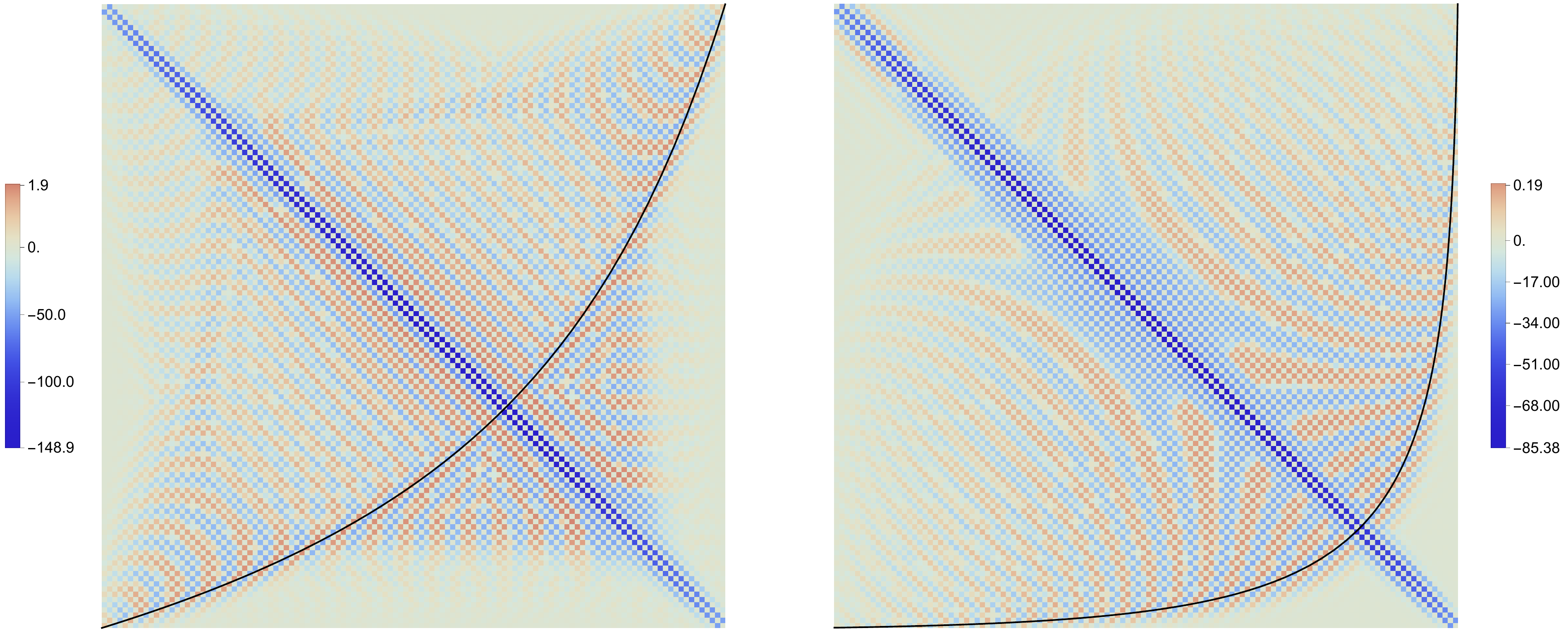}
\vspace{-.4cm}
\caption{
Elements of the matrix (\ref{peschel-formula}) characterising the entanglement Hamiltonian (\ref{EH-A-lattice})
of a block in the configuration of type (a) (see Fig.\,\ref{fig-configurations})
with $a=0$ and $b=0.8 \, L$, when the block has $L_A = 120$ sites
in a rainbow chain made by $2 L = 300$ sites,
with either $\lambda =0$ (left) or $\lambda =5$ (right). 
The black solid curve indicates the position of the conjugate point (\ref{x-conj-tilde-rainbow}).
}
\label{fig-matrixplot-case-a}
\end{figure}

\begin{figure}[t!]
\vspace{-1.cm}
\hspace{-1.8cm}
\includegraphics[width=1.22\textwidth]{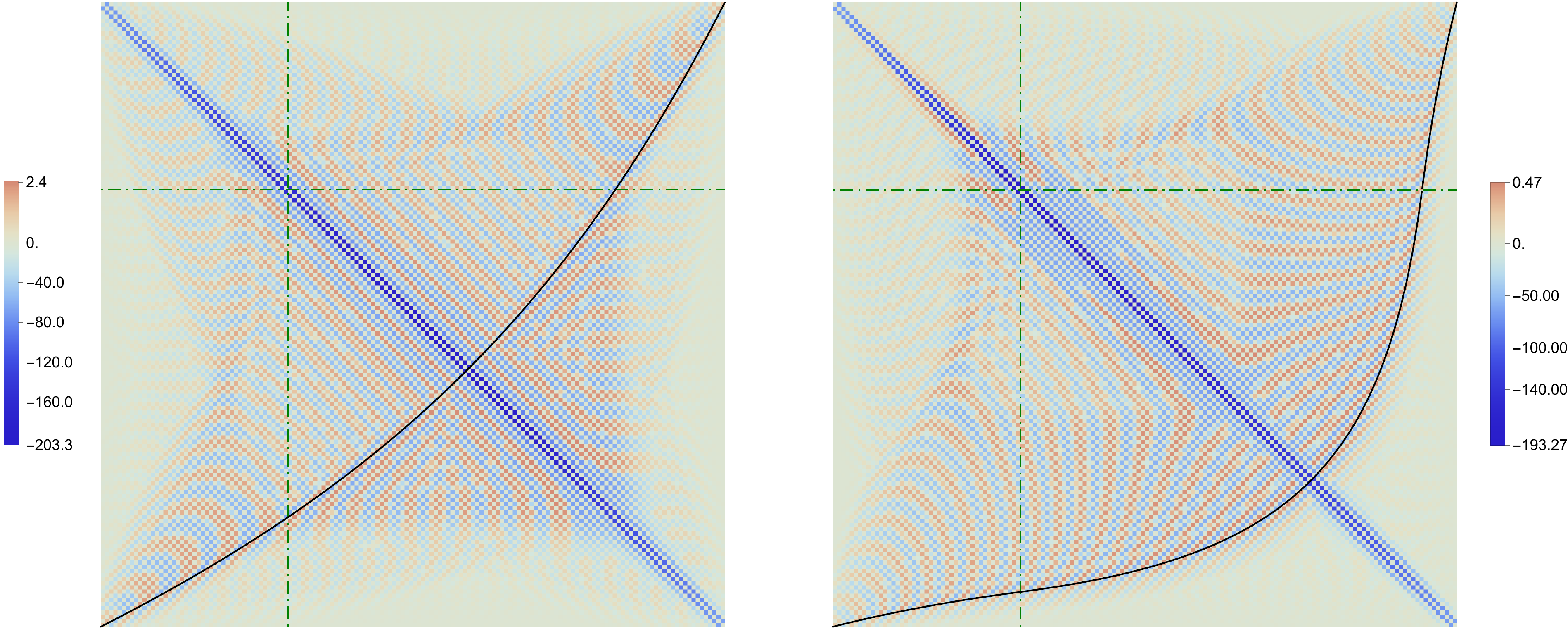}
\vspace{.6cm}
\\
\rule{0pt}{7.3cm}
\hspace{-1.885cm}
\includegraphics[width=1.22\textwidth]{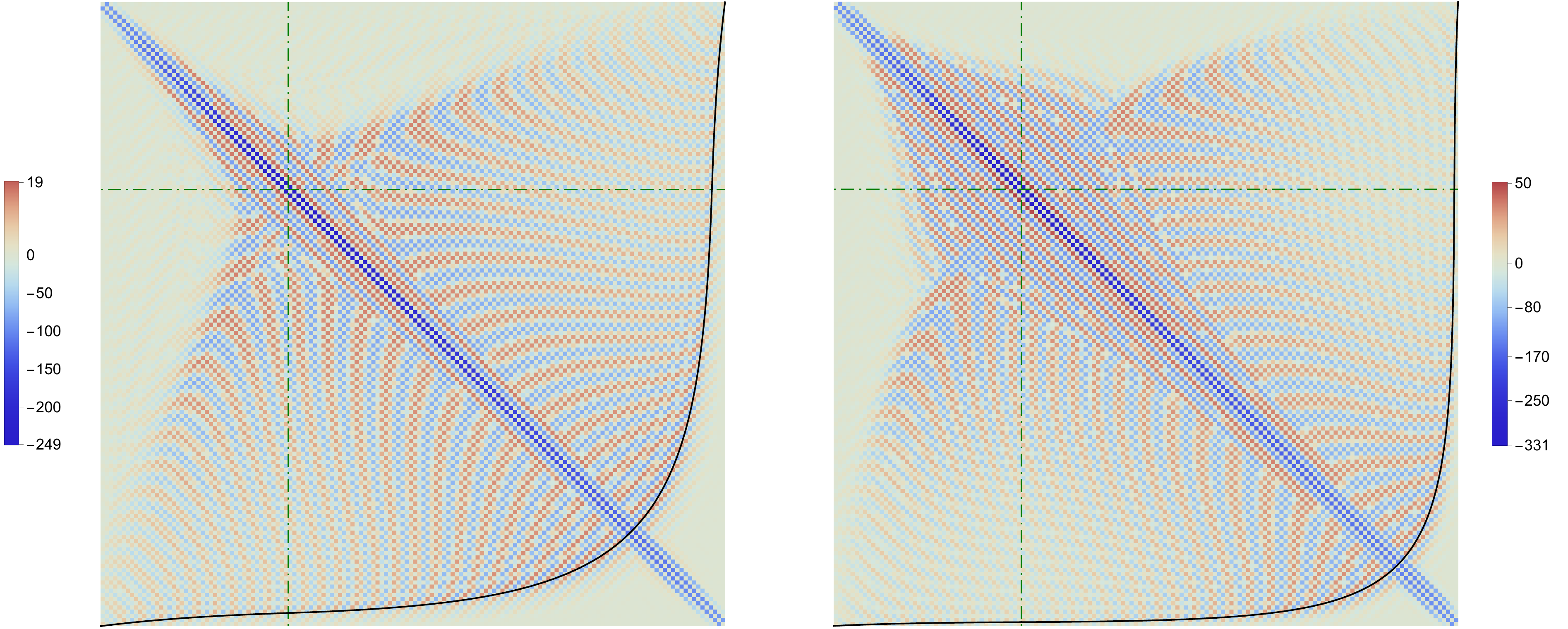}
\vspace{-.4cm}
\caption{
Elements of the matrix (\ref{peschel-formula}) characterising 
the entanglement Hamiltonian (\ref{EH-A-lattice})
of a block for a configuration of type (c) (see Fig.\,\ref{fig-configurations})
with $a=-0.3 \, L$ and $b = 0.7 \, L$,
when the block has $L_A = 150$ sites
in a rainbow chain made by $2 L = 300$ sites,
with either $\lambda =0$ (top left),
$\lambda =3$ (top right),
$\lambda =5$ (bottom left),
or $\lambda =7.5$ (bottom right).
The black solid curve indicates the position of the conjugate point (\ref{x-conj-tilde-rainbow}).
}
\label{fig-matrixplot-case-c}
\end{figure}

In Fig.\,\ref{fig-matrixplot-case-a} and Fig.\,\ref{fig-matrixplot-case-c} we show the elements of the matrix (\ref{peschel-formula}) characterising the entanglement Hamiltonian (\ref{EH-A-lattice}) in the case of a bipartition of type (a) and (c) respectively (see Fig.\,\ref{fig-configurations}), for different values of the inhomogeneity parameter $h$ (or $\lambda$ equivalently, since $L$ is kept fixed). 
The largest elements are located on the odd diagonals around the main diagonal. 
All the elements of the even diagonals, including the main diagonal, are zero. 
However, it is crucial to also understand the role played by the smaller matrix elements located away from the main diagonal in the continuum limit procedure.
For instance, the solid black curve indicates the position of the conjugate point in $A$ given by (\ref{x-conj-tilde-rainbow}),
which nicely describes the front formed by the off-diagonal matrix elements.
This front could indicate the occurrence of a bilocal term in the entanglement Hamiltonian.
The dash-dotted green segments in Fig.\,\ref{fig-matrixplot-case-c} denote the position of the centre  of the chain, which supports the singularity of the background in the continuum limit. 
This point plays a special role in the off-diagonal structure of the entries that becomes evident in the regime of large inhomogeneity, as shown 
in the bottom right panel of Fig.\,\ref{fig-matrixplot-case-c}.
The effect of the inhomogeneity in the rainbow chain on the matrix elements of $K$ in (\ref{peschel-formula}) can be observed by comparing the two panels in Fig.\,\ref{fig-matrixplot-case-a} 
or the four panels in Fig.\,\ref{fig-matrixplot-case-c}.
The matrix in the left panel of Fig.\,\ref{fig-matrixplot-case-a} is qualitatively similar to the matrix occurring in the entanglement Hamiltonian of a block in a homogeneous chain of free fermions on the half line when the block is not adjacent to the boundary
\cite{Eisler:2022rnp}, i.e. when the first site of the block does not coincide with the first site of the infinite chain on the half line. 
As the inhomogeneity grows, 
the solid black curve (\ref{x-conj-tilde-rainbow}) deforms towards the bottom right corner of the matrix plot until this front disappears. 
This can be seen from (\ref{xc-rainbow-large-h}), 
which tells us that $x_{\textrm{c}, \textrm{\tiny R}}(x) \to b$ 
in the large inhomogeneity limit 
for the bipartitions considered in Fig.\,\ref{fig-matrixplot-case-a} and Fig.\,\ref{fig-matrixplot-case-c}.

In order to test numerically the weight functions $\beta^{\textrm{\tiny (R)}}_{\textrm{\tiny loc}}  (x) $ and $\beta^{\textrm{\tiny (R)}}_{\textrm{\tiny biloc}}  (x) $ 
obtained in Sec.\,\ref{subsec-CFT-rainbow-model} for the rainbow model 
in the continuum 
(see (\ref{beta-loc-rainbow-cft})-(\ref{beta-loc-rainbow-cft-V1}) 
and (\ref{beta-biloc-rainbow}) respectively)
through the entanglement Hamiltonian matrix (\ref{peschel-formula}) 
specialised to the rainbow chain, 
we follow the procedure introduced in \cite{Arias:2016nip, Eisler:2019rnr, Eisler:2022rnp} for the homogeneous fermionic chain,
that has been developed by building also on the results of \cite{Eisler:2017cqi, Eisler:2018ugn} about the thermodynamic limit
of the diagonals of (\ref{peschel-formula}).

As for the weight function of the local term, let us introduce the combination of the matrix elements of  (\ref{peschel-formula}) given by 
\cite{Arias:2016nip, Eisler:2019rnr} 
\be
\label{beta-loc-rainbow-chain-v1}
\beta_{\textrm{\tiny loc}}^{\textrm{\tiny (R)}}(i)\, = \, -\frac{1}{\pi}  \sum_{p=0}^{ p_{\textrm{\tiny max}} } (-1)^p \, (2p+1) \, K_{i-p , i + p +1} 
\ee
where the cutoff $p_{\textrm{\tiny max}}$ corresponds to the maximal choice guaranteeing that both indices of $K_{i-p , i + p +1}$ 
belong to $A$, for each $i \in A$.
In particular, 
$p_{\textrm{\tiny max}}$ depends on $i$ and is given by 
    $p_{\textrm{\tiny max}} \equiv 
    \min \!\big\{\, i - m_a\,, m_b - (i+1) \,\big\}$,
where $m_a$ and $m_b$ label the first and last site of the block $A$ respectively
(see also \cite{Eisler:2022rnp}).
A slightly different combination of matrix elements reads
\be
\label{beta-loc-rainbow-chain-v2}
\beta_{\textrm{\tiny loc}}^{\textrm{\tiny (R)}}(i)\, = \, -\frac{1}{2\pi}  \sum_{j \,\in \, A} \sin\!\big[\pi (j-i)/2\big] \, (j-i) \, K_{i, j} 
\ee
which involves the elements along the $i$-th row.
In the case of the entanglement Hamiltonian of a block in harmonic chains, the results corresponding to different combinations of matrix elements have been compared in \cite{DiGiulio:2019cxv}.
This analysis for the rainbow chain is too cumbersome to be reported here; 
hence we just provide the choice corresponding to the best agreement with the 
expressions in the continuum limit among the ones we have considered. 
In our numerical analysis, we have employed 
\eqref{beta-loc-rainbow-chain-v1} in $A$
and 
\eqref{beta-loc-rainbow-chain-v2} in $B$,
as discussed  below.

In order to improve the agreement observed between the numerical data points and the corresponding analytic results, 
we find it convenient to introduce the following average
\begin{equation}
\label{averaging-beta}
\bar{\beta}_{\textrm{\tiny loc}}^{\textrm{\tiny (R)}}(i)
\equiv 
\frac{\beta^{\textrm{\tiny (R)}}_{\textrm{\tiny loc}}(i)}{2} + \frac{\beta^{\textrm{\tiny (R)}}_{\textrm{\tiny loc}}(i+1) + \beta^{\textrm{\tiny (R)}}_{\textrm{\tiny loc}}(i-1)}{4}
\end{equation}
as done e.g. in Eq.\,(53) of \cite{Eisler:2022rnp}.
While this averaging does not provide a relevant improvement for (\ref{beta-loc-rainbow-chain-v1}) in $A$, it becomes important for \eqref{beta-loc-rainbow-chain-v2} 
in $B$.

\begin{figure}[t!]
\vspace{-1.cm}
\hspace{-1.6cm}
\includegraphics[width=1.22\textwidth]{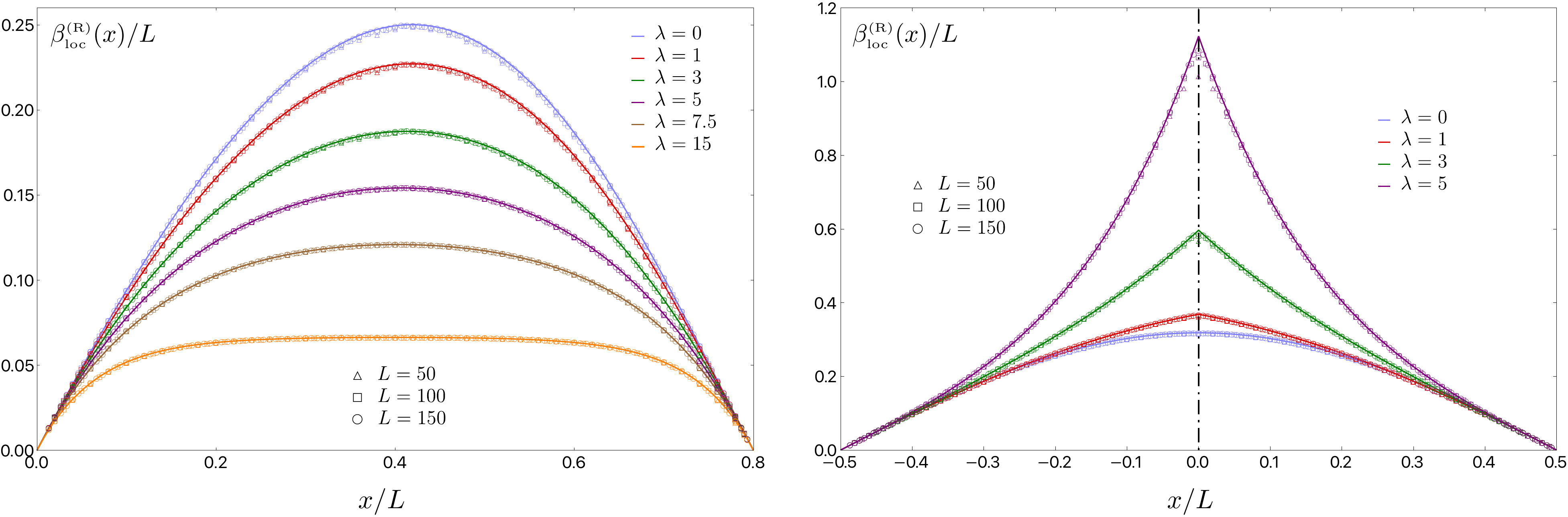}
\vspace{-.2cm}
\\
\rule{0pt}{3.3cm}
\hspace{-1.685cm}
\includegraphics[width=1.22\textwidth]{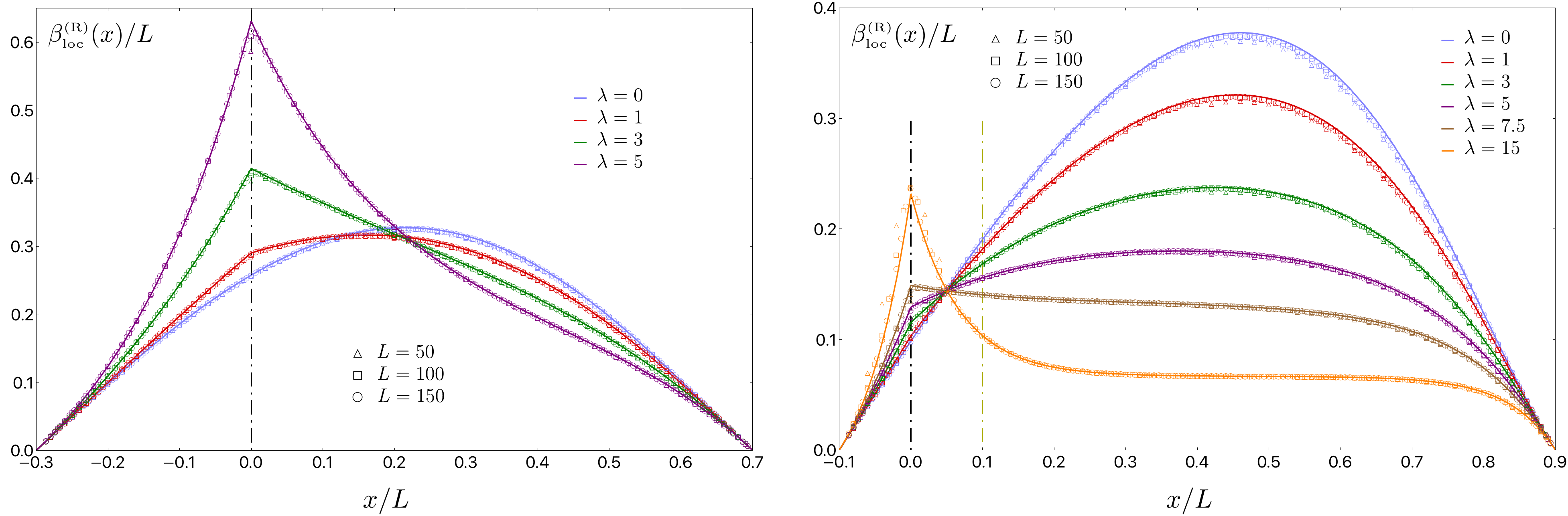}
\vspace{-.4cm}
\caption{
Weight function of the local term in $A$ 
for the rainbow chain,
obtained through (\ref{beta-loc-rainbow-chain-v1}) 
and (\ref{averaging-beta}),
for a configuration of type
(a) (top left), (b) (top right) and (c) (bottom panels).
The solid lines correspond to the analytic expression for $\beta^{\textrm{\tiny (R)}}_{\textrm{\tiny loc}}  (x) $, given in (\ref{beta-loc-rainbow-cft})-(\ref{beta-loc-rainbow-cft-V1}).
}
\label{fig-beta-loc-A}
\end{figure}

In Fig.\,\ref{fig-beta-loc-A} we compare the data points in $A$ obtained numerically through (\ref{beta-loc-rainbow-chain-v1}) and (\ref{averaging-beta})
against the analytic expression for the weight function $\beta_{\textrm{\tiny loc}}^{\textrm{\tiny (R)}}(x)$ 
reported in (\ref{beta-loc-rainbow-cft})-(\ref{beta-loc-rainbow-cft-V1}),
which provides the solid lines. 
In the top left panel, top right panel and bottom panels, 
the bipartitions of type (a), (b) and (c) have been considered, respectively. 
The results corresponding to the bipartition of type (d) have not been reported because they are qualitatively analogous to the ones for a bipartition of type (a). 
As $L$ increases, a remarkable agreement is observed between the data points obtained from the numerical calculations on the lattice 
and the corresponding analytic expression found in the continuum.
The cusp of 
$\beta_{\textrm{\tiny loc}}^{\textrm{\tiny (R)}}(x)$ at $x=0$
is also nicely observed through the lattice data points.
For large inhomogeneity,
the convergence to the expected curve is faster away from the singularity and slower around the singularity (see e.g. the data points having $\lambda=15$ in the bottom right panel of Fig.\,\ref{fig-beta-loc-A}). 
When $a< 0 <b$,
the vertical dash-dotted black line corresponds to $x=0$,
where, from (\ref{beta-loc-rainbow-cft}),  we find that 
\begin{equation}
\label{beta-loc-rainbow-cft-at-x=0}
\frac{ \beta^{\textrm{\tiny (R)}}_{\textrm{\tiny loc}}  (0) }{L}
\,=\,
\frac{ R_\lambda(1) \,\big( \tilde{b}_L^2 - 1 \big) \big( 1 - \tilde{a}_L^2 \big)}{ \pi \big( \tilde{b}_L - \tilde{a}_L\big) \big( \tilde{a}_L \,\tilde{b}_L + 1 \big)}
\end{equation}
which becomes the global maximum of the solid curve at large $\lambda$, and diverges as $\lambda \to +\infty$.

\begin{figure}[t!]
\vspace{-1.cm}
\hspace{-1.6cm}
\includegraphics[width=1.22\textwidth]{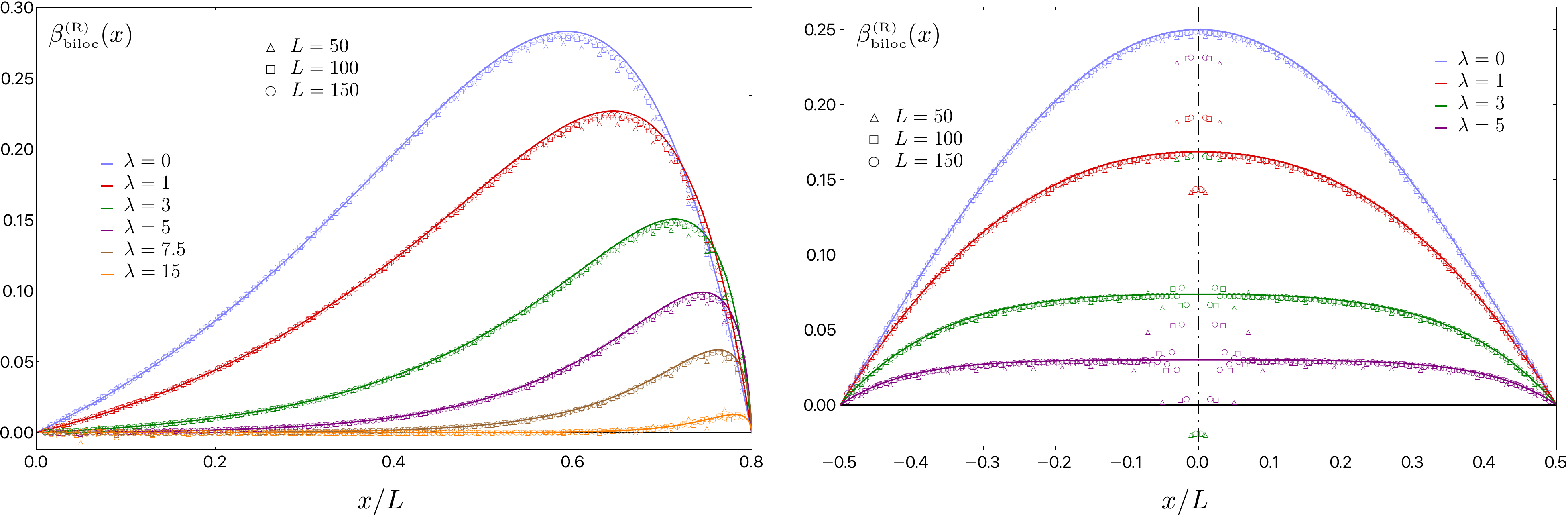}
\vspace{-.2cm}
\\
\rule{0pt}{3.3cm}
\hspace{-1.685cm}
\includegraphics[width=1.22\textwidth]{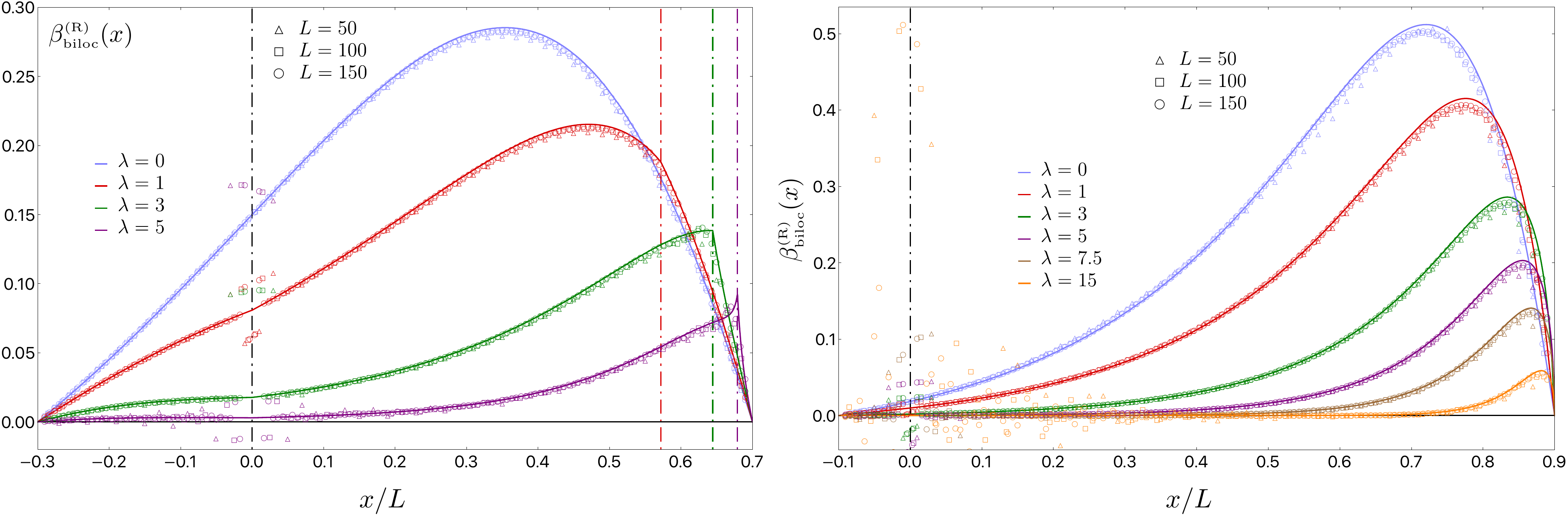}
\vspace{-.4cm}
\caption{
Weight function of the bilocal term in $A$ for the rainbow chain,
obtained from (\ref{beta-biloc-rainbow-chain}) and (\ref{averaging-beta-bilocal}),
for the same configurations considered in Fig.\,\ref{fig-beta-loc-A}.
The solid lines correspond to the analytic expression for $\beta^{\textrm{\tiny (R)}}_{\textrm{\tiny biloc}}  (x) $, given in (\ref{beta-biloc-rainbow}).
}
\label{fig-beta-biloc-A}
\end{figure}

The most interesting feature to highlight in Fig.\,\ref{fig-beta-loc-A} is the plateau with height $1/h$ occurring in $A_\ast$ 
(see the text above (\ref{beta-loc-rainbow-cft-large-lambda-case1})) when $\lambda$ is large 
(see e.g. the data points corresponding to $\lambda = 15$).
We remind that $A_\ast$, which is identified by the vertical dash-dotted yellow line at $x=-a$ in the bottom right panel, 
is given by $A_\ast = (|a|,b)$ in the bottom panels, and by $A_\ast = (a,b)$ in the top left panel. 
Instead, $A_\ast = \emptyset$ in the top right panel.
This plateau in the weight function of the local term combined with the contour function (\ref{contour-non-homo-raimbow-A})
explains the scaling of  $S_A^{(n)}$ as the length of $A_\ast$
(see the corresponding discussion in Sec.\,\ref{subsec-CFT-rainbow-model}, 
and Fig.\,\ref{fig-rainbows-types}).
In the regime of large inhomogeneity, besides  the plateau, the exponential behaviour discussed in Sec.\,\ref{subsec-CFT-rainbow-model} for the analytic expressions is confirmed by the numerical data points. 
At any value of the inhomogeneity parameter, the linear behaviour of $\beta^{\textrm{\tiny (R)}}_{\textrm{\tiny loc}}(x)$ close to the endpoints of $A$ is nicely captured by the lattice data.

The case $\lambda =0 $ corresponds to the results obtained in Sec.\,\ref{sec-EH-homo}.
Indeed, in Fig.\,\ref{fig-beta-loc-A} the weight function of the local term given in (\ref{beta-loc-homo})-(\ref{beta-loc-homo-v1}) 
has been employed (see the solid blue lines).

For the special case of $A$ adjacent to the boundary, 
these observations have been made also in \cite{Tonni:2017jom}, where a different method has been used 
to obtain the data points for the rainbow chain.
It would be insightful to compare in a more systematic way 
these two methods employed to study the weight function of the local term.

The weight function of the bilocal term found in Sec.\,\ref{subsec-CFT-rainbow-model} for the rainbow model (see (\ref{beta-biloc-rainbow})) can be obtained 
through a continuum limit procedure similar to the one described above for the weight function of the local term.
In particular, let us consider the combination of matrix elements introduced in \cite{Eisler:2022rnp}, namely
\be
\label{beta-biloc-rainbow-chain}
\beta^{\textrm{\tiny (R)}}_{\textrm{\tiny biloc}}(i) 
\, = \, \frac{1}{2\pi}  \sum_{j \,\in \, A} \cos\!\big[\pi (j+i)/2\big] \, K_{i, j} 
\ee
and also the corresponding average analogous to (\ref{averaging-beta}), namely
\be
\label{averaging-beta-bilocal}
\bar{\beta}_{\textrm{\tiny biloc}}^{\textrm{\tiny (R)}}(i)
\equiv 
\frac{\beta^{\textrm{\tiny (R)}}_{\textrm{\tiny biloc}}(i)}{2} + \frac{\beta^{\textrm{\tiny (R)}}_{\textrm{\tiny biloc}}(i+1) + \beta^{\textrm{\tiny (R)}}_{\textrm{\tiny biloc}}(i-1)}{4} \;.
\ee
This averaging plays a crucial role in obtaining a good agreement 
with the analytic expression (\ref{beta-biloc-rainbow}) 
in the continuum, 
like the averaging (\ref{averaging-beta}) for (\ref{beta-loc-rainbow-chain-v2}) 
in the case of the local term.

In Fig.\,\ref{fig-beta-biloc-A} we have reported the data points obtained from (\ref{beta-biloc-rainbow-chain}) and (\ref{averaging-beta-bilocal}), for the same configurations considered in Fig.\,\ref{fig-beta-loc-A}.
As $L$ increases, we observe 
a remarkable agreement between the numerical data points from the lattice and the solid curves provided by the analytic expression for $\beta^{\textrm{\tiny (R)}}_{\textrm{\tiny biloc}}  (x) $, given in (\ref{beta-biloc-rainbow}).
The vertical dash-dotted black line indicates the singular point at $x=0$, whenever it is contained in $A$, while the vertical dash-dotted coloured lines in the bottom left panel correspond to $x=x_{\textrm{c},  \textrm{\tiny R} }(0)$ 
(see (\ref{x-conj-0-rainbow})), which depends on $\lambda$.
Also the linear behaviour close to the entangling points 
(see (\ref{beta-biloc-non-homo-ent-pts-a})-(\ref{beta-biloc-non-homo-ent-pts-b}))
is nicely captured by the numerical data points. 
However, the singular behaviour of 
$\beta^{\textrm{\tiny (R)}}_{\textrm{\tiny biloc}}  (x) $ 
at $x=0$ is not captured by the lattice data points for the values of $L$ that we have considered. 
Nonetheless, notice that the support of the oscillations around $x=0$
decreases as $L$ increases.
We remark the vanishing of $\beta^{\textrm{\tiny (R)}}_{\textrm{\tiny biloc}}  (x) $ in the regime of large inhomogeneity, as shown in 
(\ref{beta-biloc-rainbow-cft-large-lambda-Fig2}).
For the bipartitions of type (b), 
where $a=-b$ with $b>0$ (see the top right panel in Fig.\,\ref{fig-beta-biloc-A}), we have that $x_{\textrm{c},\textrm{\tiny R}}(0) = 0$, i.e. the two singular points coincide, and, as a consequence, 
$\beta^{\textrm{\tiny (R)}}_{\textrm{\tiny biloc}}  (x) $ 
is smooth at $x=0$ (see the discussion below (\ref{gamma-weights-rainbow})).
In this case, a plateau occurs in the regime of large inhomogeneity, whose height  is given by (\ref{beta-biloc-rb-large-h}).

The case $\lambda =0 $ corresponds to the results obtained in Sec.\,\ref{sec-EH-homo}.
Indeed, the solid blue lines in Fig.\,\ref{fig-beta-biloc-A} are obtained from the analytic expression 
of the weight function of the bilocal term given in (\ref{beta-biloc-homo}), (\ref{beta-biloc-homo-v2}) 
or (\ref{beta-biloc-homo-v1}).

Besides the two weight functions $\beta^{\textrm{\tiny (R)}}_{\textrm{\tiny loc}}  (x) $ and $\beta^{\textrm{\tiny (R)}}_{\textrm{\tiny biloc}}  (x) $, 
we also find it worth exploring  
the contour functions of the entanglement entropies 
in the rainbow chain.
We consider 
$\mathcal{C}_A(i)$ and $\mathcal{C}^{(n)}_A(i)$
in (\ref{contour-functions-lattice}), 
and we compare the numerical data from the rainbow chain 
with the analytic expressions of
$C^{(n)}_{A; \textrm{\tiny (R)}}(x) $ 
and $C^{(n)}_{A,B; \textrm{\tiny (R)}}(x) $,
defined in (\ref{contour-non-homo-raimbow-A})
and (\ref{contour-non-homo-rainbow-AB}) respectively. 

As done in (\ref{averaging-beta}) 
and (\ref{averaging-beta-bilocal}) for the weight functions, 
also for the contour functions for the entanglement entropy and for the R\'enyi entropies in $A$
it is worth introducing an averaging procedure
that allows to obtain a more precise agreement 
with the analytic results in the continuum.
In particular,  we consider the combinations given respectively by 
\be
\label{averaging-contour}
\bar{\mathcal{C}}_A(i)
\equiv 
\frac{ \mathcal{C}_A(i) }{2} + \frac{ \mathcal{C}_A(i+1) + \mathcal{C}_A(i-1)  }{4}
\ee
and
\be
\label{averaging-contour-renyi}
\bar{\mathcal{C}}^{(n)}_A(i)
\equiv 
\frac{ \mathcal{C}^{(n)}_A(i) }{2} + \frac{ \mathcal{C}^{(n)}_A(i+1) + \mathcal{C}^{(n)}_A(i-1)  }{4}
\ee
which can be extended to the complementary region $B$ 
in a straightforward way.

In the case of the contour function for the entanglement entropy, 
the results of our numerical analysis
for (\ref{contour-functions-lattice}) and (\ref{averaging-contour}) in $A$ 
are reported in Fig.\,\ref{fig-contour-A},
where the same configurations explored  
in Fig.\,\ref{fig-beta-loc-A} and Fig.\,\ref{fig-beta-biloc-A} have been considered. 
Since the solid curves are proportional to those occurring in Fig.\,\ref{fig-beta-loc-A}, 
the same qualitative observations can be easily adapted to this contour function.
However, we find it worth highlighting  
the remarkable agreement 
between the analytic expressions (solid curves)
and the numerical data points as $L$ increases in this analysis.
The case $\lambda =0 $ (solid blue lines and blue data points) provides the result (\ref{contour-function-cft-homo})
for the contour function in the homogeneous background. 
It is instructive to explain the features of the entanglement entropies through the behaviour of their contour functions  $C^{(n)}_{A; \textrm{\tiny (R)}}(x)$ in (\ref{contour-non-homo-raimbow-A}).
For instance, the logarithmic divergence in $S_A$ originates from the behaviour 
of $\beta^{\textrm{\tiny (R)}}_{\textrm{\tiny loc}}(x)$ close to the entangling points,
which is given by  
(\ref{beta-loc-non-homo-ent-pts-a}) 
and (\ref{beta-loc-non-homo-ent-pts-b}).

\begin{figure}[t!]
\vspace{-1.cm}
\hspace{-1.6cm}
\includegraphics[width=1.22\textwidth]{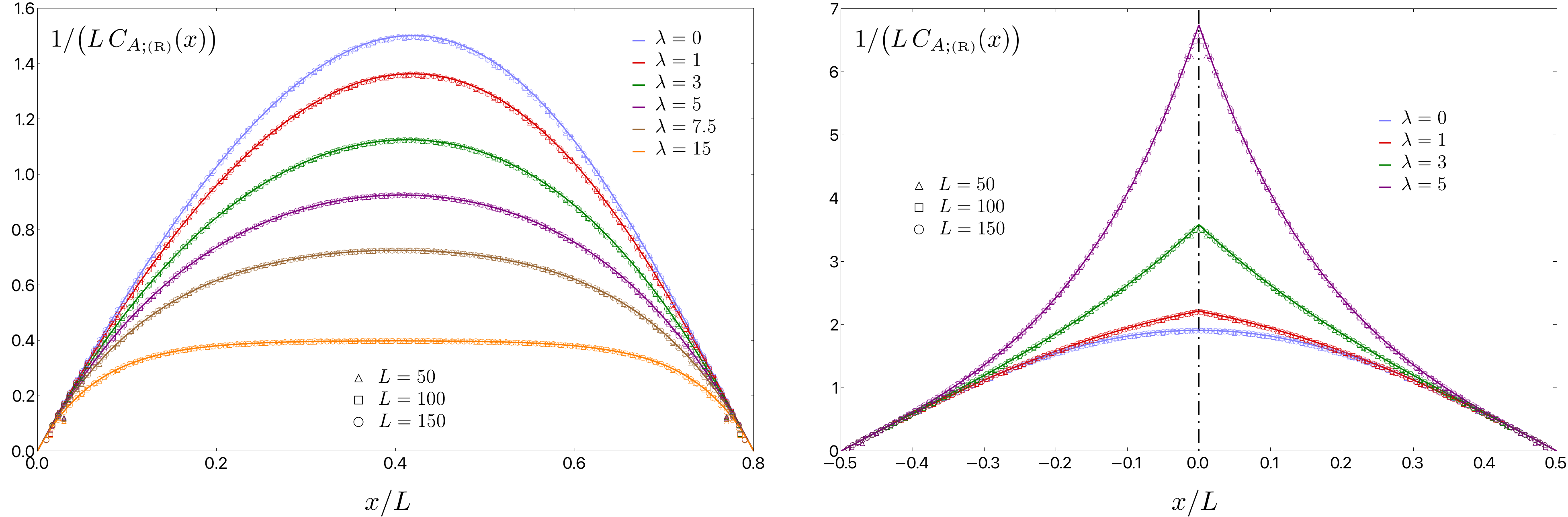}
\vspace{-.2cm}
\\
\rule{0pt}{3.3cm}
\hspace{-1.685cm}
\includegraphics[width=1.22\textwidth]{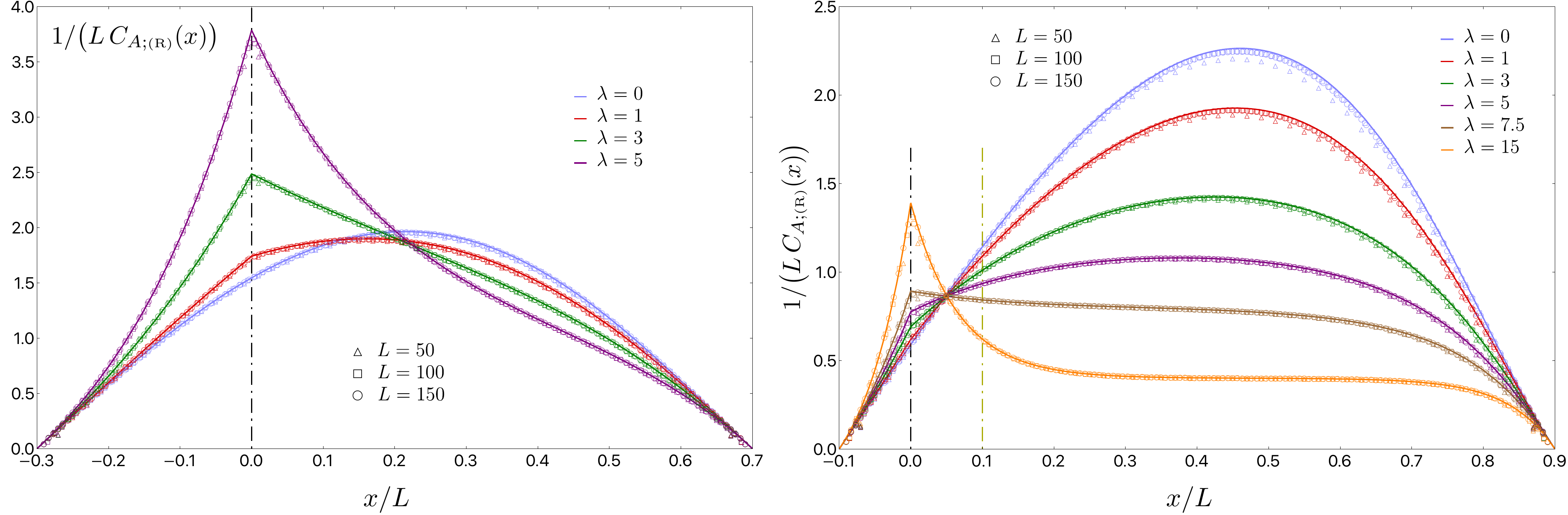}
\vspace{-.4cm}
\caption{
Contour function for the entanglement entropy in $A$ for the rainbow chain,
obtained from (\ref{contour-functions-lattice}) and (\ref{averaging-contour}),
for the same configurations considered in 
Fig.\,\ref{fig-beta-loc-A} and Fig.\,\ref{fig-beta-biloc-A}.
The solid lines correspond to the analytic expression of $C_{A; \textrm{\tiny (R)}}(x) $, given by (\ref{contour-non-homo-raimbow-A}) for $n=1$.
}
\label{fig-contour-A}
\end{figure}

The entanglement entropies in the rainbow chain have been 
largely explored in the literature \cite{Vitagliano:2010db,Ram_rez_2014,Ramirez:2015yfa,Rodriguez-Laguna:2016roi,Tonni:2017jom} through (\ref{EE-lattice}),
extending the results obtained in homogeneous fermionic chains in the presence of boundaries 
(see e.g. \cite{Affleck:1991tk, Laflorencie:2005duh, Affleck_2009, Taddia_2013, Estienne:2023tdw}). 
In the rainbow model, by also employing the results of 
\cite{Jin:2004zzi,Keating_2004,Calabrese_2010,Fagotti:2010cc},
it has been found \cite{Rodriguez-Laguna:2016roi} 
that the entanglement entropies of an interval $A$ 
in a generic position are 
(\ref{renyi-entropies-rainbow-chain})
with the $O(1)$ term given by the constant 
$\mathfrak{C}_n$ defined as follows 
\be
\label{entropies-constant-term-rainbow}
\mathfrak{C}_n \equiv
\left(1 + \frac{1}{n} \,\right) 
\Bigg\{\,
 \frac{\log(2)}{6} 
+ 
\int_0^\infty 
    \left[\,
    \frac{n^{2} }{n^{2} - 1} \, \left(\frac{1}{n \, \sinh(t/n)} - \frac{1}{\sinh(t)} \right) \, \frac{1}{\sinh(t)} - \frac{\e^{-2 t}}{6} 
    \,\right]
     \frac{\rd t}{t}\,
     \Bigg\}\;.
\ee

\begin{figure}[t!]
\vspace{-1.cm}
\hspace{-1.6cm}
\includegraphics[width=1.22\textwidth]{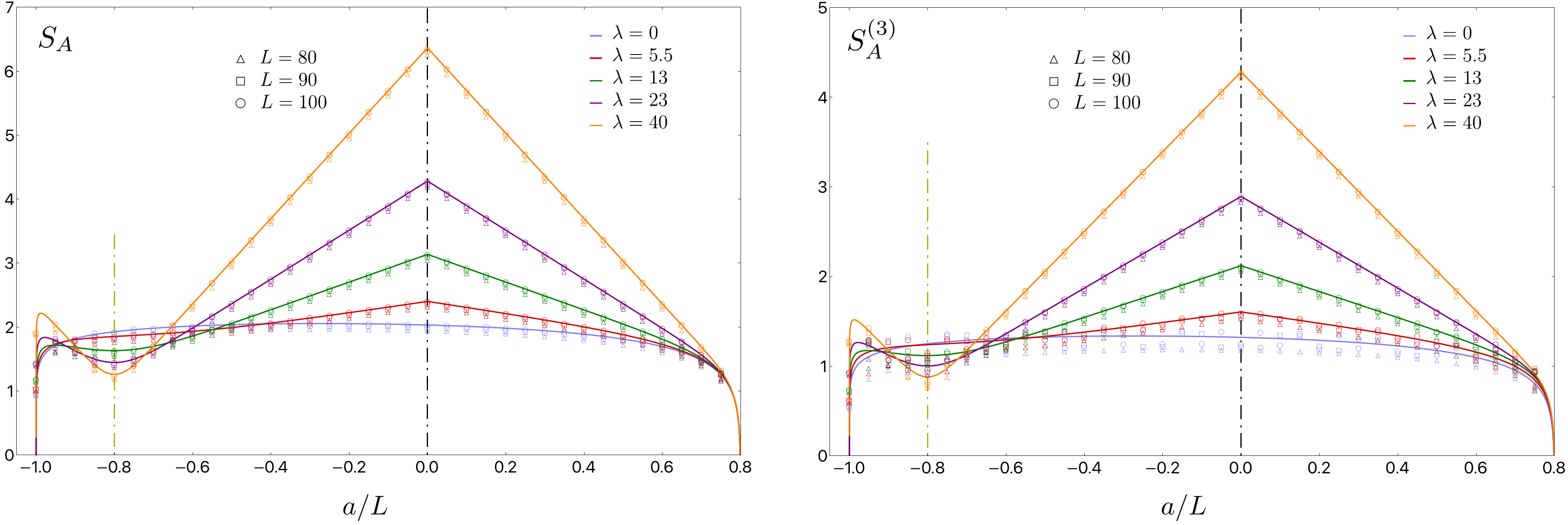}
\vspace{-.4cm}
\caption{
Entanglement entropy (left) and  R\'enyi entropy with $n=3$ (right) for the rainbow chain, obtained through (\ref{EE-lattice}).
The solid lines are given by (\ref{renyi-entropies-rainbow-chain}) combined with (\ref{entropies-constant-term-rainbow}).}
\label{fig-entropy-A}
\end{figure}

Our numerical results for 
the entanglement entropy and for the R\'enyi entropy with $n=3$ in the rainbow chain (for $b = 0.8 \, L$ and $a \in [-L, b)$)
are reported in the left and right panel of Fig.\,\ref{fig-entropy-A} respectively, for various values of $\lambda$.
The data points for $\lambda =0$ 
are obtained in the homogeneous free fermionic
chain, and the corresponding expressions in the continuum limit 
are discussed in Sec.\,\ref{sec-EH-homo}.
The solid curves correspond to the analytic expressions for the rainbow model in the continuum, given by 
(\ref{renyi-entropies-rainbow-chain}) 
and (\ref{entropies-constant-term-rainbow}) 
with UV cutoff $\epsilon$ set to $1$.
The vertical dash-dotted black and yellow lines indicate $x=0$ and $x=-b$ respectively. 
For the symmetric case where $a= -b$ with $b>0$
(i.e. the configuration (b) in Fig.\,\ref{fig-configurations}), 
we refer the reader to the results in  
\cite{Rodriguez-Laguna:2016roi}.
In the regime of large inhomogeneity, 
agreement is found with (\ref{entropies-rainbow-large-h}).
We remark that subleading oscillatory terms occur 
in the numerical data points for $S_A^{(n)}$ with $n \neq 1$, 
and are more visible for small values of $\lambda$, 
as also observed in \cite{Rodriguez-Laguna:2016roi}.
It would be very interesting to find analytic expressions 
for these subleading oscillatory terms.

Since the system that we are considering is defined on a segment, both the interval $A$ and its complement $B$ have finite volume.
Let us stress again that $B$ is made by two disjoint blocks when $A$ is separated from the boundaries of the segment. 
While in the previous analyses the various entanglement quantifiers have been considered only in $A$, in the following 
we explore them also in $B$.
Since the purity of the ground state guarantees 
that $S_A^{(n)} = S_B^{(n)}$, 
it is not useful to consider the entanglement entropies of $B$.
On the contrary, the other quantities that we investigated
exhibit interesting properties in $B$. 
The contour functions for the entanglement entropies
in $B$ can be studied by replacing $C_A$ with $C_B$ 
(introduced in the text below (\ref{f1-fn-def})) 
in the analysis performed above for $A$.
As for the weight functions (\ref{beta-loc-half-line}) and (\ref{beta-biloc-half-line}) for the interval in the half line, which play a crucial role in our analysis and 
have been derived in \cite{Mintchev:2020uom} 
for the interval $\hat{A}$,
it is not obvious that their extension to the complement of $\hat{A}$ in the half line
(that is not connected) 
provides the correct weight functions for the entanglement Hamiltonian in this subsystem.
In the following
we provide some numerical evidence supporting this assumption. 

In the numerical analyses  of the entanglement quantifiers 
in $A$ for the rainbow chain, we have always studied subsystems having $L_A \leqslant L$ 
because of the occurrence of vanishing eigenvalues when $L_A > L$.
The same observation holds for $B$.
Hence, in order to avoid vanishing eigenvalues in the entanglement spectrum either in $A$ or in $B$, 
in the following we consider bipartitions having $L_A = L_B = L$.

\begin{figure}[t!]
\vspace{-1.cm}
\hspace{-1.8cm}
\includegraphics[width=1.22\textwidth]{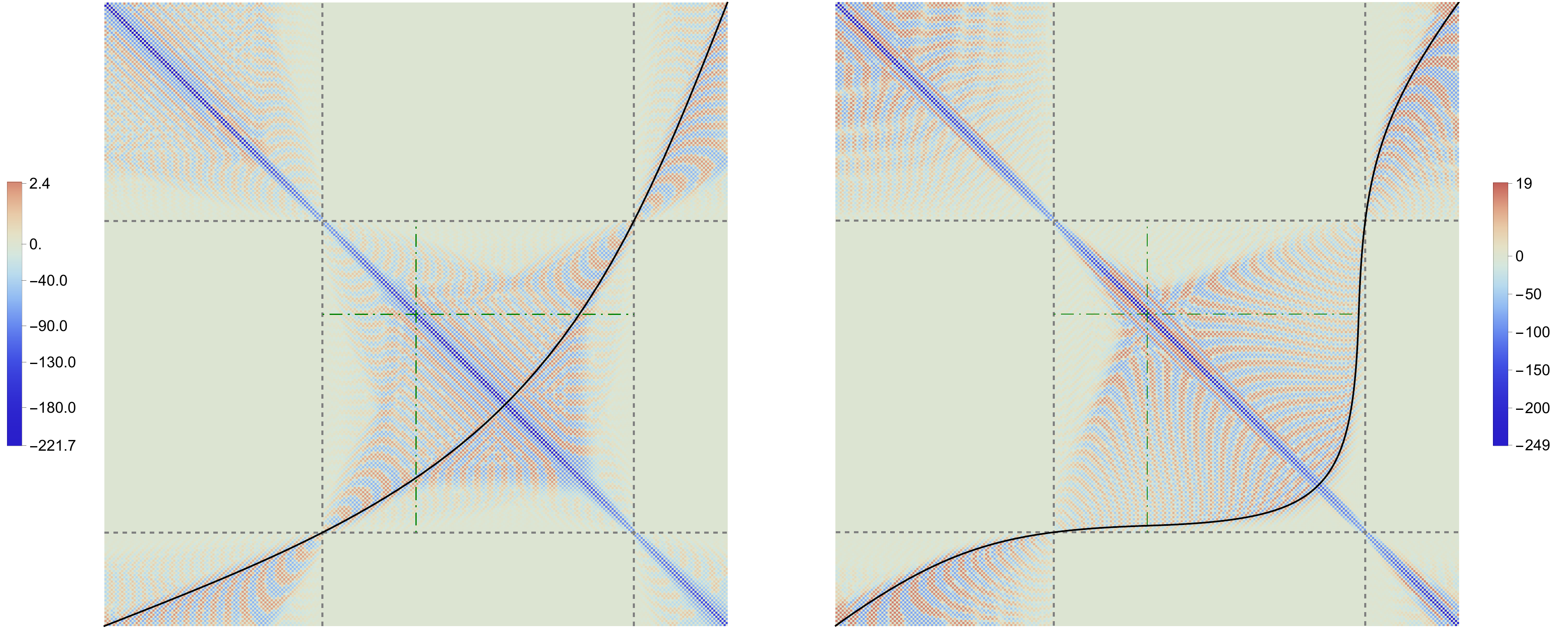}
\vspace{-.4cm}
\caption{
Matrix elements characterising the entanglement Hamiltonian of $A$ and $B$
(see (\ref{peschel-formula}) for $A$ and the analogous formula  for $B$,
where  $C_A$ is replaced by $C_B$).
The bipartition is a configuration of type (c) (see Fig.\,\ref{fig-configurations})
with $a=-0.3 \, L$ and $b = 0.7 \, L$,
and it is made by $L_A = L_B = 150$ sites.
Here, either $\lambda =0$ (left) or $\lambda =5$ (right). 
The black solid curve indicates the position of the conjugate point (\ref{x-conj-tilde-rainbow}).
See also the top left and bottom left panel of Fig.\,\ref{fig-matrixplot-case-c}, corresponding to the central block of the left and right panel respectively.
}
\label{fig-matrixplot-AB-case-cc}
\end{figure}

Let us briefly discuss here the sign of the weight functions as $x$ varies in the whole segment. 
The analytic expressions of $\beta^{\textrm{\tiny (R)}}_{\textrm{\tiny loc}}(x)$ and $\beta^{\textrm{\tiny (R)}}_{\textrm{\tiny biloc}}(x)$
(see (\ref{beta-loc-rainbow-cft})-(\ref{beta-loc-rainbow-cft-V1}) and (\ref{beta-biloc-rainbow}) respectively) 
vanish only at the entangling points 
and they are positive in $A$ and negative in $B$. 
The combinations of matrix elements introduced in 
(\ref{beta-loc-rainbow-chain-v1})-(\ref{averaging-beta}) 
and in (\ref{beta-biloc-rainbow-chain})-(\ref{averaging-beta-bilocal}) allow us to explore the continuum limit of the weight functions of the local and bilocal term respectively.
Since these combinations provide positive results, 
in the following  the numerical data points 
from the rainbow chain are compared with
$| \beta^{\textrm{\tiny (R)}}_{\textrm{\tiny loc}}(x) |$ and $| \beta^{\textrm{\tiny (R)}}_{\textrm{\tiny biloc}}(x) |$.

\begin{figure}[t!]
\vspace{-1.cm}
\hspace{-1.6cm}
\includegraphics[width=1.22\textwidth]{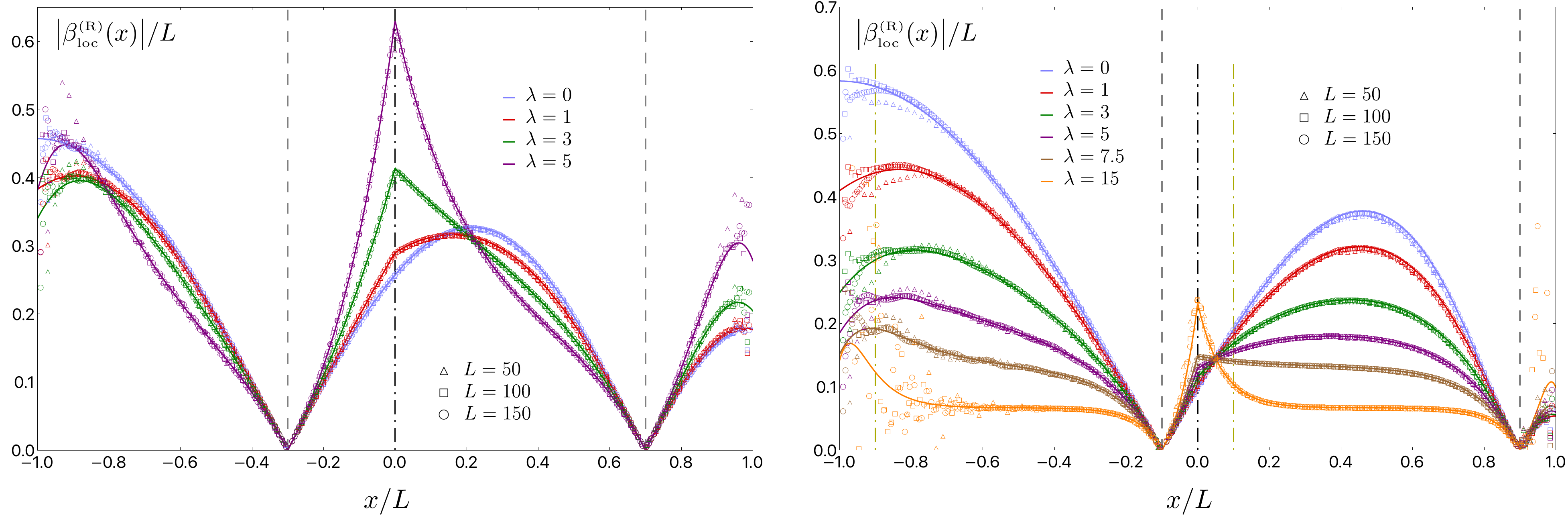}
\vspace{-.2cm}
\\
\rule{0pt}{3.3cm}
\hspace{-1.685cm}
\includegraphics[width=1.22\textwidth]{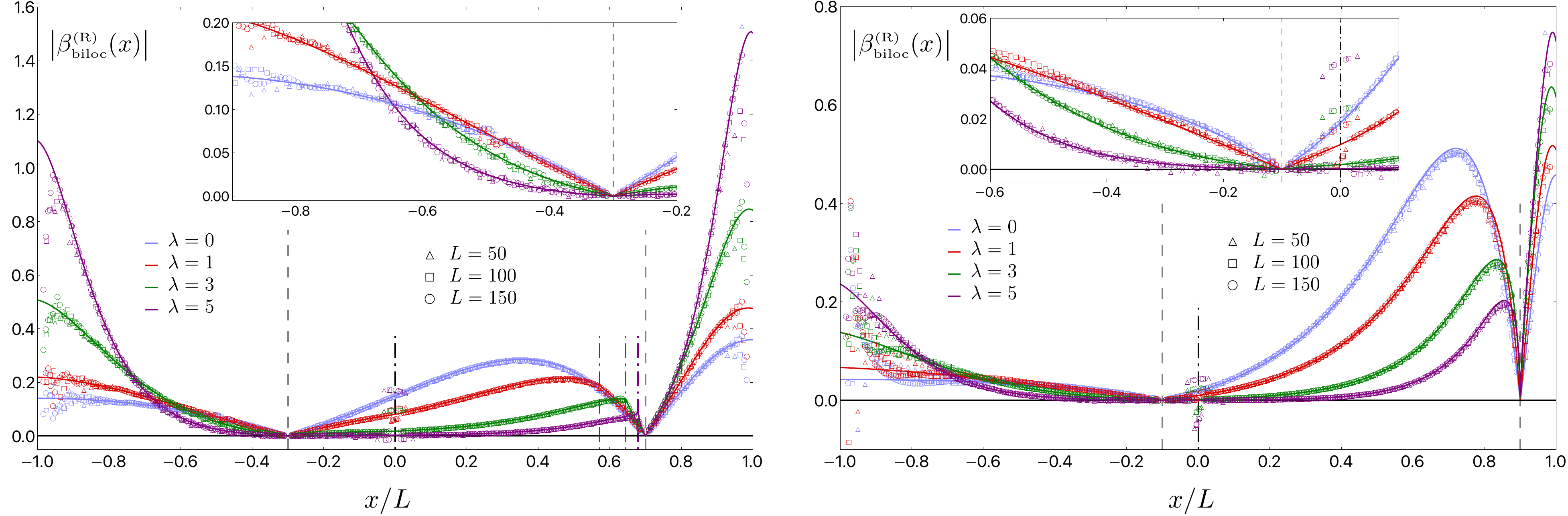}
\vspace{-.4cm}
\caption{
Weight functions of the local (top) and bilocal (bottom) term 
in $A \cup B$ for the rainbow chain,
obtained by employing (\ref{beta-loc-rainbow-chain-v1})-(\ref{averaging-beta}) 
and (\ref{beta-biloc-rainbow-chain})-(\ref{averaging-beta-bilocal}) respectively, 
also in $B$. 
The configurations (of type (c))
are the same ones considered in the bottom panels of Fig.\,\ref{fig-beta-loc-A} 
and Fig.\,\ref{fig-beta-biloc-A}.
The solid lines correspond to the analytic expressions for 
$|\beta_{\textrm{\tiny loc}}^{\textrm{\tiny (R)}}| / L$ 
and $|\beta_{\textrm{\tiny biloc}}^{\textrm{\tiny (R)}}|$ 
(see (\ref{beta-loc-rainbow-cft})-(\ref{beta-loc-rainbow-cft-V1}) and (\ref{beta-biloc-rainbow}) respectively).
The insets zoom in on the region around 
the first entangling point.
}
\label{fig-beta-AB}
\end{figure}

In Fig.\,\ref{fig-matrixplot-AB-case-cc} 
we show the elements of the matrices 
given by (\ref{peschel-formula}) for $A$ 
and by the analogous formula for $B$, where $C_A$ is replaced by $C_B$. 
The dashed grey lines separate the central block corresponding to $A$ (shown in Fig.\,\ref{fig-matrixplot-case-a} and Fig.\,\ref{fig-matrixplot-case-c})
from the other ones. 
The blocks corresponding to $B$ are the ones
whose elements are labelled  by position indices that both belong to $B$.
The remaining blocks, whose elements have one index in $A$ and the other one in $B$, have been filled with zeros. 
The dash-dotted green lines indicate  the centre  of the segment, where the singularity is located, like in Fig.\,\ref{fig-matrixplot-case-c}.
The choice of the bipartition and of the inhomogeneity parameter, described in the caption of Fig.\,\ref{fig-matrixplot-AB-case-cc}, 
implies that 
the central block in the left and right panel coincides with the top left and bottom left panel of Fig.\,\ref{fig-matrixplot-case-c} respectively. 
As we saw for $A$ in Fig.\,\ref{fig-matrixplot-case-a} and Fig.\,\ref{fig-matrixplot-case-c}, also in $B$ the largest elements are located on the odd diagonals around the main diagonal.
The position of the conjugate point (\ref{x-conj-tilde-rainbow}) corresponds to the black solid curve in Fig.\,\ref{fig-matrixplot-AB-case-cc}, and it
nicely captures a front of the off-diagonal matrix elements associated with $A$ and with $B$.
The effect of the inhomogeneity on the matrix elements of $K$ in (\ref{peschel-formula}), both for $A$ and for $B$, can be observed by comparing the two panels in Fig.\,\ref{fig-matrixplot-AB-case-cc}.
Note that, while for $x \in A$ the conjugate point (\ref{x-conj-tilde-rainbow}) tends to $b$ as the inhomogeneity grows, for $x \in B$ we have $x_{\textrm{c}, \textrm{\tiny R}}(x) \to -x$ as $\lambda \to \infty$ (see (\ref{xc-rainbow-large-h}) and Fig.\,\ref{fig-matrixplot-AB-case-cc}).

\begin{figure}[t!]
\vspace{-1.cm}
\hspace{-1.6cm}
\includegraphics[width=1.22\textwidth]{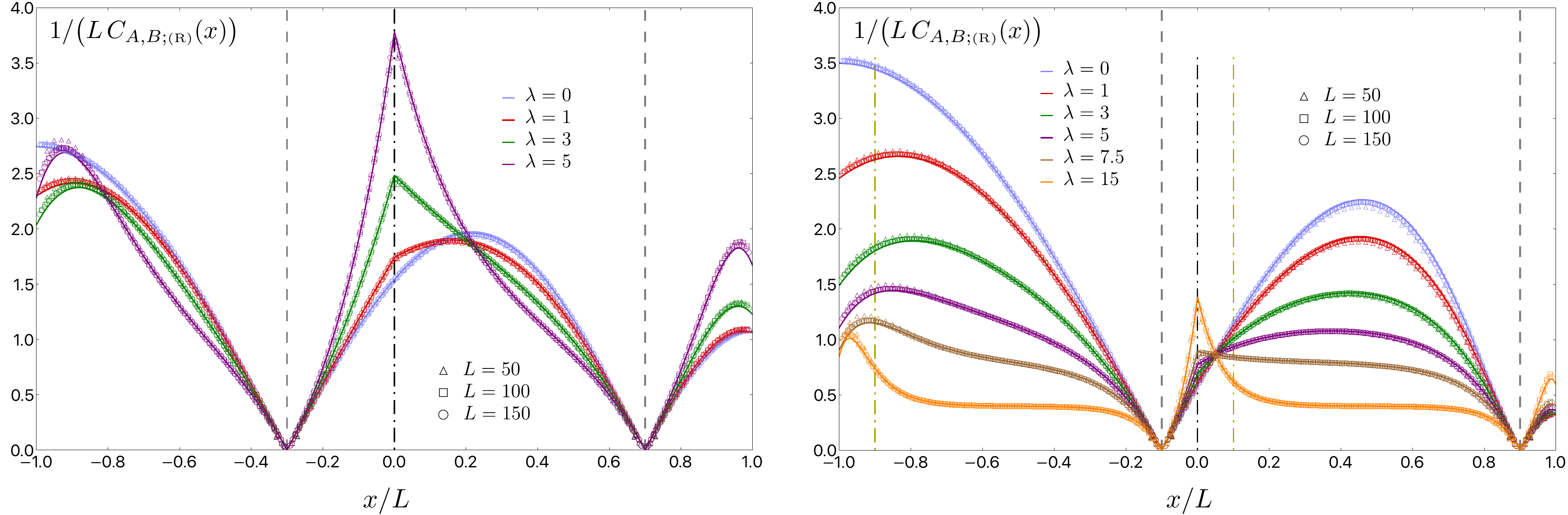}
\vspace{-.4cm}
\caption{
Contour function for the entanglement entropy in $A \cup B$ for the rainbow chain,
obtained from (\ref{contour-functions-lattice}) and (\ref{averaging-contour}) in $A$ and their straightforward extension to $B$,
for the configurations considered in Fig.\,\ref{fig-beta-AB}.
The results in $A$ are the ones in the bottom panels of Fig.\,\ref{fig-contour-A}.
The solid lines are provided by the analytic expression of $C_{A, B; \textrm{\tiny (R)}}(x) $, given by (\ref{contour-non-homo-rainbow-AB}) for $n=1$.
}
\label{fig-contour-AB}
\end{figure}

Regarding the weight functions of the entanglement Hamiltonian, 
in Fig.\,\ref{fig-beta-AB} we have reported some results obtained by 
extending to $B$ the expressions 
in (\ref{beta-loc-rainbow-chain-v1})-(\ref{averaging-beta}) 
for the local term (top panels)
and in (\ref{beta-biloc-rainbow-chain})-(\ref{averaging-beta-bilocal}) 
for the bilocal term (bottom panels), 
which have been previously employed for $A$.
In particular, in the top panels, 
the best agreement between 
the data points and the analytic predictions for the local term has been obtained by employing 
\eqref{beta-loc-rainbow-chain-v1} 
with \eqref{averaging-beta} in $A$
and \eqref{beta-loc-rainbow-chain-v2} 
with  \eqref{averaging-beta} in $B$.
Notice that the configurations of type (c) (see Fig.\,\ref{fig-configurations}) considered 
in the left and right panels of Fig.\,\ref{fig-beta-AB} 
are the same ones analysed in the bottom panels of Fig.\,\ref{fig-beta-loc-A} and Fig.\,\ref{fig-beta-biloc-A}; hence the same colour code has been adopted. 
The vertical dashed grey lines correspond to the entangling points  at $x=a$ and $x=b$,
while the vertical dash-dotted black line denotes the centre  
of the segment.
The dash-dotted yellow lines in the top right panel identify $A_\ast$ and $B_\ast$,
and the coloured dash-dotted lines in the bottom left panel have been already introduced in the bottom left panel of Fig.\,\ref{fig-beta-biloc-A}.
The solid lines in the top and bottom panels are given respectively by 
$|\beta_{\textrm{\tiny loc}}^{\textrm{\tiny (R)}}| / L$ 
from (\ref{beta-loc-rainbow-cft})-(\ref{beta-loc-rainbow-cft-V1})
and by $|\beta_{\textrm{\tiny biloc}}^{\textrm{\tiny (R)}}|$ 
from (\ref{beta-biloc-rainbow}), which are functions of $x/L$.
The curves and the data points for $\lambda =0$ correspond to the homogeneous case considered in Sec.\,\ref{sec-EH-homo}.
The agreement between the data points from the rainbow chain as $L$ increases and the analytic curves for $|\beta_{\textrm{\tiny loc}}^{\textrm{\tiny (R)}}| / L$ and 
$|\beta_{\textrm{\tiny biloc}}^{\textrm{\tiny (R)}}|$ is very satisfactory (see also the insets in Fig.\,\ref{fig-beta-AB}, 
that highlight 
the nice collapse of the data on the solid curves around the first entangling point), except for the regions close the boundaries of the segment, where the support of the unstable behaviour of the data points should vanish for larger values of $L$. 
We find it worth remarking that in the regime of large inhomogeneity, 
the expected plateau given by (\ref{beta-loc-rainbow-cft-large-lambda-case1}) is observed also in $B_\ast$ for large enough values of $\lambda$ (see also the bottom left panel of Fig.\,\ref{fig-rainbows-types}), 
as shown e.g. in the top right panel of Fig.\,\ref{fig-beta-AB} for  $\lambda = 15$.

We find it very instructive to also consider the contour functions for the entanglement entropies of the rainbow chain in the entire segment. 
These quantities have been obtained by using (\ref{contour-functions-lattice}), (\ref{averaging-contour}) and (\ref{averaging-contour-renyi}) in $A$
(see Fig.\,\ref{fig-contour-A} for the $n=1$ case),
and here we employ the analogous expressions adapted to $B$ in a straightforward way, as previously discussed. 
In the continuum limit, (\ref{contour-non-homo-rainbow-AB})  is the natural candidate 
for these quantities, for a generic value of the R\'enyi index.
In Fig.\,\ref{fig-contour-AB} we show the contour function for the entanglement entropy 
in the case of the configurations already considered in Fig.\,\ref{fig-beta-AB} 
(also the same colour code has been employed);
hence, in $A$ the data points coincide with the ones reported in the bottom panels of Fig.\,\ref{fig-contour-A}.
The agreement between the data points for the rainbow chain and the analytic expression for $C_{A, B; \textrm{\tiny (R)}}(x)$ (i.e. (\ref{contour-non-homo-rainbow-AB}) for $n=1$) is excellent 
along the entire segment. 
Notice that the oscillations of the data points observed in the top panels of Fig.\,\ref{fig-beta-AB} close to the endpoints of the segment
for the weight function of the local term 
do not occur for the contour function of the entanglement entropy. 
We remark that the average (\ref{averaging-contour}) 
plays a crucial role in finding 
such remarkable agreement with the analytic expressions in the continuum. 
Indeed, in the results for the contour function 
reported in \cite{Tonni:2017jom} for the case of a block adjacent to the boundary (where (\ref{averaging-contour}) was not employed),
oscillations are observed in the vicinity of the boundary.

\begin{figure}[t!]
\vspace{-1.cm}
\hspace{-1.6cm}
\includegraphics[width=1.22\textwidth]{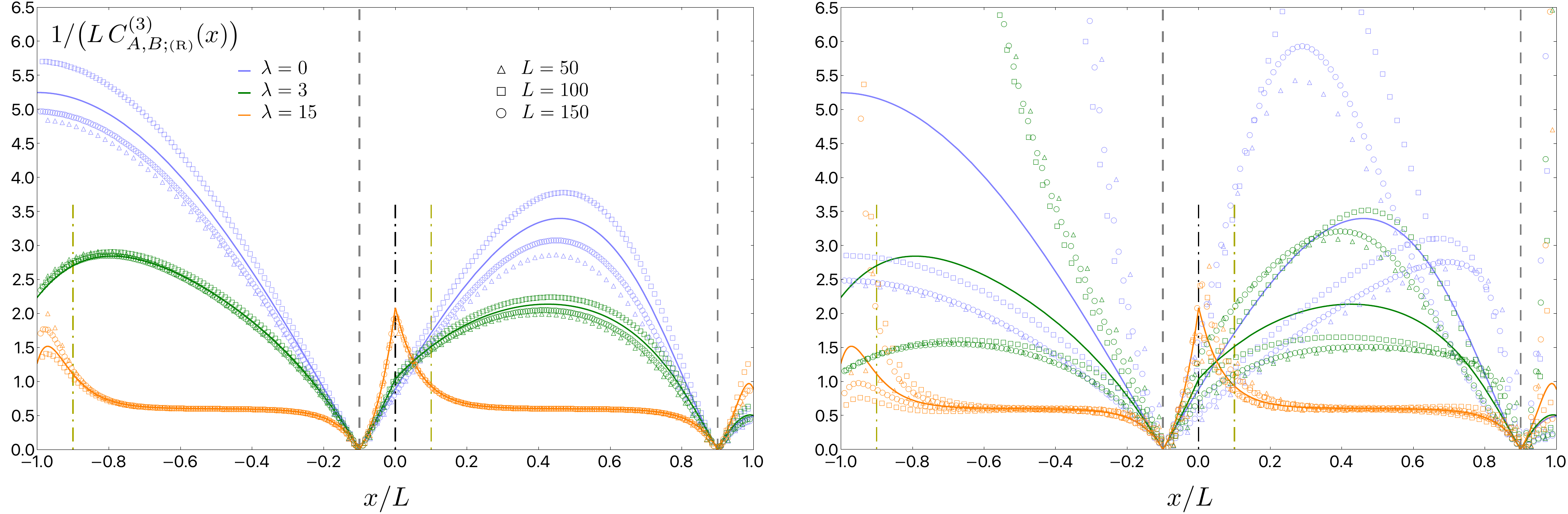}
\vspace{-.4cm}
\caption{
Contour function for the R\'enyi entropy with $n=3$   in $A \cup B$ for the rainbow chain,
for the same configuration 
considered in the right panel of Fig.\,\ref{fig-contour-AB}.
The quantities $\bar{\mathcal{C}}^{(3)}_A(i)$ (see (\ref{averaging-contour-renyi})) and $\mathcal{C}^{(3)}_A(i)$ (see (\ref{contour-functions-lattice})) 
for $i \in A \cup B$ are shown in the left and right panel respectively. 
The solid lines correspond to the analytic expression in (\ref{contour-non-homo-rainbow-AB}) for $n=3$.
}
\label{fig-contour-n3-AB}
\end{figure}

The crucial role of the averaging procedure given by (\ref{averaging-contour}) and (\ref{averaging-contour-renyi}) 
is evident when the contour functions of the R\'enyi entropies are considered. 
In Fig.\,\ref{fig-contour-n3-AB} we report our numerical data points for the contour function of the R\'enyi entropy with $n=3$.  
In particular, in the left and right panel of this figure
we show $\bar{\mathcal{C}}^{(3)}_A(i)$ 
(see (\ref{averaging-contour-renyi})) and $\mathcal{C}^{(3)}_A(i)$ 
(see (\ref{contour-functions-lattice})) respectively, 
extended to the entire segment in the 
straightforward way. 
The bipartition is the same one
considered in the right panel of Fig.\,\ref{fig-contour-AB} to facilitate the comparison 
with the $n=1$ case. 
This gives some insights into the $n$-dependence of the contour functions for the entanglement entropies.
Large oscillations occur in the data points for $\mathcal{C}^{(n)}_A(i)$ with $n \neq 1$, 
whose amplitude significantly increases with $n$, especially in the regions close to the boundaries of the segment. 
This has been observed also in 
Fig.\,11 of \cite{Tonni:2017jom}
for the case of a block adjacent to the boundary.
By comparing the two panels of 
Fig.\,\ref{fig-contour-n3-AB}, it is evident that the averaging procedure introduced in (\ref{averaging-contour-renyi}) 
leads to a significant improvement in finding agreement 
with the analytic expression (\ref{contour-non-homo-rainbow-AB}).
For both $\mathcal{C}^{(n)}_A(i)$ and $\bar{\mathcal{C}}^{(n)}_A(i)$, the oscillations increase 
towards the boundaries and for increasing values of $n$, but they are suppressed as the inhomogeneity grows. 
Indeed, the largest oscillations are observed in the homogeneous case.
It would be very insightful 
to find analytic results for these oscillations.

\section{Conclusions}
\label{sec-conclusions}

In this paper we explored the entanglement Hamiltonian and 
the contour functions for the entanglement entropies 
for the massless Dirac field on a segment
with the same boundary condition imposed at both its endpoints,
in the class of spatially inhomogeneous backgrounds given by (\ref{metric})
and in its ground state,
when the segment is bipartitioned by an interval $A$
in a generic position (see Fig.\,\ref{fig-interval-segment}).
The analytic results for these quantities
are reported in Sec.\,\ref{sec-EH-homo} 
for the homogeneous case 
and in Sec.\,\ref{subsec-generic-background} for a generic background.
The latter expressions have been specialised to the rainbow model in Sec.\,\ref{subsec-CFT-rainbow-model}, 
and in Sec.\,\ref{sec-EH-rainbow-chain}
they have been checked against 
the continuum limit of the corresponding exact numerical results 
in the rainbow chain (\ref{ham-rainbow-chain}), 
for the configurations in Fig.\,\ref{fig-configurations}.
In the limiting case where $A$ is adjacent to the boundary, 
the results of \cite{Tonni:2017jom} are recovered.

The massless Dirac field on a segment is a BCFT with $c=1$. 
Since the same boundary condition is imposed at both endpoints,
we have employed the entanglement Hamiltonian of an interval in the half line 
found in \cite{Mintchev:2020uom}.
Our results for  the homogeneous case are discussed
in Sec.\,\ref{sec-EH-homo}.
Combining the conformal map (\ref{map-strip-to-rhp}) 
with the weight functions found in \cite{Mintchev:2020uom}, 
we have obtained the entanglement Hamiltonian 
(\ref{EH-segment-homo}) with the weight functions 
(\ref{beta-loc-homo})-(\ref{beta-loc-homo-v1}) 
and (\ref{beta-biloc-homo}) 
(or (\ref{beta-biloc-homo-v2}) or (\ref{beta-biloc-homo-v1}) equivalently)
for the local and bilocal term respectively. 
From the weight function of the local term, we have obtained 
the contour function for the entanglement entropies 
in $A$ and in the entire segment
(see (\ref{contour-function-cft-homo}) 
and (\ref{contour-function-cft-homo-AB}) 
respectively).
The corresponding entanglement entropies are reported in (\ref{renyi-entropies-homo}).

These analytic results for the homogeneous case have been employed to find 
the same quantities for the class of inhomogeneous backgrounds defined by (\ref{metric}).
In particular, the entanglement Hamiltonian of the interval is (\ref{EH-segment-in-homo}), where the weight functions are given by either
(\ref{beta-loc-non-homo}) or (\ref{beta-loc-non-homo-v1}) for the local term 
and by (\ref{beta-biloc-non-homo-V1})-(\ref{beta-biloc-non-homo-V2}) 
for the bilocal term.
In the case of these inhomogeneous backgrounds, we have written 
a candidate expression for the contour functions for the entanglement entropies 
by employing the weight function of the local term
(see (\ref{contour-non-homo-A}) in $A$ 
and (\ref{contour-non-homo-sigma-AB}) in the entire segment).
This provides the expression (\ref{renyi-entropies-non-homo}) for 
the entanglement entropies,
first found in \cite{Rodriguez-Laguna:2016roi} 
through the twist fields method.

In the second part of the paper, 
our analytic results have been specialised 
to the rainbow model (see Sec.\,\ref{subsec-CFT-rainbow-model} and Sec.\,\ref{sec-EH-rainbow-chain}), 
whose Weyl factor is given by (\ref{sigma-rainbow}).
In this model, the weight functions 
for the local and bilocal term  become
(\ref{beta-loc-rainbow-cft}) 
(or equivalently (\ref{beta-loc-rainbow-cft-V1}))
and (\ref{beta-biloc-rainbow}) respectively. 
Considering the bipartition in Fig.\,\ref{fig-interval-segment}
of the rainbow chain (\ref{ham-rainbow-chain}) in its ground state, 
in Sec.\,\ref{sec-EH-rainbow-chain} we studied the continuum limit of the entanglement Hamiltonian 
and of the contour functions for the entanglement entropies, 
finding a remarkable agreement with the analytic expressions 
obtained in Sec.\,\ref{subsec-CFT-rainbow-model}.
The matrix (\ref{peschel-formula}) determining the entanglement Hamiltonian 
(\ref{EH-A-lattice}) on the lattice
displays the inhomogeneous and the non-local nature of this operator,
as shown in Fig.\,\ref{fig-matrixplot-case-a} and Fig.\,\ref{fig-matrixplot-case-c} for $A$  and in Fig.\,\ref{fig-matrixplot-AB-case-cc} for the entire segment.
The method that we have employed to study the continuum limit 
is based on the analysis performed in \cite{Eisler:2022rnp} 
for entanglement Hamiltonians in homogeneous fermionic chains 
(either on the line or on the half line),
which contain also a bilocal term. 
Our numerical results for the weight function of the local and bilocal term are shown in Fig.\,\ref{fig-beta-loc-A} 
and Fig.\,\ref{fig-beta-biloc-A} respectively, 
and those for the contour functions 
and the corresponding entanglement entropies 
in Fig.\,\ref{fig-contour-A} 
and Fig.\,\ref{fig-entropy-A} respectively. 
These quantities have been explored also in the entire segment, obtaining
the results reported in Fig.\,\ref{fig-beta-AB}, Fig.\,\ref{fig-contour-AB} and Fig.\,\ref{fig-contour-n3-AB}.
The agreement between the analytic expressions and the numerical data for these quantities in $B$ provides evidence that the analytic formulas 
derived in Sec.\,\ref{sec-EH-homo} and Sec.\,\ref{sec-EH-non-homo} are also valid when $x \in B$. 
As for the numerical approach to the continuum limit,  
it would be interesting to further explore the differences between 
(\ref{beta-loc-rainbow-chain-v1}) and (\ref{beta-loc-rainbow-chain-v2}) 
for the weight function of the local term in $x \in B$. 
It would also be insightful to understand why 
the lattice formula for the contour function for the entanglement entropy
captures the CFT prediction in a very nice way, 
whereas the ones employed for the weight function of the local term 
display oscillations in $B$ near the boundaries of the segment 
(see Fig.\,\ref{fig-contour-AB} and the top panels of Fig.\,\ref{fig-beta-AB}).

Our results confirm the interesting relation between 
the reduced density matrix of some bipartitions 
obtained from the ground state of the rainbow chain 
in the regime of large inhomogeneity 
and the density matrix of a thermal state in the subsystem
\cite{Vitagliano:2010db,Ram_rez_2014,Ramirez:2015yfa,
Rodriguez-Laguna:2016roi} (see Fig.\,\ref{fig-rainbows-types}).
Indeed, the weight function of the local term can develop a plateau (\ref{beta-loc-rainbow-cft-large-lambda-case1}) in this regime,
as observed in bottom right panels of 
Fig.\,\ref{fig-beta-loc-A} and Fig.\,\ref{fig-contour-A},
in the top right panel
of Fig.\,\ref{fig-beta-AB},
and in the right panels 
of Fig.\,\ref{fig-contour-AB} and Fig.\,\ref{fig-contour-n3-AB}.
Interestingly, the weight function of the bilocal term 
can also display a plateau in the regime of large inhomogeneity
(see e.g. (\ref{beta-biloc-rb-large-h}) 
and the top right panel of Fig.\,\ref{fig-beta-biloc-A}),
which however does not admit a thermal interpretation, as discussed in Sec.\,\ref{subsec-CFT-rainbow-model}.
In Fig.\,\ref{fig-contour-n3-AB},
where the contour function for the R\'enyi entropy with $n=3$ is considered, 
the improvement provided by the averaging defined 
in (\ref{averaging-contour-renyi}) is highlighted. 
The same figure shows the relevant role played by oscillatory terms 
in the lattice data points, and it would be insightful to describe them analytically. 
We remark that the numerical methods employed in this work 
and in \cite{Tonni:2017jom}
are qualitatively different; 
hence, it could be interesting to perform 
a systematic comparison between them. 

The results discussed in this work suggest various directions for future studies.

It would be interesting to apply the method discussed in \cite{Arias:2016nip,Eisler:2017cqi, Eisler:2018ugn,DiGiulio:2019cxv,Eisler:2019rnr,Eisler:2022rnp}
to study the entanglement Hamiltonians in
other inhomogeneous free models \cite{Eisler:2017iay,Gruber:2019mtr,Eisler:2019gnr,Gruber:2020afu,Eisler:2021uzt,Capizzi:2022igy,Eisler:2022hej,Eisler:2023yys,Bonsignori:2024gky,Eisler:2024okk,Bernard:2024pqu,
Crampe:2019upj,Finkel:2020lgf,Finkel:2021gji,Cramp_2021,Bernard:2022dik,Bernard:2024ieo,Blanchet:2024hzb,Bernard:2024eoj,Bernard:2025mnr, Dubail_2017,Brun:2017aam,Brun_2018,Bastianello:2019ovv},
considering with particular attention
the entanglement Hamiltonians  
that contain also non-local terms
\cite{Casini:2009vk,Blanco:2019xwi,Fries:2019ozf,
Mintchev:2020jhc, Mintchev:2020uom,Eisler:2020lyn, Bostelmann:2022yvj, Baranov:2024vru,Gentile:2025koe,Rottoli:2022plr} 
and the limiting regimes where 
they become local.
The extension of these analyses to interacting models 
would be of great relevance 
(see e.g. the numerical results in \cite{ParisenToldin:2018uzz}).
In the case of free fermionic chains, 
it is important to explore also the role played by 
the filling parameter in the entanglement Hamiltonian 
(see e.g. \cite{Eisler:2022rnp, Mula:2022lsj}).
In these setups, it would also be worth investigating  
other interesting quantities related to the entanglement Hamiltonian.
These include e.g. the entanglement spectra \cite{Peschel:1999xeo,Chung_2000,Chung:2001zz,Peschel:2003rdm,Peschel:2004qkm,Cardy:2016fqc,Lauchli:2013jga,Poilblanc_2010,L_uchli_2012,Schliemann_2012,Calabrese:2008iby,Alba:2017bgn,Surace:2019mft},
important measures of bipartite entanglement in mixed states such as the logarithmic negativity \cite{Peres:1996dw, Vidal:2002zz, Plenio:2005cwa, Calabrese:2012ew, Calabrese:2012nk, Calabrese:2014yza, Calabrese:2013mi, Eisler:2015tgq, Coser:2015eba, Coser:2015mta, Shapourian:2016cqu, Eisler:2016bfd, DeNobili:2016nmj}, 
and other quantities probing various features of entanglement \cite{Mula:2020udv,Mula:2022lsj,Mula:2022akc, Langlett:2021efq}.
The latter class includes also alternative expressions for the spatial density of the entanglement entropy, like e.g. the one discussed in \cite{SinghaRoy:2019urc}.
The aforementioned entanglement Hamiltonians containing 
non-local terms should be explored 
also in more complicated models (see e.g. \cite{Caraglio:2008pk, Furukawa:2008uk, Calabrese:2009ez, Calabrese:2010he, Coser:2013qda, Grava:2021yjp, Agon:2013iva, DeNobili:2015dla} for the entanglement entropies).
It is important to study entanglement Hamiltonians 
for inhomogeneous models
also  in higher dimensions 
\cite{Javerzat:2021hxt,Ramirez:2015yfa,Boada:2010sh, 
Huerta:2022tpq, Huerta:2023dqt, Eisler:2023yys,Bernard:2023ijj,deBuruaga:2025kja}
and in out of equilibrium scenarios
\cite{Cardy:2016fqc, Wen:2018svb, DiGiulio:2019lpb, Rottoli:2022ego, Bonsignori:2025whl}.
Finally, pivotal insights could arise from the analysis of entanglement Hamiltonians and related quantities in the context of the gauge/gravity correspondence
(see e.g. \cite{Ryu:2006bv, Ryu:2006ef,Hubeny:2007xt,Freedman:2016zud, Agon:2018lwq, MacCormack:2018rwq, Tonni18:bariloche-talk, Kudler-Flam:2019oru, Mintchev:2022fcp, Caggioli:2024uza}).

\vskip 20pt 
\centerline{\bf Acknowledgments} 
\vskip 5pt

It is our pleasure to thank 
Viktor Eisler, Mihail Mintchev
and Javier Rodr\'iguez-Laguna 
for useful discussions.  
We are especially grateful to Sara Pasquetti for insightful conversations 
and thoughtful mentorship of ST during his master’s thesis project.
ET acknowledges the Isaac Newton Institute (Cambridge) 
for hospitality and financial support 
during the last part of this work,
within the program {\it Quantum field theory with boundaries, impurities, and defects}.
This work was funded by the European Union - NextGenerationEU, Mission 4, Component 2, Inv.1.3, in the framework of the PNRR Project National Quantum Science and Technology Institute (NQSTI) PE00023; CUP: G93C22001090006.
The project that gave rise to these results received the support of a fellowship from ``la Caixa” Foundation (ID 100010434), awarded to ST, with code LCF/BQ/DI24/12070002.

\newpage

\bibliographystyle{nb}

\bibliography{refsEHrainbow}

\end{document}